\documentclass[a4paper,11pt]{article}
\usepackage{geometry}
\usepackage{aas_macros}
\usepackage{ragged2e}
\usepackage{pifont}
\usepackage{doi}
\newcommand{\cmark}{\ding{51}}%
\newcommand{\checked}{\rlap{$\square$}{\raisebox{2pt}{\large\hspace{1pt}\cmark}}%
\hspace{-2.5pt}}

\usepackage{graphicx}
\usepackage{amssymb}
\usepackage{amsmath}    % Advanced maths commands
\usepackage{epstopdf}
\usepackage{xcolor}
\usepackage{eurosym}
\usepackage{tabularx}
\usepackage{paralist}
\usepackage{hyperref}
\usepackage[font={small}]{caption}
\usepackage[T1]{fontenc}
\usepackage{sfmath}
\usepackage{textgreek}
\usepackage{multicol}
\usepackage{paralist}
\usepackage{helvet}
\usepackage{microtype}
\usepackage{xspace}
\usepackage[version=4]{mhchem}

\hypersetup{colorlinks=true,linkcolor=blue,citecolor=blue,filecolor=blue,urlcolor=blue}

\usepackage{titlesec}
\titlespacing{\section}{0pt}{*1}{*0}
\titlespacing{\subsection}{0pt}{*1}{*0}
\titlespacing{\subsubsection}{0pt}{*1}{*0}
\titlespacing{\paragraph}{0pt}{*0}{*0}

\usepackage[square,numbers,sort&compress,comma]{natbib}
\newcommand{\HII}{H\,\textsc{ii}\xspace}
\newcommand{\eg}{e.\hspace{0.125em}g.\xspace}

\newcommand{\Gaia}{\textit{Gaia}\ }

\begin{document}

%----------------------------------------------------------------------------------------
%       TITLE PAGE
%----------------------------------------------------------------------------------------

\begingroup
\thispagestyle{empty}

\centerline{\huge A Collection of German Science Interests in the  }
\vspace*{0.5\baselineskip}

\centerline{\huge Next Generation Very Large Array }\par
% Book title
\vspace*{0.5\baselineskip}

%\vspace*{1cm}
\centerline{\today ~(Version 2.0)}

\vspace*{1cm}

\noindent 
{\Large M. Kadler$^{1}$, D.~A. Riechers$^{2}$, J. Agarwal$^{3}$, A.-K. Baczko$^{4,24}$, H. Beuther$^{5}$, F. Bigiel$^{6}$, T. Birnstiel$^{7}$, B. Boccardi$^{4}$, D. J. Bomans$^{8}$, L. Boogaard$^{5}$, T. T. Braun$^{8}$, S. Britzen$^{4}$, M. Brüggen$^{9}$, A. Brunthaler$^{4}$, P. Caselli$^{10}$, D. Elsässer$^{11}$, S. von Fellenberg$^{4}$, M. Flock$^{5}$, C. M. Fromm$^{1}$, L. Fuhrmann$^{12}$, P. Hartogh$^{13}$, M. Hoeft$^{14}$, R. P. Keenan$^{5}$, Y. Kovalev$^{4}$, K. Kreckel$^{15}$, J. Livingston$^{4}$, A. P. Lobanov$^{4}$, H. Müller$^{4}$,  E. Ros$^{4}$, P. Schilke$^{2}$, M. De Simone$^{16}$, L. Spitler$^{4}$, T. Ueda$^{5}$, E. Vardoulaki$^{14}$, S. Vegetti$^{17}$, K. Weis$^{8}$, C. Wendel$^{1}$, M. H. Xu$^{18}$, G.-Y. Zhao$^{4}$, M. Albrecht$^{12}$, A. Basu$^{14}$, J. Becker Tjus$^{8}$, S. Bernhart$^{19,4}$, J. Blum$^{3}$, E. Bonnassieux$^{1}$, C. Bredendiek$^{12}$, M. van Delden$^{8}$, G. Di Gennaro$^{9}$, A. Enders$^{8}$, F. Eppel$^{1,4}$, H. Hase$^{20}$, D. Hoang$^{9}$, U. Hugentobler$^{21}$, M. Kaasinen$^{16}$, N. Krupp$^{13}$, E. Kun$^{8,25,26}$, M. Laubach$^{12}$, Y. Lin$^{10}$, K. Mannheim$^{1}$, K. M. Menten$^{4}$, R. Perkuhn$^{12}$, N.  Pohl$^{8,12}$, D. M. Powell$^{17}$, L. Rezzolla$^{22}$, L. Ricci$^{1}$, E. Schinnerer$^{5}$, K. Schmidt$^{11}$, J.  Schöpfel$^{8}$, S. Stanko$^{12}$, M. Stein$^{8}$, N. Sulzenauer$^{4}$, S. Taziaux$^{8}$, A. Tursunov$^{4}$, F. Walter$^{5}$, A. Weiß$^{4}$, G. Witzel$^{4}$, S. Wolf$^{23}$, J. A. Zensus$^{4}$
}

\vspace*{\baselineskip}
\begin{minipage}{0.9\textwidth}
\noindent 
{\small $^{1}$JMU, Würzburg,
$^{2}$Univ. zu Köln, 
$^{3}$TU Braunschweig,
$^{4}$MPIfR, Bonn,
$^{5}$MPIA, Heidelberg,
$^{6}$AIfA, Univ. Bonn,
$^{7}$LMU, Munich,
$^{8}$RUB, Bochum,
$^{9}$Univ. Hamburg,
$^{10}$MPE, Garching,    
$^{11}$TU Dortmund,
$^{12}$Fraunhofer FHR,
$^{13}$MPS Göttingen,
$^{14}$TLS, Tautenburg,
$^{15}$Univ. Heidelberg,
$^{16}$ESO, Garching,
$^{17}$MPA, Garching,
$^{18}$GFZ, Potsdam,
$^{19}$Reichert GmbH, Bonn,
$^{20}$BKG Wettzell, Bad Kötzting,
$^{21}$TUM Munich,
$^{22}$Goethe Univ. Frankfurt,
$^{23}$Univ. Kiel
}
\end{minipage}

%\vspace*{\baselineskip}

\newpage
\noindent 
\centerline{\normalsize
With international contributions by }
%\vspace*{0.5\baselineskip}

\noindent 
{\Large
A. Mus$^{27}$, L. V. Toth$^{28}$,  A. Alberdi$^{29}$, M. Benisty$^{30}$,  P. Cox$^{31}$, J. C. Guirado$^{29}$, M. D. Johnson$^{32}$, M. Juvela$^{33}$, M. Neeleman$^{34}$, I. N. Pashchenko$^{35,36}$, M. Á. Pérez Torres$^{29}$, K. Perraut$^{37}$, M. Zaja\v{c}ek$^{38}$
}

\vspace*{\baselineskip}
\begin{minipage}{0.9\textwidth}
\noindent 
{\small 
$^{24}$Chalmers Univ. of Technology, Gothenburg, Sweden,
$^{25}$Konkoly Obs.,  Hungary,
$^{26}$CSFK, Hungary,
$^{27}$Univ. de Val\`encia,
$^{28}$Eötvös Univ. Budapest,
$^{29}$Inst. de Astrof\'{\i}sica de Andaluc\'{\i}a, Spain,
$^{30}$IPAG, France,
$^{31}$IAP, France,
$^{32}$CfA Harvard \& Smithsonian, USA,
$^{33}$Univ. Helsinki,
$^{34}$National Radio Astronomy Observatory, USA,
$^{35}$Lebedev Physical Institute, Russia,
$^{36}$Moscow Inst. of Physics and Technology, Russia,
$^{37}$Univ. Grenoble Alpes, France,
$^{38}$Masaryk Univ., Czech Republic
}
\end{minipage}

%\newpage

%\setcounter{tocdepth}{2}
%\tableofcontents

\newpage

\pagenumbering{arabic}
\setcounter{page}{1}

\newpage 
\section{A Foreword by the Curators of this Community Paper}
\noindent
{\textbf{Matthias Kadler (JMU W\"urzburg) \& Dominik Riechers (Univ. K\"oln)}}

\vspace*{2\baselineskip}

\noindent
 The Next Generation Very Large Array (ngVLA) project (Fig.~\ref{fi1}),\footnote{Reference design:\ {\tt https://ngvla.nrao.edu/page/projdoc}} developed by the international astronomy community under the lead of the National Radio Astronomy Observatory (NRAO; \cite{murphy18}), has recently been identified as one of the top-priority new ground-based observatories in the 2020 US decadal report "Pathways to Discovery in Astronomy and Astrophysics for the 2020s"\footnote{\tt https://nap.nationalacademies.org/resource/26141/interactive/}. As currently designed, it is expected to be a transformative scientific instrument and a key facility for the US astronomical community to be constructed by the next decade.  In terms of scientific capabilities, the ngVLA will be a key complement to facilities targeting lower radio frequencies such as the Square Kilometre Array (SKA). Presently, the project is looking to provide opportunities for international cooperation. Given the long-standing tradition of significant usage of NRAO facilities by the German community (see Fig.~\ref{fi2}), Germany would be a natural partner, but any such potential engagement requires further exploration. \\
 
 In response to this opportunity, an exploratory workshop\footnote{\tt https://events.mpifr-bonn.mpg.de/indico/event/306/}  for members of the German community interested in the ngVLA project was held at the Max-Planck-Institute for Radio Astronomy (MPIfR) on 14 December 2022 in Bonn. This workshop was met with great interest and has brought forward a number of interesting ideas. A second, equally stimulating workshop\footnote{\tt https://events.mpifr-bonn.mpg.de/indico/event/339/} was held at the Max-Planck-Institute for Mathematics in the Sciences in Leipzig on 27-28 September 2023. 
 
 Given the considerable participation by the German community in both workshops, it was decided to collect the science interests of this community through short contributions. These contributions are collected in this volume. 
  The main purpose of this volume is to demonstrate the breadth of the current interest across the German community in the facility as a basis for further discussion.
 %This document is supposed to be a `living document', i.e. it can and should be updated with additional science cases in the future.\footnote{Contributors are asked to use this template:\  \tt https://www.overleaf.com/project/651bde92dbe7c6e9e8308574} Interested authors should contact us if additional science topics should be included that are not represented in this volume. Also, 
 This document is not meant to be a "Science Book". It is rather a consolidation of goals and ideas. Among the science cases in the existing ngVLA Science Book, 18 already involved German researchers, a subset of which are also represented in this document. We encourage authors to expand any new ideas from this document  into full science cases for the ngVLA Science Book, in collaboration with our US colleagues. E.g., it was noted that the German community has a particular interest in very-long-baseline science cases that are currently not captured prominently in the ngVLA Science Book. 
 %At the same time, we expect future contributions in additional science areas included in the ngVLA Science Book, but not yet represented in this volume.
 %On the other hand, additional science cases with leading contributions of German researchers are already presented in the ngVLA Science Book\footnote{11 science cases involving German researchers.} but not replicated in the present version of this document.
% As such, the main purpose of this volume is to demonstrate the breadth of the current interest across the German community in the facility  as a basis for further discussion.

 The current timeline for the ngVLA, as presented at the 2023 workshop by NRAO, foresees delivery of the first prototype antenna, built by the Germany-based company {\em mtex antenna technology,} at the testing site in New Mexico in 2024. Construction of the array is anticipated to start in 2027/28, with Early Science opportunities in the early 2030s, and completion of construction and Full Science Operations around 2037. \\

\noindent
 \textsl{This is version 2.0 of the German ngVLA science-interests community paper. It contains 7 additional science cases that were not yet presented in version 1 (published on Nov 16, 2023).}
 
 \begin{figure}[h]
    \centering
	\includegraphics[angle=0.0,width=1.0\textwidth]{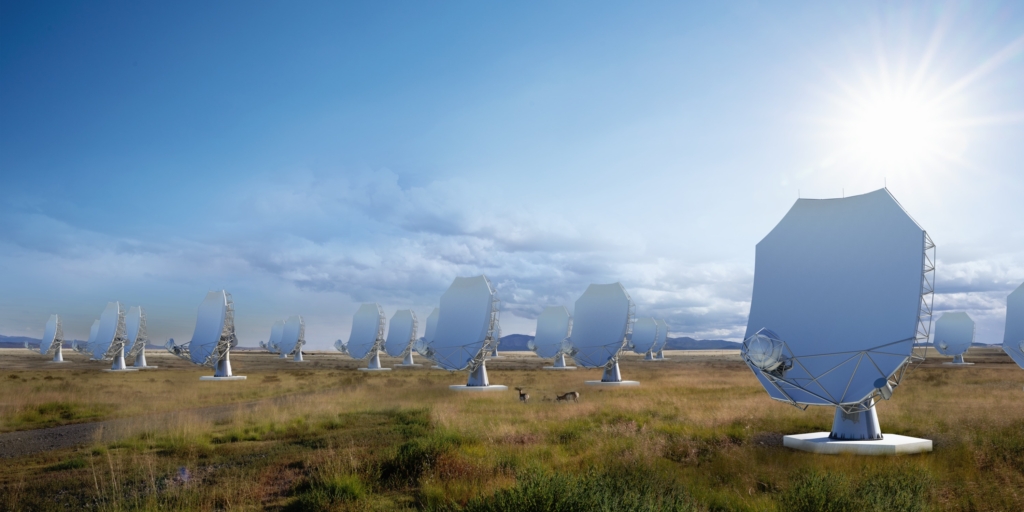}
    \vspace{-6mm}
	\caption{Artist's conception of the {\em mtex} design for the ngVLA prototype antenna (credit: Sophia Dagnello, NRAO/AUI/NSF). The array will consist of three components:\ a main interferometric array with 214 $\times$ 18\,m reflector antennas covering baselines up to 1068\,km, a short baseline array with 19 $\times$ 6\,m antennas covering the shortest baselines, and a long baseline array with 30 $\times$ 18\,m antennas covering baselines up to 8860\,km. It will be deployed in the southern US and northern Mexico, with long baselines reaching up to Canada and down to Puerto Rico and the US Virgin Islands. This will be sufficient to achieve sub-milliarcsecond resolution. The array will operate between wavelengths of 21\,cm and 3\,mm.}
	\label{fi1}
\end{figure}

\begin{figure}[h]
    \centering
	\includegraphics[angle=0.0,width=1.0\textwidth]{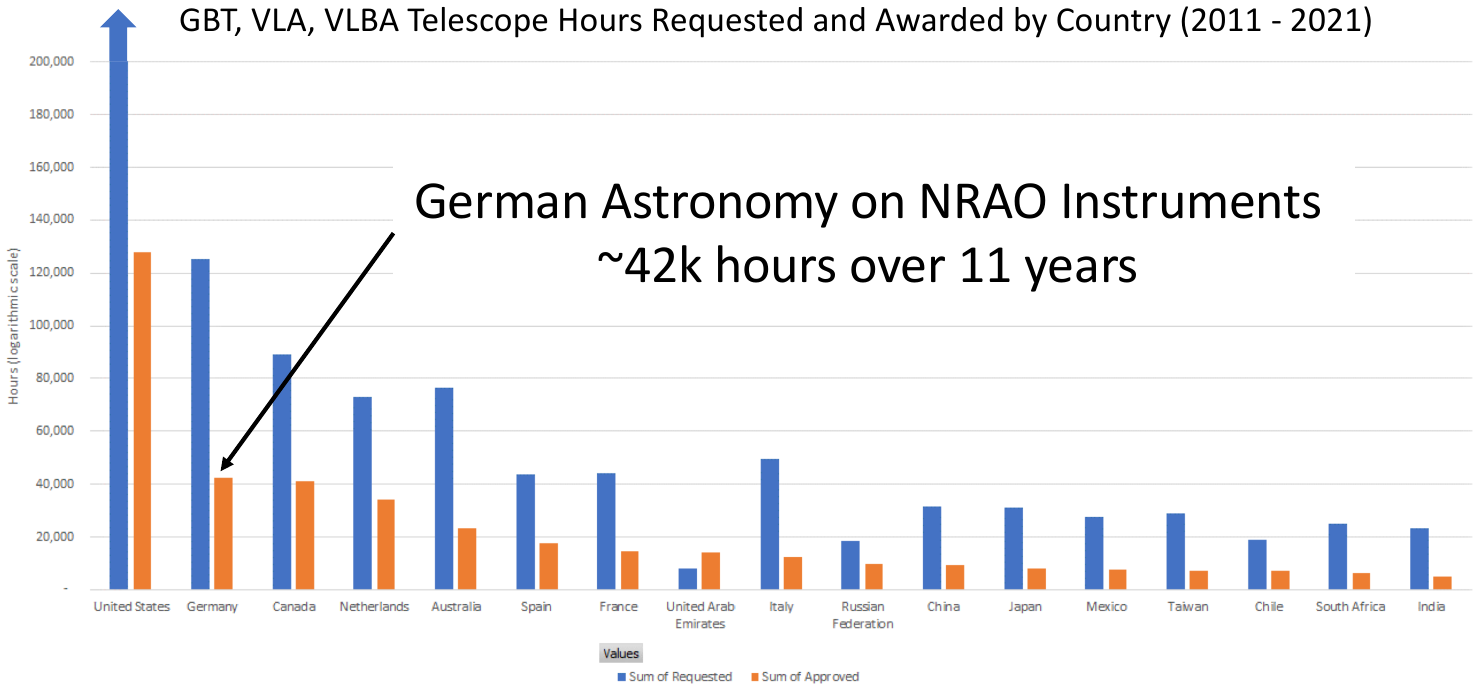}
    \vspace{-6mm}
	\caption{International usage of NRAO facilities for the years 2011--2021, kindly provided by NRAO (figure sourced from a presentation given by Tony Beasley at the 2023 workshop). Astronomers located in Germany are the most prolific users outside the US through the "open skies" agreement, demonstrating the major interest in these facilities in the German community. The ngVLA will be the natural evolution of both the Karl G. Jansky Very Large Array (VLA) and the Very Long Baseline Array (VLBA), and thus, is of major interest to astronomers located in Germany.}
	\label{fi2}
\end{figure}

\clearpage
\newpage

\section{Science Cases}
\small 

\noindent
\textbf{Topical Areas of Science Cases}

\noindent
1: Stellar Astrophysics,
2: Solar System, Planetary Systems and Habitability,
3: Circuit of Cosmic Matter (incl. star formation),
4: The Galaxy and the Local Group,
5: Galaxies and AGN,
6: Cosmology, Large Scale Structure and Early Universe,
7: Extreme conditions in the cosmos, fundamental astrophysics,
8: Interdisciplinary research and technology 

\vspace*{\baselineskip}

   \noindent 
\begin{tabular}{@{}l@{~}p{8.7cm}c@{}c@{}c@{}c@{}c@{}c@{}c@{}c@{}c@{}}
  \hline
 Principal & Science Case    & Page & \multicolumn{8}{c}{Topical Area} \\
   Author & &  &  1 & 2 & 3 & 4 & 5 & 6 & 7 & 8\\
  \hline
Agarwal & From cometary dust to planetary atmospheres: solar system astronomy at radio wavelengths &  \pageref{agarwal01} & & \checkmark &  & &  &\\
Baczko & AGN jets at all scales & \pageref{baczko01} & & &  & & \checkmark &\\  
Beuther & Small-scale structure of the warm and cold neutral medium (WNM \& CNM) with HI emission and HI self-absorption (HISA) & \pageref{beuther01} &  & & \checkmark & \checkmark & \checkmark & \\
Bigiel & HI Studies of Nearby Galaxies & \pageref{bigiel01} & & & \checkmark & \checkmark & \checkmark & \\
Birnstiel & Imaging Young Solar System Analogues in Formation & \pageref{birnstiel01} & & \checkmark & \\
Boccardi & The acceleration and collimation zone in AGN jets & \pageref{boccardi01} & & & & & \checkmark &  \\
Bomans & Compact Starburst Dwarf Galaxies & \pageref{bomans01} & & & \checkmark & & \checkmark & \checkmark  \\
Boogaard & Redshift Measurements & \pageref{boogaard01} & & & & & \checkmark & \checkmark  \\
Boogaard & Kinematics of cold gas in high-redshift galaxies & \pageref{boogaard02} & & & \checkmark & & \checkmark & \checkmark  \\
Braun & Integrated Frequency Synthesizers for the Band 6 ngVLA Receiver & \pageref{braun01} & & & & &  &  & & \checkmark  \\
Britzen & Detecting Supermassive Binary Black Holes & \pageref{britzen01} & & & & & \checkmark & \checkmark & \checkmark  \\
Brüggen & Radio studies of the cosmic web   & \pageref{brueggen01} &  & & & & & \checkmark &  \\
Brunthaler & Galactic Structure   & \pageref{brunthaler01} &  & & & \checkmark &  \\
Caselli & Chemical and physical conditions at the dawn of star and planet formation   & \pageref{caselli01} &  & \checkmark & & &  \\
Elsässer & Gravitational-wave triggers and other energetic transient events  & \pageref{elsaesser01} & & &  & & \checkmark & & \checkmark \\
von Fellenberg & Hyper Compact Radio sources in the Galactic Center & \pageref{vonFellenberg01} & \checkmark & &  & \checkmark & & & \checkmark \\
Flock & Substructures of dust rings in protoplanetary disks & \pageref{flock01} & & \checkmark & \\
Fromm & Probing SgrA$^\star$ flares & \pageref{fromm01} & &  & & \checkmark &  & & \checkmark\\
Fromm & Jet Launching and Radiation Microphysics & \pageref{fromm02} & &  & &  & \checkmark & & \checkmark\\
Fuhrmann & Bi-/Multistatic Radar for Space Situational Awareness and Near-Earth Object Studies & \pageref{fuhrmann01} & & \checkmark & &  &  & & & \checkmark\\
Hartogh & Constraining the general circulation of the terrestrial planet's atmospheres by Doppler wind measurements & \pageref{hartogh01} & & \checkmark & &  &  & & \\
Hoeft & Galaxy clusters  & \pageref{hoeft01} & & &  & & \checkmark & \checkmark &\\
Kadler & Resolving the Doppler Crisis  & \pageref{kadler01} &  & & & & \checkmark & & \checkmark \\
Kadler & VLBI Probes of Neutrino Emission Processes in Blazars  & \pageref{kadler02} &  & & & & \checkmark & & \checkmark \\
  \hline
\end{tabular}
  \centerline{\textit{to be continued on next page}}
\normalsize

\newpage
   \noindent \small 
\begin{tabular}{@{}l@{~}p{8.7cm}c@{}c@{}c@{}c@{}c@{}c@{}c@{}c@{}c@{}}
  \hline
 Principal & Science Case    & Page & \multicolumn{8}{c}{Topical Area} \\
   Author & &  &  1 & 2 & 3 & 4 & 5 & 6 & 7 & 8 \\
  \hline
  Keenan & Low-$J$ CO Line Excitation & \pageref{keenan01} & & & \checkmark &  & \checkmark & \\
  Kovalev & Radio-optical link: AGN-based VLBI and \Gaia fundamental reference systems and their application for astrometry and astrophysics    & \pageref{kovalev01} &  & & & & \checkmark &  \\
Kovalev & Blazars as neutrino candidates: complete sample studies    & \pageref{kovalev02} &  & & & & \checkmark &  & \checkmark\\  
Kreckel & Dense gas in nearby galaxies & \pageref{kreckel01} &  & &  \checkmark &  & \checkmark\\ 
Livingston & Wide-band polarization AGN studies, covering scales from parsecs to kiloparsecs, measuring extreme Faraday rotation and complex Faraday effects    & \pageref{livingston01} &  & & & & \checkmark &  & \checkmark\\
Lobanov & Astrophysical searches for physics beyond the Standard Model    & \pageref{lobanov01} &  & & & &  &  & \checkmark\\  
Lobanov & Physics and astrometry with photon rings   & \pageref{lobanov02} &  & & & &  &  & \checkmark\\ 
Müller \& Mus & Scattering mitigation of the Galactic Center & \pageref{mueller_mus01} & & &  & \checkmark & \checkmark &  & \checkmark \\ 
Riechers & The Cosmic Density and Excitation of Cold Molecular Gas  & \pageref{riechers01} & & &  & & \checkmark & \checkmark &\\  
Riechers & Molecular Line Absorption Against the CMB as a Probe of Cosmological Parameters & \pageref{riechers02} & &  & & & &\checkmark & \checkmark\\
Ros & VLBI probes of supernovae, supernova factories and Starburst galaxies & \pageref{ros01} & \checkmark &  &  & &\\
Schilke & Zooming into Feedback Engines & \pageref{schilke01} & &  & \checkmark & &\\
de Simone & The chemical characterisation of young protostars: reveal the heritage of forming planets  & \pageref{deSimone01} & & \checkmark & &\\
Spitler & Studies of fast radio bursts & \pageref{spitler01} & & &  &  & & & \checkmark \\ 
Toth & Study of the Dense ISM  & \pageref{toth01} & &  & \checkmark & \checkmark &\\
Ueda & Star and Planet Formation  & \pageref{ueda03} & & \checkmark & \checkmark  & &  & &  \\
Ueda & Planet Formation at the Innermost Region of Protoplanetary Disks  & \pageref{ueda01} & & \checkmark &  & &  & &  \\
Ueda & From near-infrared to centimeter: comprehensive understanding of the inner planet-forming region  & \pageref{ueda02} & & \checkmark &  & &  & &  \\
Vardoulaki & Big Data, surveys, classification  & \pageref{vardoulaki01} & & &  & & \checkmark & & & \checkmark  \\ 
Vegetti & Constraining the nature of dark matter with strong gravitational lensing   & \pageref{vegetti01} & & &  & & & \checkmark  & \checkmark   \\ 
%Walter &    & \pageref{walter01} & & &  & & & \checkmark  & \checkmark   \\ 
Weis & Circumstellar Nebulae and Mass Loss  & \pageref{weis01} & \checkmark & & \checkmark & \checkmark &  & & \\ 
Wendel & Sub-Structures in Blazar Jets and Synergetic TeV Observations  & \pageref{wendel01} & & &  & & \checkmark & & \checkmark \\ 
Xu & Antenna positions with a sub-mm repeatability  & \pageref{xu01} & & &  & & \checkmark & & & \checkmark \\ 
Zhao & High-precision $\lambda$-astrometry at mm-wavelengths & \pageref{zhao01} & & &  & & \checkmark & &  \\ 
  \hline
\end{tabular}
\normalsize

\subsection{From cometary dust to planetary atmospheres: solar system astronomy at radio wavelengths}\RaggedRight\label{agarwal01}
\vspace*{\baselineskip}

\noindent \textbf{Thematic Areas:} \linebreak $\square$ Stellar Astrophysics \linebreak $\checked$ Solar System, Planetary Systems and Habitability \linebreak
$\square$ Circuit of Cosmic   Matter (incl. star formation) \linebreak $\square$ The Galaxy and the Local Group \linebreak
  $\square$   Galaxies and AGN \linebreak $\square$  Cosmology, Large Scale Structure and Early Universe \linebreak
  $\square$    Extreme conditions in the cosmos, fundamental astrophysics    \linebreak
    $\square$ Interdisciplinary research and technology \linebreak
  
\textbf{Principal Author:}

Name:	Jessica Agarwal
 \linebreak						
Institution:  Technische Universit\"at Braunschweig
 \linebreak
Email: j.agarwal@tu-braunschweig.de
 \linebreak
 
\textbf{Co-authors:}\\
J\"urgen Blum (TU Braunschweig),
Paul Hartogh, Norbert Krupp (Max Planck Institute for Solar System Research, G\"ottingen)
  \linebreak

Our home planetary system is the only such system that we can investigate in situ by sending spacecraft even to its more distant regions and by returning samples to laboratories on Earth. Both methods yield information at a level of detail that is unrivaled by astronomical telescopes. The price of this detail is comprehensiveness. Spacecraft operate for limited periods of time and each can only be in one place at a time. Also, space missions tend to have limited flexibility when it comes to reacting to unexpected events of short duration that nevertheless can be of high scientific interest. Therefore, astronomy with ground- and space-based telescopes plays a crucial role in exploring the solar system and allows us to interpret the data returned by space probes in a system-wide context.

In addition to the Sun and the four terrestrial and four giant planets, the solar system comprises three reservoirs of small bodies (the asteroid belt between the orbits of Mars and Jupiter, the Transneptunian region beyond the orbit of Neptune, and the Oort Cloud at the very limit of the Sun’s gravitational influence). These regions are the sources of those small body populations whose members are on planet-crossing orbits and therefore have dynamical lifetimes short compared to the age of the solar system, such as Near-Earth asteroids, comets, and Centaurs. Finally, the interplanetary space is filled with dust (mainly originating from the small bodies), solar wind plasma and the solar and planetary magnetic fields.

Radio astronomy at the wavelengths and sensitivity of the ngVLA can make significant contributions to understanding diverse aspects of our solar system and will also offer strong synergies with various space missions in which German scientists are involved.

\textbf{Large dust particles from comets --} Due to their formation and storage in the outer solar system, beyond the solar system’s “snowline”, cometary nuclei are rich in a wide range of species that sublimate when they enter the inner solar system. The resulting gas carries along refractory material (“dust”) which subsequently forms the comet’s tail under the influence of solar radiation pressure and is visible to us in scattered sunlight. For typical cometary dust size distributions, the scattering cross-section is mainly provided by small particles, while the mass of the ejected refractory material is concentrated in the largest particles (e.g., mm- to metre-sized). The total mass ejected from comets is relevant to understand their input to the solar system’s zodiacal cloud and their lifetimes. The size and properties of the largest particles emitted may also reflect those of the building blocks of comets, if they formed from “pebbles” in a streaming instability \citep[e.g., ][]{youdin-goodman2005, blum-gundlach2017, blum-bischoff2022}. Since dust particles are inefficient emitters at wavelengths longer than their sizes, we expect the thermal radiation at ngVLA wavelengths to mainly originate from the mm-sized and larger particles whereas shorter wavelengths probe smaller sizes. Radio continuum maps of cometary coma, tails and debris trails (the latter containing the largest chunks of material accumulating along the comet's projected orbit) therefore give us access to this otherwise elusive dust population. The polarisation of this thermal emission will add further constraints to the structure and shapes of the emitting particles.

\textbf{High-resolution gas mapping in comets --} The most abundant volatile species in comets are H$_2$O, CO$_2$, and CO, but there is a whole “zoo” of minor species ranging from noble gases to complex organics \citep{altwegg-balsiger2019}. Most molecules dissociate due to solar photons and solar wind particles, and it is primarily the dissociation products that emit radiation at radio wavelengths from rotational transitions. The inner comae of comets cover a few 10 to 100 kilometres, depending on definition and the process of interest. Seen from a distance of 1\,au, 100\,km correspond to 140\,mas, hence a resolution of well smaller scales is needed to map the distribution of gas species in the coma. Such maps at high spatial and temporal resolution (typical rotation periods are of order several hours) will constrain the dissociation processes and the distribution of parent species across the surface. Combined with the mapping of the dust size distribution, this will make a significant contribution to understanding the processes behind cometary activity and indirectly the structure and potential heterogeneity of the cometary nucleus, which are still ill-constrained but of key relevance for our understanding of comet (and, by extension, planet) formation \citep{cordiner2018}.

\textbf{Sub-surface thermal properties of asteroids, atmosphere-less planets and moons --} Typical surface temperatures of atmosphere-less bodies vary with heliocentric distance, $r_h$, roughly as 280\,K $r_h^{-0.5}$, which puts the maximum of thermal emission from asteroids into the mid-infrared (10-20$\mu$m). Thermal emission at mm-cm-wavelengths can to first order be described by the Rayleigh-Jeans law. The measured thermal emission typically originates from the uppermost near-surface layers of material, down to a depth of a few times the wavelength probed. Hence, longer wavelengths probe deeper layers of the surface. The measured integrated spectrum is then a superposition of the thermal emission from various depths, and the deviation from the $\lambda^{-4}$ dependency of the Rayleigh-Jeans law will inform us on the temperature gradient with depth. This temperature gradient in turn reveals the thermal inertia of the material in the uppermost metres below the surface which relates to the material properties such as thermal conductivity, heat capacity, and porosity. Through dedicated thermophysical modelling, also the optical constants of the subsurface regolith material at the wavelength observed can be constrained \citep{buerger-glissmann2023}, which can be of high relevance also beyond solar system science, e.g. to better understand rocky exoplanets.

The thermophysical and optical properties of the regolith and their variation between and across objects (as probed by rotationally resolved spectra) will give us access to the “deep” layers that are not easily probed even with spacecraft and will permit us to infer on the processes that shape the evolution of this material, such as collisions and reaccumulation, space weathering including micrometeorite bombardment, and thermal fatigue \citep{depater-butler2018}. 

\textbf{Albedo-independent size measurement of asteroids --} Since the temperature of an asteroid can be estimated more reliably than its albedo (despite the details described above), the cross-section derived from an asteroid’s thermal emission is also more reliable than that derived from its scattered sunlight. Hence infrared and radio wavelength brightness distribution measurements of asteroids are essential not only to probe the size distribution of the asteroid population, but also to infer (from combination with visible light data) their albedo distribution, which constrains the composition.
 
\textbf{Interstellar small bodies --} The above aspects apply also to interstellar asteroids and comets. Until now, two such interlopers (asteroid-like 1I/'Oumuamua and comet-like 2I/Borisov) have been detected while crossing our solar system. These objects are ideal to understand how our solar system compares to other planetary systems., and their discovery rate is expected to rise with the launch of the LSST. Given their extremely hyperbolic orbits, these objects appear and disappear on short notice, and are observable for weeks or months only. Hence, telescope observations will keep playing a crucial role in studying these rare visitors. 

\textbf{ESA's Comet Interceptor mission --} The continued mapping of gas and dust in cometary comae will be complementary to the results expected from the ESA-lead Comet Interceptor mission to a yet-to-be determined comet from the Oort cloud or beyond. Since this is a flyby mission, Earth-based telescope observations will have to provide the context data to understand the time evolution of the mission target and how it compares to the wider population of such objects. 

\textbf{The Jupiter system --} German institutions are heavily involved in ESA’s Juice mission that will arrive at Jupiter in 2031 and carry out a wide variety of measurements in the Jupiter system with a focus on the icy moons. Given that the time- and space dependency of these data will be linked by Juice’s trajectory, a more global picture at high spatial and temporal resolution and on a timescale extending the mission duration (of nominally 4 years) will be crucial to understand the context of Juice’s data. This is particularly relevant for the magnetic field structure (the PI institution of the magnetometer is TU Braunschweig) and electron distribution (CoI-involment of MPS G\"ottingen in the particle instrument PEP) that manifest at ngVLA wavelengths through synchrotron radiation \citep{depater-butler2019}, and for understanding the context of the data from the sub-mm wave instrument (SWI, PI institution-MPS) targeting the elemental and isotopic composition, structure and general circulation of Jupiter’s atmosphere and the thermophysical and electrical properties of the Galilean satellites' surfaces. 
\subsection{AGN jets at all scales}\RaggedRight\label{baczko01}
\vspace*{\baselineskip}

\noindent \textbf{Thematic Areas:} \linebreak $\square$ Stellar Astrophysics \linebreak $\square$ Solar System, Planetary Systems and Habitability \linebreak
$\square$ Circuit of Cosmic   Matter (incl. star formation) \linebreak $\square$ The Galaxy and the Local Group \linebreak
  $\checked$   Galaxies and AGN \linebreak $\square$  Cosmology, Large Scale Structure and Early Universe \linebreak
  $\square$    Extreme conditions in the cosmos, fundamental astrophysics    \linebreak
  $\square$ Interdisciplinary research and technology \linebreak
  
\textbf{Principal Author:}

Name: Anne-Kathrin Baczko	
 \linebreak						
Institution:  Chalmers University of Technology, Gothenburg Sweden; MPIfR Bonn, Germany
 \linebreak
Email: anne-kathrin.baczko@chalmers.se
 \linebreak
 
\textbf{Co-authors:}  Eleni Vardoulaki, Thüringer Landessternwarte Tautenburg TLS, Etienne Bonnassieux, JMU W\"urzburg; LESIA, Observatoire de Paris, France; INAF, Italy 
  \linebreak

The unified model for active galactic nuclei (AGN) \citep{Urr95} (see Fig.~\ref{fig:agn_sketch}), specifically the physical description of the formation and evolution of extragalactic jets,  remains incomplete. One reason being, our current understanding of jet evolution and the acceleration and collimation zone (ACZ) up to pc scales is strongly biased towards bright, easily-detectable blazars \citep[compare e.g.][]{Kov20}. Overcoming this limitation will require at the same time predictions from theoretical models and observational constraints from the formation to the energy dissipation of jets, including the fainter AGN jet populations. Not only will such constraints need to cover as much of the emitting spectrum as possible, but also study as diverse a population of AGN as possible at as many physically-relevant scales as possible.

The ngVLA \citep{Sel18} has the potential to provide the necessary observational constraints, and unlock a new era in our understanding of the physics of AGN jets and their interactions with both the host galaxy and the multi-scale environment. This is due to three factors: 
(1) an unprecedented dynamic range for continuum and spectral line observations, allowing to probe the complete populations of jets, and thereby enabling to study the AGN/SFG (Star formation Galaxy) dichotomy like never before; 
(2) the homogeneity of the ngVLA array antenna design at multiple frequencies (1--116\,GHz), crucial for morphological and continuum spectral studies over several scales; 
(3) sensitive polarization observations, enabling us to trace the strength and distribution of the jets magnetic field, relevant for the physical modeling of jet formation and acceleration.

Present Very Long Baseline Interferometry (VLBI) arrays can reach mas resolution at GHz frequencies, but have a very limited field of view (FoV) \citep[compare e.g.][]{Lis18}, providing valuable constraints on the gravitational and magneto-hydrodynamic processes describing the ACZ (\cite{Bla82,Bla77}). 
Non-VLBI arrays such as ALMA or VLA, on the other hand, provide a large FoV but at the cost of limited resolution, providing constraints on the jet propagation and growing instabilities \citep{Tch16} towards energy dissipation in the outer lobes. ALMA, the high-frequency VLA and SKA pathfinders such as the ILT have attempted to achieve sub-arcsec resolutions with multiple-degree FoVs \citep{Mor22}. These operate at very different frequency regimes, however, and therefore trace different regions and physics of AGN jets. Connecting between these regimes is impossible with current instruments, which is crucial when, e.g., studying jet expansion as shown by \cite{Boc21}.

The ngVLA will be key in solving this issue. The sensitivity to emission up to arcsec scales will provide simultaneous insight into the regions not only of formation and acceleration (sub pc--10pc scales) or of expansion (kpc scales), but also into the energy dissipation region (Mpc jet, which the shortest baselines of the ngVLA are sensitive to up to band 4). For the first time this will enable an inclusive physical description of the transition between relativistic/ballistic physics and energy deposition into surrounding media.

Summarizing, the ngVLA will enable us to access a much larger population of AGN jets at all scales than currently available. In so doing, the current blazar-dominated populations serving as the basis of our present understanding of AGN jets will be balanced by healthy populations of non-blazar sources, building a solid foundation for a complete description of jets physics in AGN. 

\begin{figure}
    \centering
    \includegraphics[width=\linewidth]{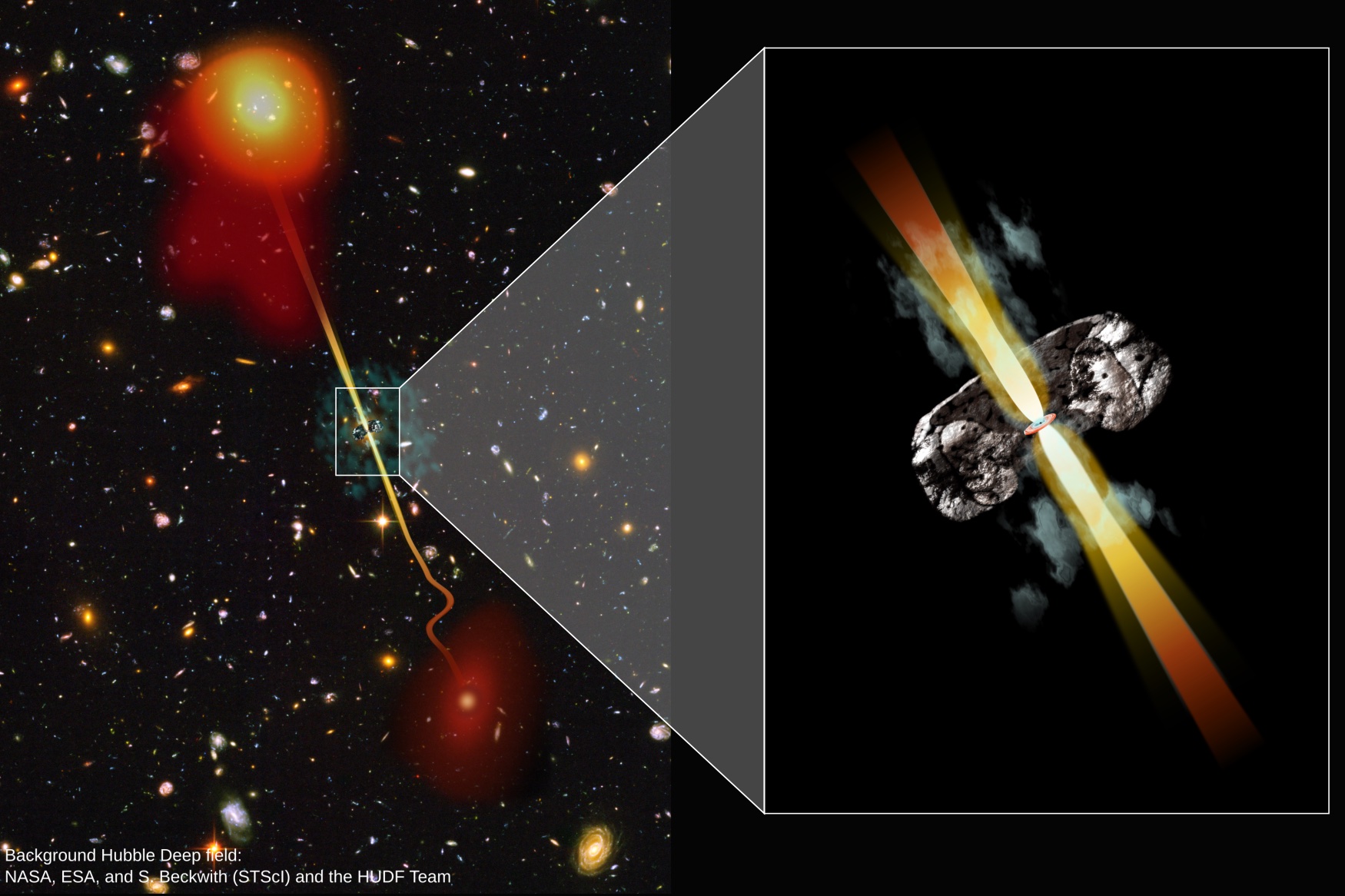}
    \caption{Sketch of an AGN at large (\textit{left}) and small (\textit{right}) scales. The region of formation, acceleration and collimation of the jets as well as properties of absorbing structures as the torus can be studied with VLBI, whereas the large scale instabilities and interactions with the surrounding media, e.g. shocks along the jet or radio lobes, can currently be studied only with telescopes as ILT, VLA or ALMA. The ngVLA will for the first time allow to bridge the gap between both scales.}
    \label{fig:agn_sketch}
\end{figure}

\subsection{Small-scale structure of the warm and cold neutral medium (WNM \& CNM) with HI emission and HI self-absorption (HISA)}
\RaggedRight\label{beuther01}
\vspace*{\baselineskip}

\noindent \textbf{Thematic Areas:} \linebreak $\square$ Stellar Astrophysics \linebreak $\square$ Solar System, Planetary Systems and Habitability \linebreak
$\square$ Circuit of Cosmic   Matter (incl. star formation) \linebreak $\checked$ The Galaxy and the Local Group \linebreak
  $\square$   Galaxies and AGN \linebreak $\square$  Cosmology, Large Scale Structure and Early Universe \linebreak
  $\square$    Extreme conditions in the cosmos, fundamental astrophysics    \linebreak
  $\square$ Interdisciplinary research and technology \linebreak
  
\textbf{Principal Author:}

Name:	Henrik Beuther
 \linebreak						
Institution:  Max Planck Institute for Astronomy
 \linebreak
Email: beuther@mpia.de
 \linebreak
 
%\textbf{Co-authors:} (names and institutions)
%  \linebreak

The structure and physical properties of the atomic gas are key parameters of the interstellar medium. The transitions from the warm neutral medium (WNM) to the cold neutral medium (CNM) and then to molecular gas are essential steps for the formation of stars as well as shaping galaxy structures as a whole. Over the last decades, the VLA has been one of the major contributors to the understanding of HI properties in the Milky Way and external galaxies (e.g., \cite{stil2006,walter2008,beuther2016}). While the classical 21\,cm emission line typically probes a mixture of the WNM and CNM, specially for Milky Way clouds, one can isolate the CNM by studying HI self-absorption (HISA) (e.g., \cite{gibson2000,wang2020b,syed2023}, see Fig.~\ref{syed2023_fig3} for an example). Such kind of studies allowed us to characterize the physical properties of the different phases of the atomic gas on Milky Way scales.\\[0.3cm]

Nevertheless, the studies so far suffered from severe limitations. These are largely related to insufficient spatial and spectral resolution achievable with current observatories and instrumentation. The HI/OH/Recombination line survey of the Milky Way (THOR) conducted with the VLA is the best HI survey of our Milky Way so far \cite{beuther2016}. However, based on brightness sensitivity limitations, the final HI data have only a spatial resolution of $40''$ (although the VLA C-array data in principle would allow higher spatial resolution) \cite{wang2020a}. Furthermore, the spectral resolution of the final datacubes is also comparably coarse around $\sim$1.5\,km\,s$^{-1}$ \cite{wang2020a}. Hence, given observational limitations omitted in-depth studies of the small-scale spatial structure as well as detailed kinematic analysis of the densest HI structures that are most closely related to molecular cloud and star formation processes.\\[0.3cm]

The ngVLA will boost the sensitivity for HI observations and by that allow us deep insights into the small-scale spatial and kinematic properties of the atomic gas during its conversion processes into the molecular phase. With the vastly increased collecting area of the ngVLA, the achievable brightness sensitivity will allow us to create maps of the WNM and CNM at spatial scales below $10''$ that correspond at typical Galactic distances of 4\,kpc to linear resolution elements $\leq$0.2\,pc. Furthermore, while for a few nearby clouds narrow HISA has been detected (so-called HINSA \cite{li2003}), such studies were not possible on Milky Way scales so far. Measured line-width of HINSA in these nearby clouds are typically below 1\,km\,s$^{-1}$, significantly narrower than the spectral resolution of existing Milky Way-scale surveys (e.g., THOR \cite{wang2020a}). Hence, again the vast sensitivity improvements of the ngVLA with allow HI studies at $\sim$0.1\,km\,s$^{-1}$ spectral resolution, disentangling the velocity structures within atomic and molecular clouds.\\[0.3cm]

To summarize, the ngVLA will allow us for the first time to investigate on Milky Way scales the small scale spatial and kinematic atomic gas properties that lead to molecular clouds and ultimately to the formation of stars.

\begin{figure}[htb]
\includegraphics[width=0.99\textwidth]{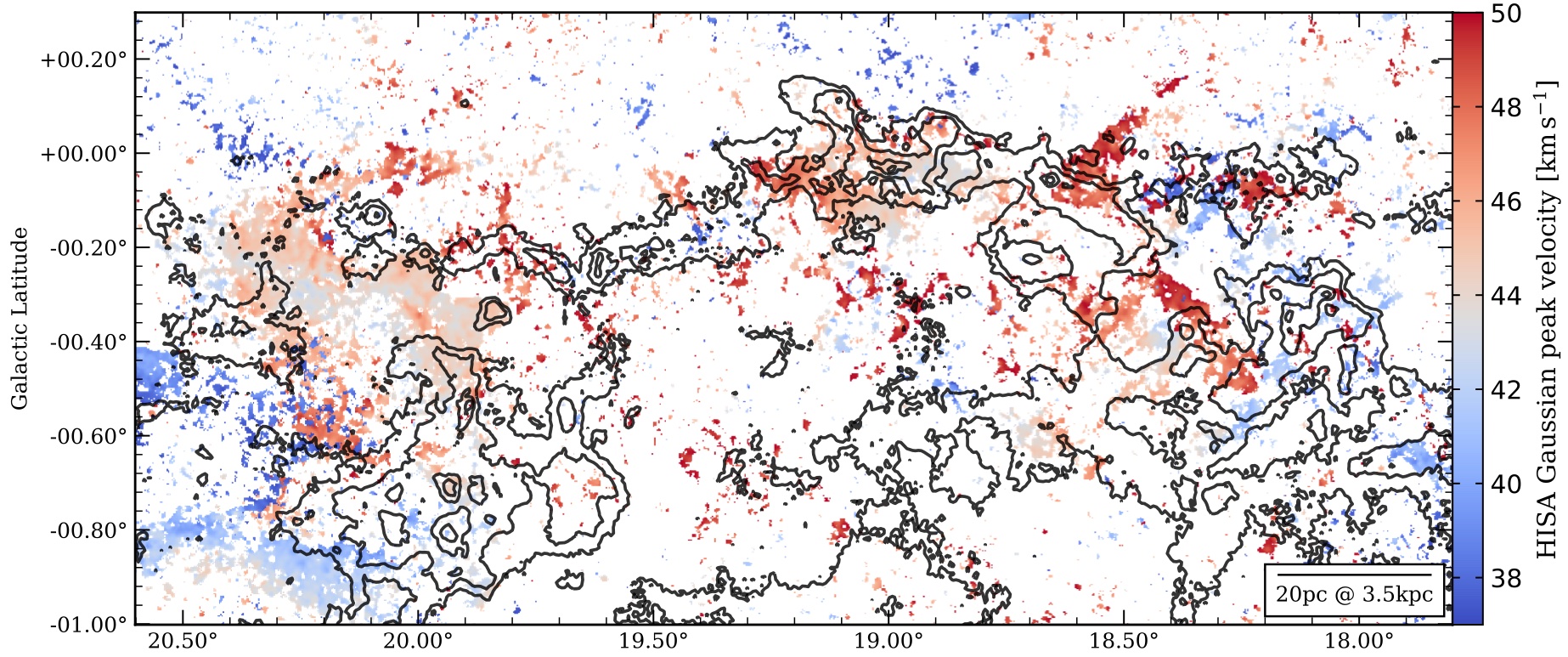}
\caption{HISA peak velocities are presented in color-scale for one large cloud in the Milky Way (presented in Galactic longitude and latitude \cite{syed2023}). The contours show the corresponding integrated $^{13}$CO(1--0) emission from the Galactic ring survey \cite{jackson2006}.}
\label{syed2023_fig3} 
\end{figure}   
\subsection{HI Studies of Nearby Galaxies}
\RaggedRight\label{bigiel01}
\vspace*{\baselineskip}

\noindent \textbf{Thematic Areas:} \linebreak $\square$ Stellar Astrophysics \linebreak $\square$ Solar System, Planetary Systems and Habitability \linebreak
$\checked$ Circuit of Cosmic   Matter (incl. star formation) \linebreak $\checked$ The Galaxy and the Local Group \linebreak
  $\checked$   Galaxies and AGN \linebreak $\square$  Cosmology, Large Scale Structure and Early Universe \linebreak
  $\square$    Extreme conditions in the cosmos, fundamental astrophysics    \linebreak
  $\square$ Interdisciplinary research and technology \linebreak
  
\textbf{Principal Author:}

Name:	Frank Bigiel
 \linebreak						
Institution:  AIfA, Univ. Bonn
 \linebreak
Email: bigiel@astro.uni-bonn.de
 \linebreak
 
\textbf{Co-authors:} Fabian Walter (MPIA)
  \linebreak

Studies of HI emission in and across galaxies address a wide range of science questions: HI emission probes the galactic potential out to very large radii which, via mass modelling, constrains the density structure of the dark matter halo \citep[e.g.][]{deblok2008}, extended HI responds to gravitational perturbations from galaxy-galaxy interactions \citep[e.g.][]{braun2004}, kinematics constrain non-circular motions linked to gas in- and outflows in the plane of the disk \citep[e.g.][]{eibensteiner2023}, which in combination with extraplanar gas clouds provides important constraints on present day gas accretion fueling star formation. The superior resolution and sensitivity of the ngVLA will shed light on the nature and physical properties of atomic gas clouds and the overall diffuse, extended HI disks of galaxies at column densities $\ll10^{18}\,\textrm{cm}^{-1}$, where a break in the exponential gas profiles is expected \citep[e.g.][]{braun2012}. Such sensitive observations will also offer key points of comparison to cosmological simulations \citep[e.g.][]{shen2013}, regarding gas accretion rates in the present day universe, the "missing dwarfs" problem by providing a sensitive census of "dark" dwarf galaxies and, given the exquisite spatial resolution, in particular regarding the dark matter profile in galaxy centers ("cusp vs. core"). \newline

Atomic hydrogen is the raw material for star formation, but how exactly the conversion to H$_2$ is regulated is not settled on kpc-scales \citep[e.g.][]{foyle2010} and essentially (except for the Magellanic Clouds) unprobed on the relevant cloud scales of a few 10\,pc (corresponding to around 1$"$) across nearby galaxies. Measuring the coupling between dynamics (line widths), angular momentum around molecular clouds, comparing them to those of the CO tracing the molecular gas, as well as probing HI columns, power spectra and structure functions on molecular cloud scale will all represent major new avenues. How all of these properties are ultimately regulated by ISM physical conditions (radiation field, metallicity) and galactic properties and how they impact the conversion of HI to H$_2$ are further central topics. While these measurements all exist for the CO at such scales \citep[e.g.][]{leroy2021}, the excellent surface brightness sensitivity of the ngVLA will enable matched HI observations for a diverse and significant sample of local galaxies by pushing such studies beyond the local group. In addition, the superior sensitivity will also allow to probe alternative molecular gas tracers such as OH over significant areas of local galaxies. Arcsecond-scale 21\,cm mapping with the ngVLA will also probe scales where HI absorption features against bright sources can be isolated, giving access to HI spin temperature measurements so far only feasible at decent signal-to-noise in the local group \citep[e.g.][]{braun1997}. \newline

Overall, while current facilities (in particular MeerKAT and the VLA) do reach required sensitivities and resolution, mapping any reasonable area outside the local group would still require an enormous or even prohibitive amount of observing time (cf.\ the ongoing 2000\,hr LGLBS survey at the VLA). On top of that, this is currently an "either / or" situation; either mapping at high sensitivity but approaching arcminute resolution or pushing to arcsecond-scale which means basically giving up mapping and/or pushing to required column density sensitivities. The ngVLA will be an indispensable tool for sensitive, molecular cloud- / arcsecond-scale HI mapping of prominent northern sky local galaxies with significant complementary data coverage at matched resolution from the UV into the mid-IR from JWST.\newline

\begin{figure}[h]
\centering
  \includegraphics[width=\textwidth]{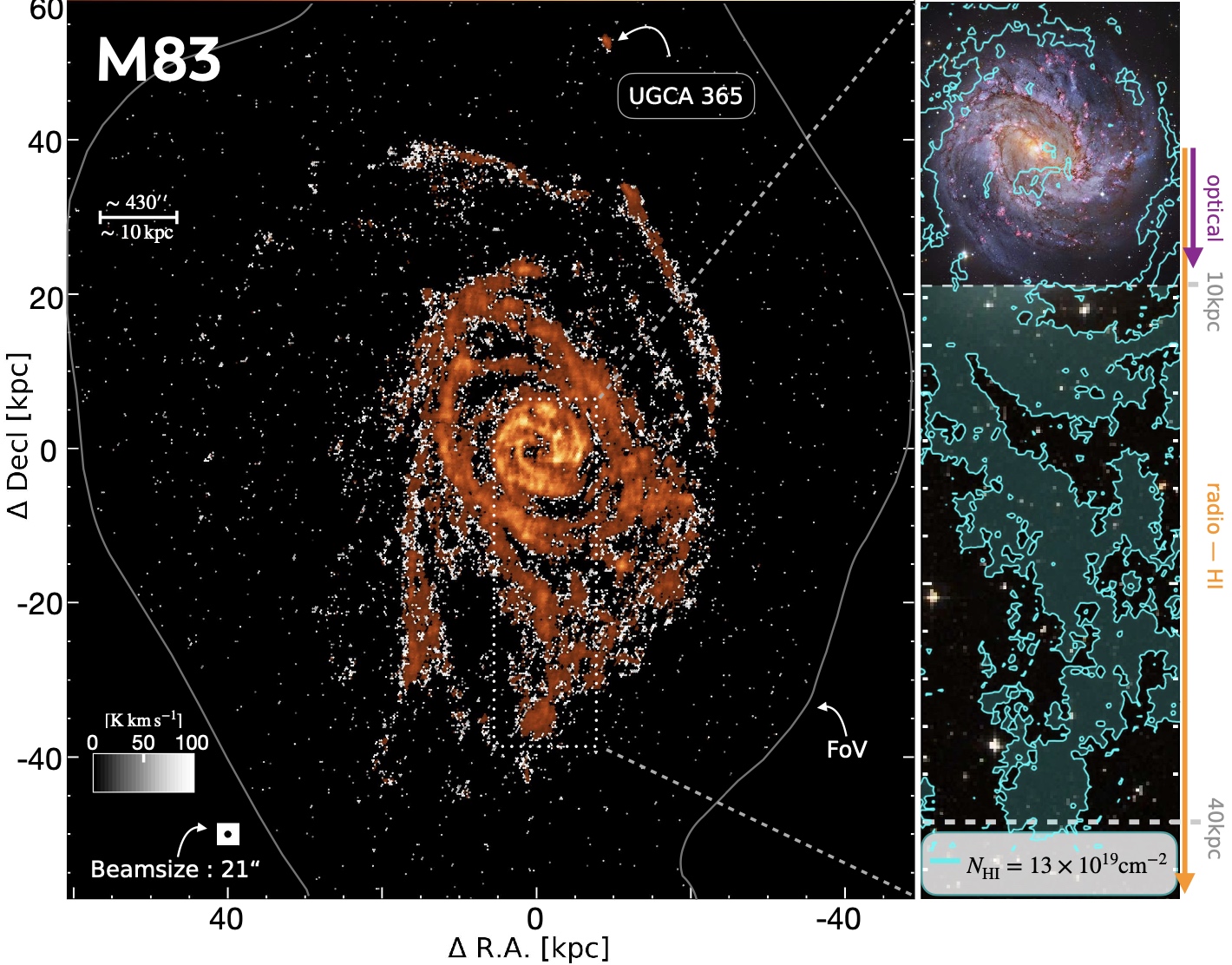}
\caption{VLA HI observations of the local spiral galaxy M83 from \citet[][PI: F. Bigiel]{eibensteiner2023}, illustrating several of the ngVLA HI science cases: extended HI as a probe of the galactic potential, (kinematic) search for gas accretion and radial transport as well as star formation in different galactic environments. Pushing such observations to arcsecond, molecular cloud scale and recovering the low-column density gas filling by far most of the area at large radii is new observational phase space the ngVLA will unlock.}
\end{figure}
  
\subsection{Imaging Young Solar System Analogues in Formation}
\RaggedRight\label{birnstiel01}
\vspace*{\baselineskip}

\noindent \textbf{Thematic Areas:} \linebreak $\square$ Stellar Astrophysics \linebreak $\checked$ Solar System, Planetary Systems and Habitability \linebreak
$\square$ Circuit of Cosmic   Matter (incl. star formation) \linebreak $\square$ The Galaxy and the Local Group \linebreak
  $\square$   Galaxies and AGN \linebreak $\square$  Cosmology, Large Scale Structure and Early Universe \linebreak
  $\square$    Extreme conditions in the cosmos, fundamental astrophysics    \linebreak
  $\square$ Interdisciplinary research and technology \linebreak
  
\textbf{Principal Author:}

Name: Til Birnstiel
 \linebreak						
Institution: LMU Munich
 \linebreak
Email: til.birstiel@lmu.de
 \linebreak
 
\textbf{Co-authors:} Takahiro Ueda, Max Planck Institute for Astronomy,
Mario Flock, Max Planck Institute for Astronomy,
Sebastian Wolf, Kiel University
  \linebreak

  Planets can form quickly in substructures in protoplanetary disks (see \citep{Lau2022}, and \autoref{fig:lau2023} which depicts a simulated ngVLA observation of a gas+dust hydrodynamic simulation of a young solar system analogue). At the same time, substructures are also considered signposts of planets \citep{Teague2018,Pinte2020}. This pushes the chicken-or-egg question of how planets form, and whether they are the cause or consequence of substructure towards the earliest stages of disk evolution. However, young disks are dense and small and therefore highly optically thick. Observing solar system analogues requires long wavelength observations (several mm to cm) at high angular resolution (1~au at 140~pc corresponds to $\sim$7~mas). The ngVLA is the only instrument that can provide this combination of sensitivity, angular resolution, and wavelength coverage.

  Observing these disks and understanding the processes governing planet formation within them are essential to study the fundamental mechanisms that shaped our own solar system and the diversity of exoplanetary systems. The exceptional spatial resolution of the ngVLA will be a key to understand the complex structures of young disks. These disks exhibit intricate sub-structures, including rings, gaps, and spirals, that are indicative of planet formation processes.
 
  An important feature of planet formation on scales of only few au are the timescales: over periods of few years, possibly entire orbits of the planets can be observed which will push radio continuum observations into the time domain. This will enable studies of the dust dynamics and its collisional evolution at unprecedented detail. ngVLA's multi-wavelength capabilities, spanning wavelength from millimeters to a decimeter, will enable a comprehensive view of dust properties via the spatial distribution of the spectral index (see \autoref{fig:lau2023}). These studies will strongly benefit if polarization measurements will be available as well.

  In summary, ngVLA's unique combination of sensitivity, spatial resolution, and wavelength coverage will make it an indispensable tool in the study of planet formation in young disks. It will enable time domain studies, will resolve azimuthal substructures, probe previously optically thick regions, and will push towards the realms of terrestrial planet formation near the snow line. These advancements will revolutionize our understanding of the formation and diversity of planetary systems.

    \begin{figure}[bht]
    \centering
    \includegraphics[width=0.49\textwidth]{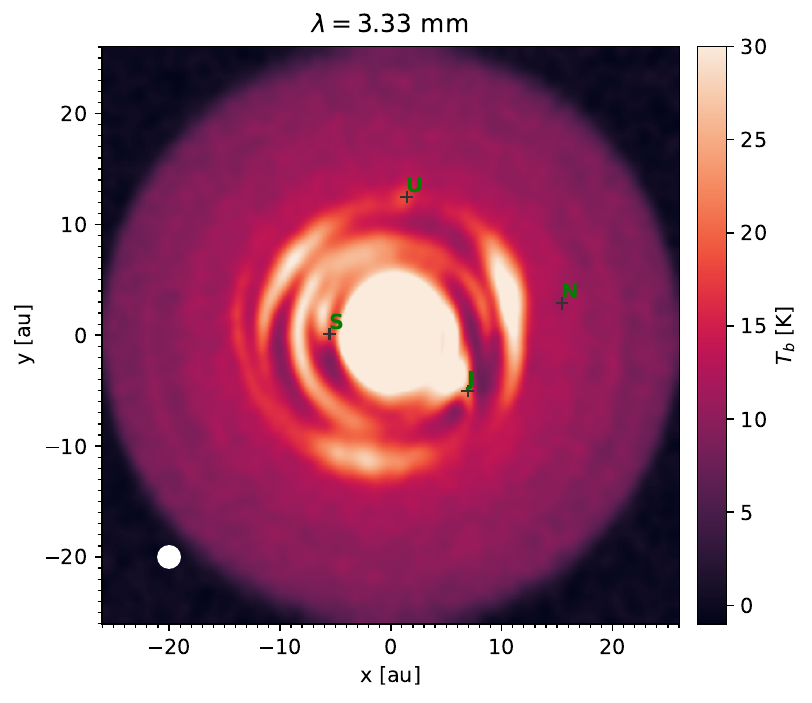}
    \includegraphics[width=0.49\textwidth]{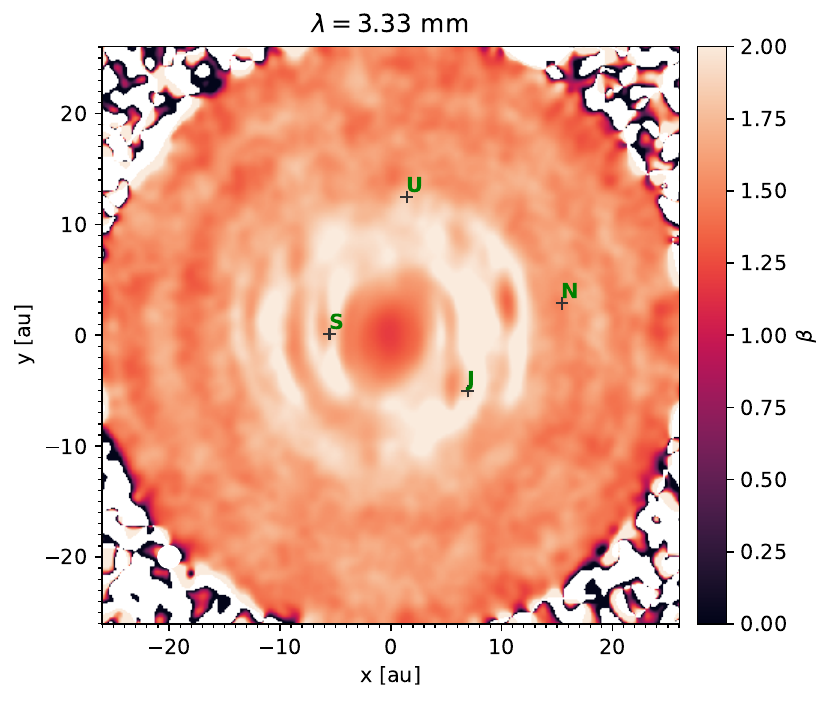}
    \caption{The ngVLA will be able to resolve structures in young disks at high angular resolution. The image shows a simulated observation at 3~mm with the ngVLA (6~mas beam, 0.43~$\mu$Jy noise level, following \citep{Ricci2018}). The disk is resolved at around 1~au resolution, revealing the complex sub-structures and vortices within the disk. The disk and planet setup is based on sequential  planet formation models of Lau, Birnstiel, et al. (submitted). The figure on the right depicts the dust spectral index of the same simulation.}
    \label{fig:lau2023}
  \end{figure}

\subsection{The acceleration and collimation zone in AGN jets}
\RaggedRight\label{boccardi01}
\vspace*{\baselineskip}

\noindent \textbf{Thematic Areas:} \linebreak $\square$ Stellar Astrophysics \linebreak $\square$ Solar System, Planetary Systems and Habitability \linebreak
$\square$ Circuit of Cosmic   Matter (incl. star formation) \linebreak $\square$ The Galaxy and the Local Group \linebreak
  $\checked$   Galaxies and AGN \linebreak $\square$  Cosmology, Large Scale Structure and Early Universe \linebreak
  $\square$    Extreme conditions in the cosmos, fundamental astrophysics    \linebreak
  $\square$ Interdisciplinary research and technology \linebreak
  
\textbf{Principal Author:}

Name: Bia Boccardi
 \linebreak						
Institution: MPIfR, Bonn
 \linebreak
Email: bboccardi@mpifr.de
 \linebreak
 
\textbf{Co-authors:} Anne Baczko, Chalmers University Gothenburg,
Matthias Kadler, JMU Würzburg, 
Yuri Kovalev, MPIfR Bonn,
Luca Ricci, JMU Würzburg
  \linebreak

%The formation of relativistic jets in active galactic nuclei (AGN) has been subject of intense study in recent years. 
Very-Long-Baseline-Interferometry (VLBI) observations
%achieving angular resolution down to tens of $\rm \mu$\,$\rm as$, 
have been able to access the jet launching region in active galactic nuclei (AGN), zooming in the immediate proximity of the black hole \citep{2019ApJ...875L...1E}. In several nearby sources, these have revealed that jets undergo acceleration and collimation on scales of $10^3-10^7$ Schwarzschild radii \citep{Kov20, Boc21}, finally attaining bulk Lorentz factors of the order of tens %\textcolor{orange}{[LR: maybe a ref for the Lorentz factors? e.g., Hovatta+, 2009 and Lister+, 2016.]} 
and opening angles of few degrees on parsec scales \citep{Pushkarev2017}. The physical mechanisms driving these processes are still a matter of debate.
%\textcolor{orange} {Namely, the acceleration in jets may be both thermally and magnetically driven.} 
The co-spatiality of the acceleration and collimation zone (ACZ) observed in a handful of objects \citep[see the case of NGC\,315 in Fig. 1 - left, reported by][]{Ricci2022} is in agreement with theoretical predictions for mainly magnetically-driven jets \citep{Komissarov2007}.
%\textcolor{orange} {nonetheless not excluding thermal acceleration, whose contribution is still up to debate}. 
Concerning the jet confinement, however, it is unclear whether the ambient medium also plays an important role, or if self-collimation due to the magnetic field permeating the jet is sufficient. The former hypothesis was first suggested by \cite{Asada2012} based on their study of M87, showing an approximate coincidence between the parabolic jet region and the Bondi radius.
%, which delimits the theoretical extent of a spherical hot accretion flow in the AGN. 
This coincidence is however not always observed \citep{Ricci2022, Yan2023}. Moreover, the very nature of the ambient medium in which the jet initially propagates is uncertain. For instance, winds produced by the accretion disk could contribute to the jet confinement as well \citep{Globus2016}. The results collected so far do hint at a relation between the AGN accretion mode (radiatively efficient vs. inefficient) and the properties of the ACZ, which tends to be more extended in sources powered by radiatively efficient, thin disks \citep[][Fig. 1 - right]{Boc21}. However, the number of objects where the ACZ could be imaged is still too low to make any solid statement in this regard.

The ngVLA will be the prime observing facility in this research area. Indeed, while providing maximum angular resolution similar to that of current global mm-VLBI arrays
%in its LBA configuration 
%\textcolor{red}{[MK: NRAO is transitioning to the term 'ngVLA LONG' to avoid confusion with the Australian LBA]; you could say 'with its good uv coverage on the longest baselines 
{(\href{https://library.nrao.edu/public/memos/ngvla/NGVLA_105.pdf}{NGVLA Memo 105})'}, it will be characterized by orders-of-magnitude improvements in sensitivity. This implies that the number of sources suitable for such studies will dramatically rise, extending to the vast and unexplored population of faint radio galaxies where the ACZ is expected to be well resolved \citep[e.g.,][]{Rama2023}. The increased image fidelity and observing cadence enabled by the ngVLA will strongly facilitate the determination of the true jet shape in stacked images, as well as the investigation of the jet acceleration based on kinematic studies and jet-to-counter jet ratio profiles. Finally, the unprecedented capability of combining (sub)mas-scale resolution with sensitivity to thermal emission will give us a complete view of the outflow stratification properties, probing the presence and properties of accretion disk winds and other material in the nuclear environment. The improved line sensitivity and broader spectral line window coverage of the ngVLA will also be fundamental for tracing the molecular jet environment down to mm-wavelengths.

\begin{figure}[h]\centering
\includegraphics[width=0.43\textwidth]{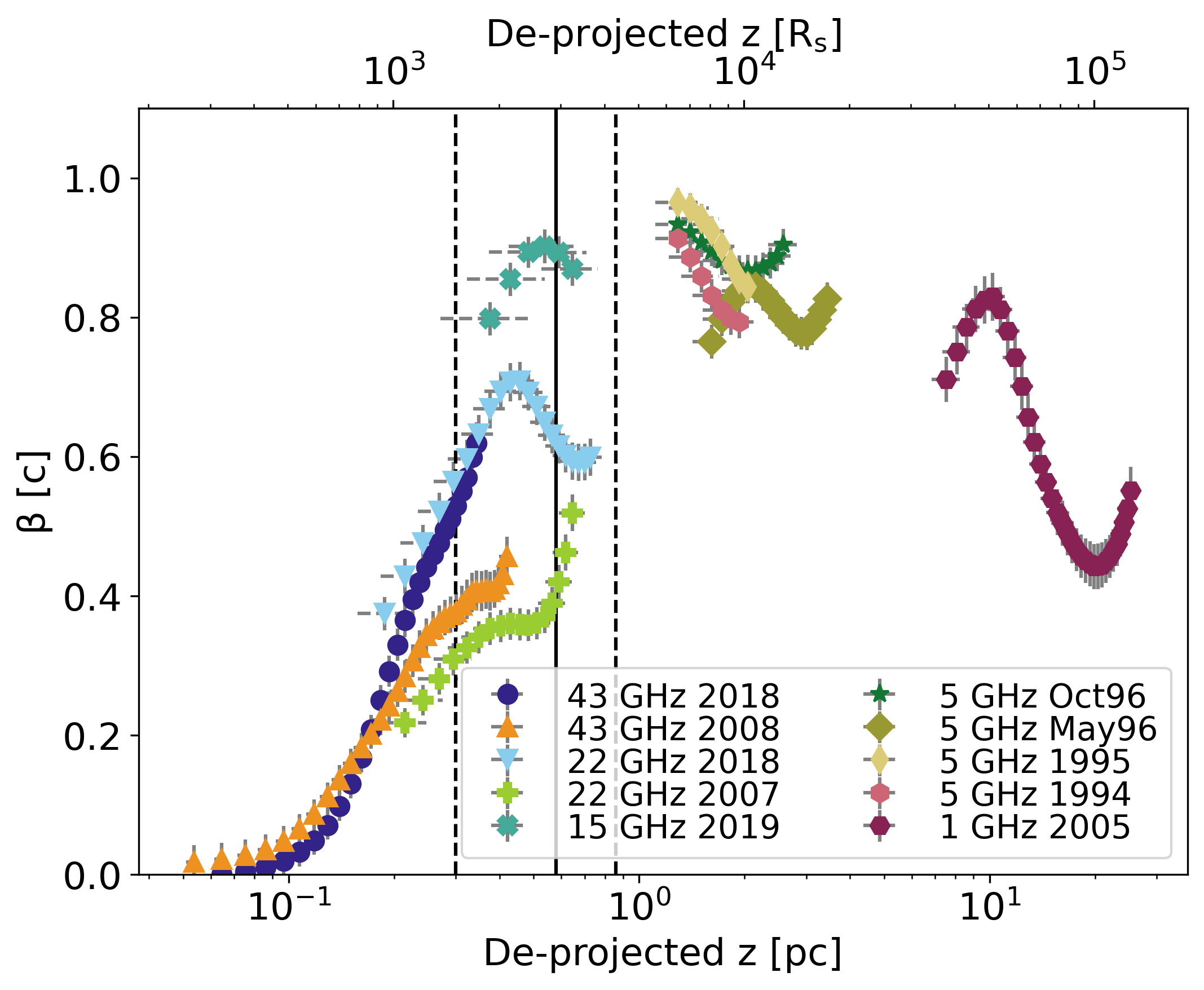}
\includegraphics[width=0.48\textwidth]{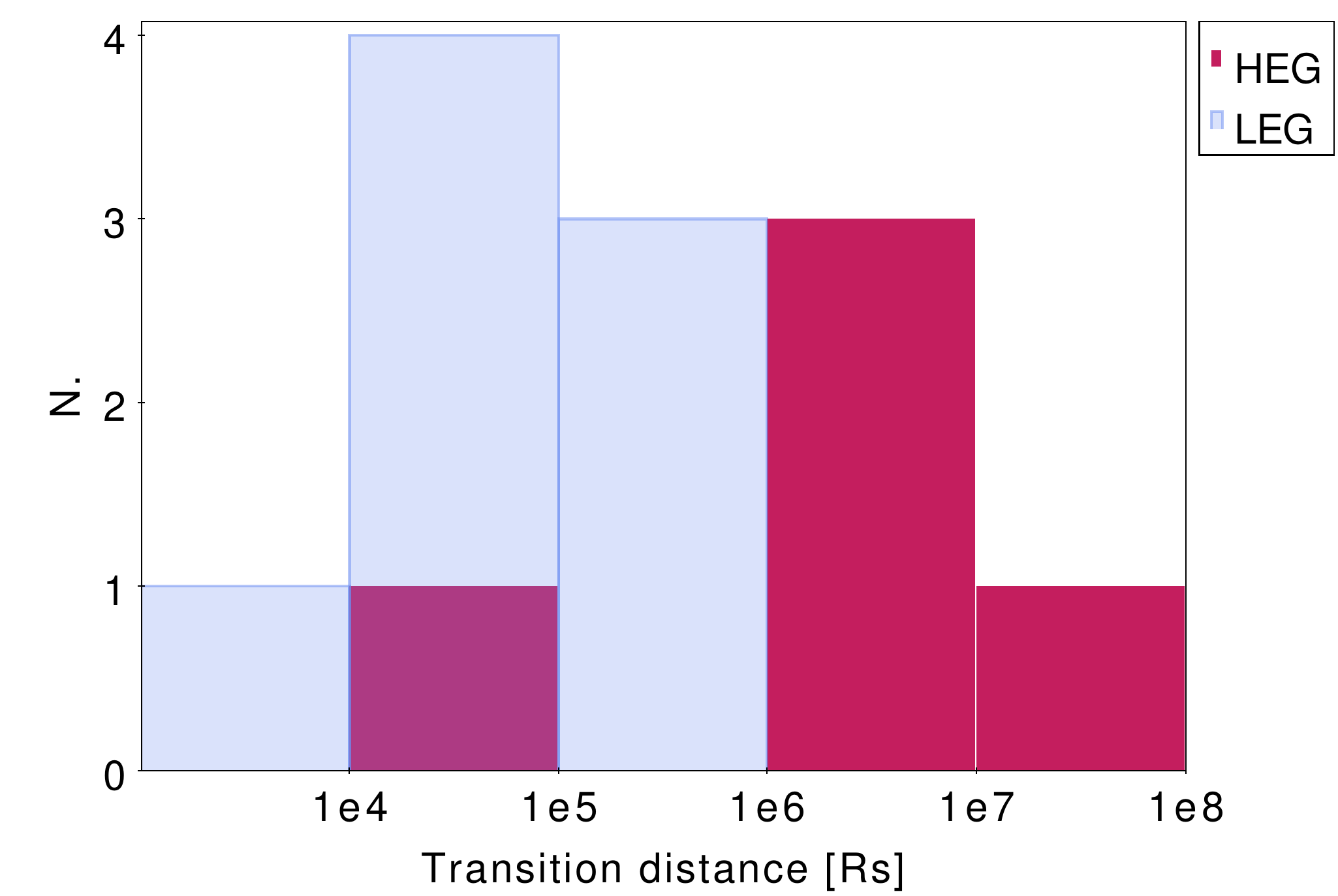}
\caption{Left: Intrinsic speed profile derived for NGC315 based on the jet-to-counter jet ratio in different VLBI images. The vertical line marks the end of the collimation region \citep{Ricci2022}. Right: ACZ extent in sources classified as high-excitation (HEG) or low-excitation (LEG) galaxies \citep{Boccardi2021}. The former are powered by radiatively efficient standard disks, the latter by radiatively inefficient hot accretion flows.}
\end{figure}
\subsection{Compact Starburst Dwarf Galaxies}
\RaggedRight\label{bomans01}
\vspace*{\baselineskip}

\noindent \textbf{Thematic Areas:} \linebreak $\square$ Stellar Astrophysics \linebreak $\square$ Solar System, Planetary Systems and Habitability \linebreak
$\checked$ Circuit of Cosmic   Matter (incl. star formation) \linebreak $\square$ The Galaxy and the Local Group \linebreak
  $\checked$   Galaxies and AGN \linebreak $\checked$  Cosmology, Large Scale Structure and Early Universe \linebreak
  $\square$    Extreme conditions in the cosmos, fundamental astrophysics    \linebreak
  $\square$ Interdisciplinary research and technology \linebreak
  
\textbf{Principal Author:}

Name:	
Dominik J. Bomans \linebreak						
Institution:  
Ruhr University Bochum, Faculty of Physics and Astronomy, Astronomical Institute (AIRUB)
\linebreak
Email: bomans@astro.rub.de
 \linebreak
 
\textbf{Co-authors:}
Kerstin Weis (AIRUB), Sam Taziaux (AIRUB), Adam Enders (AIRUB), Michael Stein (AIRUB)
\linebreak

%Insert your white paper text here (max of 1.5 pages including figures and references).

Local and intermediate redshift (z up to 1) low mass starburst galaxies have 
shallow gravitational potentials, low metallicities, often quite
hard radiation fields, and often exhibit galactic outflows/winds.  Despite HST, JWST, and the SKA precursors detailed, spatially resolved studies of starburst regions, galactic outflows and winds, and triggered SF is not obtainable. 
Low and intermediate redshift dwarfs can serve as laboratories for the detailed physical processes during galaxy formation \citep{Breda2022}, and evolution, as well as reionization  \citep{Enders2023}, and magnetization of 
the universe  \citep{Bertone2006}.

Low mass starburst systems span a large zoo of different classes, many characterized by very high equivalent width of strong lines, e.g.\ [OIII] and H$\alpha$. Examples of this zoo are Blue Compact Dwarf Galaxies (BCDG), extreme emission lines galaxies (EELG),
extremely metal poor galaxies (XMPG), Green Pea galaxies, and Blueberry galaxies.
Due to the bright emission lines from their young starburst region, they are nice targets for UV, optical, and NIR studies, but pose problems for radio continuum observations. It is difficult to disentangle the thermal and non-thermal 
contributions to the radio continuum flux even at low frequencies.  Young starbursts with the often high ionization and high gas densities lead to strong thermal emission from these galaxy sized HII regions. A two band spectral index will show a large scatter, and multi-frequency observations as well as careful analysis of the radio
'spectrum' (or SED) is required to disentangle thermal and non-thermal contributions, the effects of spectral aging of the relativistic electrons, 
and self absorption at low frequencies \citep{Klein2018}.  It is important to note, that optical and NIR observations of recombinations lines do not solve the problem due to different spatial 
scales of the dust absorbing clouds and the need for exact star formations histories to determine 
the stellar absorption line contribution to the Hydrogren recombination lines.  Both effects limit 
the quality of thermal flux estimatate from optical/NIR lines.

The parameters of ngVLA are perfectly suited to tackle the problems.  High sensitivity over a wide 
spectral range up to high frequency and the high spatial resolution will allow detailed studies of 
the radio spectra of compact starbursts and even spatial resolved analyses of the low and intermediate
redshift compact starburst galaxies. With such high resolution and high sensitivity maps of thermal 
and non-thermal emission one can study the structure of the starburst and the base of the outflows/wind, 
trace the outflow/wind morphology and energetic, studies the impact of magnetic fields and the dominant 
cosmic ray transport mechanisms.  Together with ngVLA HI observations, the impact on the environment, 
ionization channels and the therefore conditions for Lyman continuum photon leaking can be studies.

The other missing aspect is detailed magnetic field structure of these galaxies, their outflows/winds 
and of their circumgalactic medium.   
Here, observation of the polarization are critical.  The calibration of the ngVLA system will be 
likely somewhat tricky due to the offset Gregorian design and hopefully given significant attention. Again 
the unique combination of high resolution and high sensititivity in L and S band will open the possibility to study of ordered magnetic fields and their impact on galaxy evolution and the magnetization of 
the universe by observing these low and intermediate redshift proxies for the early universe.

The nearby (distance about 20 Mpc) BCD II Zw 71 can serve as an example of a few points raised here:  
Fig.\ref{fig_IIZw71} shows multi-wavelength data.
%the continuum subtracted H$\alpha$ image \citep{GildePaz2003} and the LOFAR LoTSS 150 Mhz 
%\citep{Shimwell2022} 
%image left, and the GALEX FUV und the B-R color map right. 
The complexity of the galaxy is apparent, Th B-R color map shows the star forming knots and dust patches, as well as possible interaction/merger signs, the GALEX FUX and the H$\alpha$ shows the widespread starburst, 
diffuse gas above and below the disk, and possible dust scattered light, and the large diffuse 150 MHz emission 
points at a magnetized outflow.  The complexity of the system requires multi-frequency data and polarization 
measurements to be disentangled.  The presence of a companion galaxy (II Zw 70, also a dwarf starburst) 4 arcmin 
(about 23 kpc) and the possible polar ring signatures \citep{Cox2001} also underlines the need for high quality HI data to fully understand the galaxies and their circumgalactic medium. The necessary data will be all obtainable with the ngVLA and perfectly complenents LOFAR observations.

\begin{figure}[h]
    \centering
	\includegraphics[angle=0.0,width=0.85\textwidth]{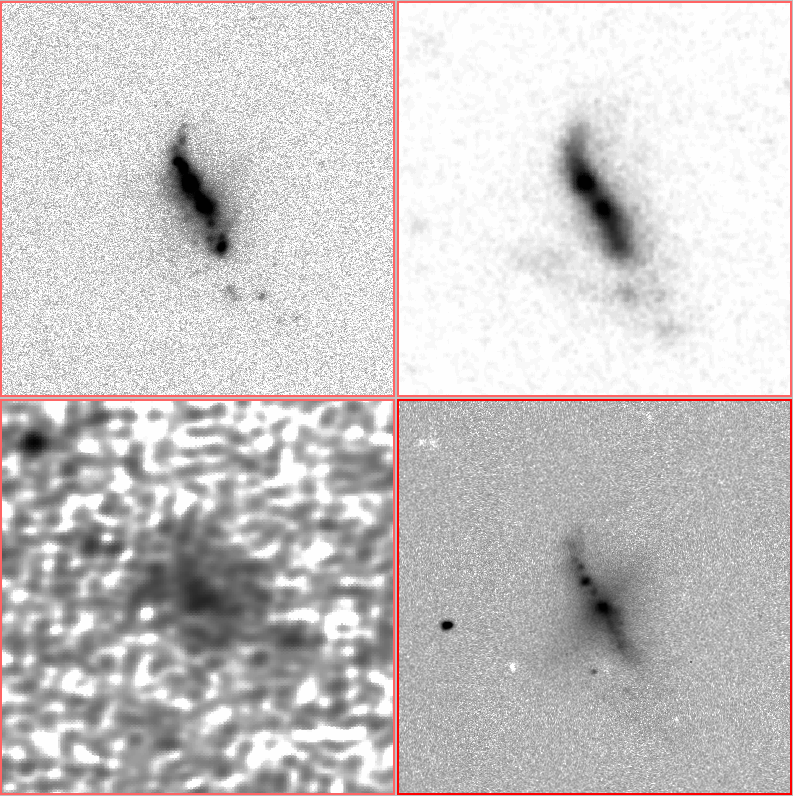}
	\caption{4 views of the nearby dwarf starburst galaxy II Zw 71: top left: H$\alpha$ line image,  top right: GALEX FUV image, bottom left: LOFAR 150 MHz image, bottom right: B-R color map.  All images are 
	$2.5 \times 2.5$ arcmin in size and use an logarhithmic intensity stretch.}
	\label{fig_IIZw71}
\end{figure}

{\bf Acknowledgements}
\linebreak
DJB, KW, AE, ST, and MS acknowledge funding from the German Science Foundation DFG, via the Collaborative Research Center SFB1491 ‘Cosmic Interacting Matters – From Source to Signal’.

\subsection{Redshift Measurements}
\RaggedRight\label{boogaard01}
\vspace*{\baselineskip}

\noindent \textbf{Thematic Areas:} \linebreak $\square$ Stellar Astrophysics \linebreak $\square$ Solar System, Planetary Systems and Habitability \linebreak
$\square$ Circuit of Cosmic   Matter (incl. star formation) \linebreak $\square$ The Galaxy and the Local Group \linebreak
  $\checked$   Galaxies and AGN \linebreak $\checked$  Cosmology, Large Scale Structure and Early Universe \linebreak
  $\square$    Extreme conditions in the cosmos, fundamental astrophysics    \linebreak
  $\square$ Interdisciplinary research and technology \linebreak
  
\textbf{Principal Author:}

Name:	Leindert A. Boogaard
 \linebreak						
Institution:  Max-Planck-Institut f\"{u}r Astronomie (MPIA), Heidelberg
 \linebreak
Email: boogaard@mpia.de
 \linebreak
 
\textbf{Co-authors:} Dominik Riechers (Universit\"{a}t zu K\"{o}ln), Fabian Walter (MPIA)
  \linebreak

The (sub)mm/radio regime provides unique means to measure spectroscopic redshifts from molecular, atomic, and ionised gas (fine-structure) emission lines for galaxies throughout cosmic time, into regimes where UV/optical/near-IR facilities cannot.  Spectral scan surveys, both in the context of molecular deep fields \citep{gonzalez-lopez2019} as well as targeted efforts \citep{reuter2020, bouwens2022} have demonstrated this out to the highest-redshift galaxies \citep{hashimoto2018}, and have the potential to provide a wealth of lines for high-redshift galaxies \cite{riechers2021}.  Key for efficient redshift measurements through spectral scans are broad instantaneous bandwidths ($\Delta \nu/\nu$) and a large field-of-view, coupled with high sensitivity.  The ngVLA baseline receiver configuration from 1.2-116\,GHz features large instantaneous bandwidths that encompass the full bandwidth of Bands 1-5 from 1.2 up to 50\,GHz $(\Delta \nu/\nu = 1)$ and 20\,GHz in Band 6 at 70--116\,GHz ($\Delta \nu / \nu \sim 0.43$), covering a broad range of spectral features out to cosmic dawn (Fig.~\ref{fig}).  Coupled with a significant increase in sensitivity compared to current interferometers and a larger field-of-view at low frequencies than ALMA can provide, these capabilities open a wealth of opportunities:  Molecular deep fields and CO excitation studies are explored in further detail in other the chapters in this volume.  In addition, the [CI] lines may prove an relevant alternative to measure cold gas masses in low metallicity (CO-dark) systems \citep{bolatto2013, hunt2017} with the ngVLA, in particular at higher redshift \citep{coogan2019, boogaard2021}.  The order-of-magnitude increase in sensitivity compared to ALMA implies that a dense gas tracers can be detected on the same timescale as CO with current facilities \citep[][]{decarli2018}, implying rich spectra can be obtained for gas-rich galaxies out to high redshift (Fig.~\ref{fig}).  These topics highlight the broad range of opportunities for redshift measurements and spectral line studies enabled by the ngVLA.

\begin{figure}[h]
    \centering
	\includegraphics[angle=0.0,width=0.44\textwidth]{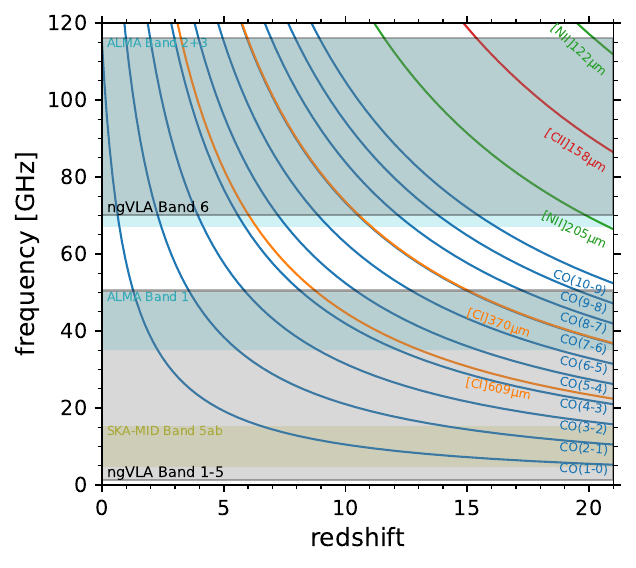}
 	\includegraphics[angle=0.0,width=0.55\textwidth]{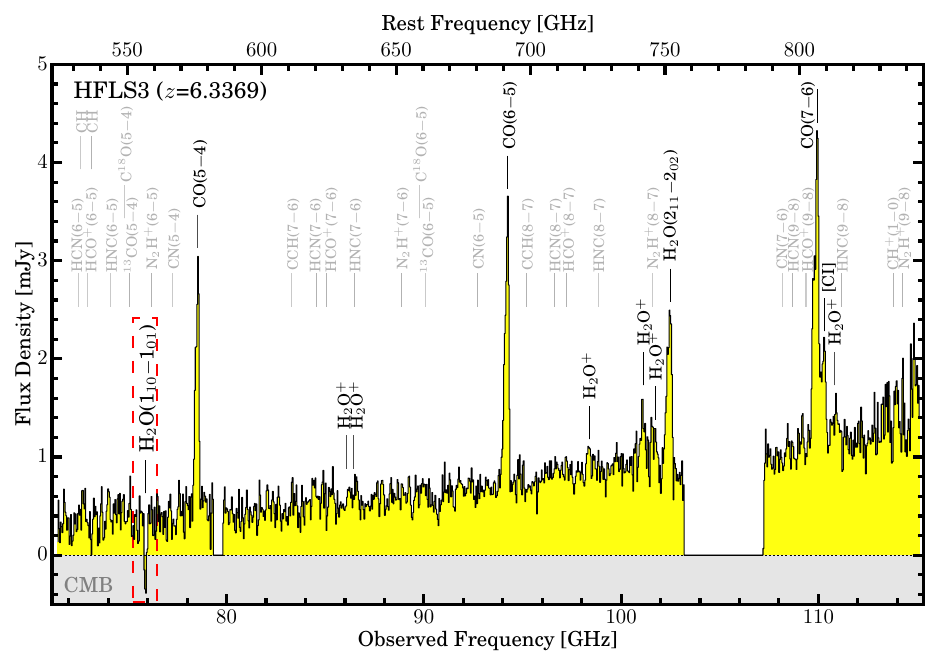}
    \vspace{-6mm}
	\caption{\textbf{Left:} Frequency vs. redshift for the ngVLA Bands 1--6 (shown in gray), highlighting the coverage of the strongest lines (CO, [CI] and fine structure lines) up to $z=20$.  Also shown is the (planned) coverage of ALMA (cyan) and SKA (yellow). \textbf{Right:} Broad-band spectrum of the dusty starburst galaxy HFLS3 at redshift 6.3, showing a wealth of features from CO, [CI], H$_2$O and fainter species, including H$_2$O absorption against the CMB at 75.9\,GHz,  obtained with a NOEMA spectral scan spanning 45\,GHz at 3\,mm (ngVLA Band 6) \citep{riechers22}.}
	\label{fig}
\end{figure}

\subsection{Kinematics of cold gas in high-redshift galaxies}
\RaggedRight\label{boogaard02}
\vspace*{\baselineskip}

\noindent \textbf{Thematic Areas:} \linebreak $\square$ Stellar Astrophysics \linebreak $\square$ Solar System, Planetary Systems and Habitability \linebreak
$\checked$ Circuit of Cosmic   Matter (incl. star formation) \linebreak $\square$ The Galaxy and the Local Group \linebreak
  $\checked$   Galaxies and AGN \linebreak $\checked$  Cosmology, Large Scale Structure and Early Universe \linebreak
  $\square$    Extreme conditions in the cosmos, fundamental astrophysics    \linebreak
  $\square$ Interdisciplinary research and technology \linebreak
  
\textbf{Principal Author:}

Name:	Leindert A. Boogaard
 \linebreak						
Institution:  Max-Planck-Institut f\"{u}r Astronomie (MPIA), Heidelberg
 \linebreak
Email: boogaard@mpia.de
 \linebreak
 
\textbf{Co-authors:} Marcel Neeleman (NRAO), Fabian Walter (MPIA), Melanie Kaasinen (ESO), Dominik Riechers (Univ.\ zu K\"oln)
%, Chris Carilli (NRAO), 
  \linebreak

The dynamical properties of distant galaxies provide key insights into their structure and formation mechanisms.  The kinematics of the cold gas can give direct insight into, for example, the presence of disks, accretion and mergers, the dynamical support (rotation, random motions, turbulence), and the dynamical substructure of the galaxies and how these precipitate star formation \citep[e.g.,][]{hodge2012, neeleman2020, rizzo2023}.  The rotation curves, tracing the gravitational potential, provide important constraints on the (dark) matter distribution and content \citep[e.g.,][]{price2021, lelli2023}.  Measuring kinematics is therefore essential to constrain theoretical models of star formation, the formation mechanism of disks at high redshift, the  physical drivers of turbulence and feedback \citep[][]{zolotov2015, krumholz2018}.  The ngVLA will be capable of performing sub-kpc scale kinematics of cold neutral and molecular gas for star-forming galaxies during the peak of cosmic star formation, and potentially the brightest first galaxies at $z\geq15$ via the fine structure lines, with a sensitivity that is several times that of ALMA,  % sensitivty that is $6\times$ that of alma
as shown by simulations \citep[][see Fig.~\ref{fig}]{memo103}.  The ngVLA will be able to trace disk dynamics to the large radii needed to constrain the radial profile of both the rotation and velocity dispersion.  These measurements will provide important constraints on the dark matter content of early galaxies and reveal the inner substructure and density variations of galaxies (such as spiral arms, clumps, bars, rings), that are key to understand the dynamical processes that are driving the formation of stars during the peak epoch of cosmic star formation.

\begin{figure}[h]
    \centering
	\includegraphics[angle=0.0,width=\textwidth]{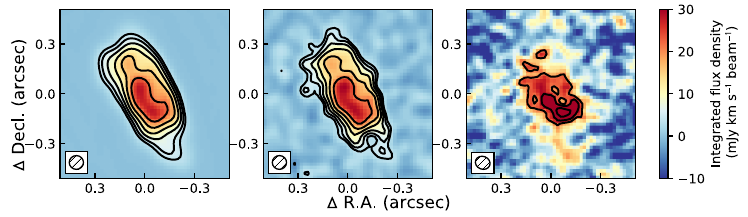}
     \includegraphics[angle=0.0,width=\textwidth]{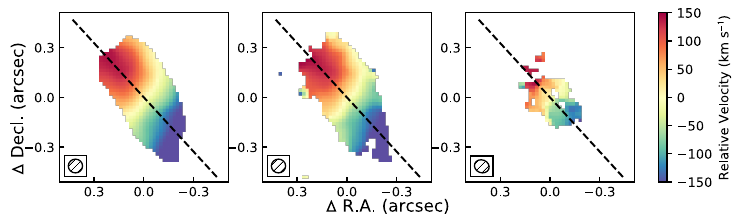}
     \includegraphics[angle=0.0,width=\textwidth]{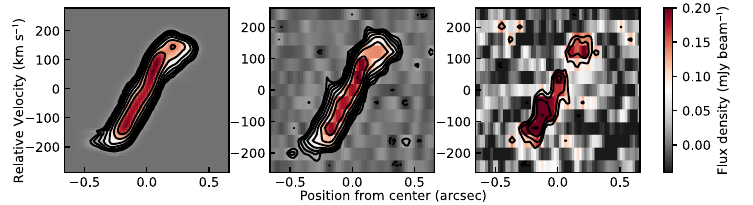}
    \vspace{-6mm}
	\caption{Simulated resolved kinematics in 20\,h of observations for the ngVLA compared to ALMA \citep{memo103}.  In all rows, the panels show the input model (\textbf{left}; based on NGC\,2366, convolved to the target resolution) and the simulated observations with the ngVLA (\textbf{center}) and ALMA (\textbf{right}).   The simulations assume 1.2\,mJy (peak), 400\,km\,s$^{-1}$ line emission at 86\,GHz, representing, for example, CO(2--1) in a $z=1.7$ main-sequence galaxy or CO(3--2) in a starburst galaxy at $z=3.0$. \textbf{Top and middle}: velocity integrated emission and velocity field, respectively, at a resolution of 85\,milli-arcseconds. The ngVLA can recover the source (velocity) structure and surface brightness profile out to 0.4'' radius, while ALMA only recovers the bright peaks and structure in the inner disk within 0.2''. \textbf{Bottom}: Position-velocity diagram along the major axis of the galaxy.  Several departures from pure disk rotation (e.g., the kink and extra emission in the north part) are clearly visible in the ngVLA data but not detectable with ALMA.}
	\label{fig}
\end{figure}

\subsection{Integrated Frequency Synthesizers for the Band 6 ngVLA Receiver}
\RaggedRight\label{braun01}
\vspace*{\baselineskip}

\noindent \textbf{Thematic Areas:} \linebreak $\square$ Stellar Astrophysics \linebreak $\square$ Solar System, Planetary Systems and Habitability \linebreak
$\square$ Circuit of Cosmic   Matter (incl. star formation) \linebreak $\square$ The Galaxy and the Local Group \linebreak
  $\square$   Galaxies and AGN \linebreak $\square$  Cosmology, Large Scale Structure and Early Universe \linebreak
  $\square$    Extreme conditions in the cosmos, fundamental astrophysics    \linebreak
  $\checked$ Interdisciplinary research and technology \linebreak
  
\textbf{Principal Author:}

Name:	Tobias T. Braun
 \linebreak						
Institution:  Ruhr University Bochum, Germany
 \linebreak
Email: tobias.t.braun@rub.de
 \linebreak
 
\textbf{Co-authors:} 
  \linebreak
  Marcel van Delden (Ruhr University Bochum), Christian Bredendiek (Fraunhofer FHR), Jan Schöpfel (Ruhr University Bochum), Nils Pohl (Ruhr University Bochum, Fraunhofer FHR)
 \linebreak

To allow for new scientific discoveries, the ngVLA aims to advance the capabilities of the VLA by significantly increasing the number of antennas and baseline length. Furthermore, it extends the coverage of its predecessor with the introduction of receiver band 6, addressing the additional frequency range of 70--116\,GHz. To achieve the desired sensitivity, the required hardware with its instrumental delay and phase noise must not degrade the overall system signal-to-noise ratio (SNR) by more than 1\,\% \cite{Selina2020}. Thus, innovative solutions for hardware components such as the local oscillator (LO) signal generation are required.
The degradation $\mathcal{D}$ can be estimated as $\mathcal{D}=1-\frac{1}{2}\langle\varphi_{mn}^2\rangle$ with the phase degradation $\varphi$ on baseline $mn$ in radians \cite{Thompson2017}. Accounting for the maximum observing frequency, a $\varphi$=188\,fs was determined. For independent contributions from each antenna, its respective limitations are therefore 188\,fs/$\sqrt{2}$=132\,fs. This total system variation itself summarizes that of the LO, digitizer clock, and physical jitter of the 18-m antenna structure. Equal contributions in a root sum square sense, leave just 76\,fs of jitter for the LO phase noise \cite{Selina2020}. Additionally, this accounts for integration over more than nine decades of offset frequencies from 1\,Hz to 2.9\,GHz.

Therefore, a high-quality phase-locked loop (PLL) at each antenna is required to generate the LO signals essential for the reception of the band 6 sky-signals. Traditionally, those demanding requirements necessitate solutions based on discrete electronics, making them expensive, bulky, and fragile. However, on multiple occasions, we have demonstrated the capabilities of integrated SiGe-PLLs to synthesize frequencies with ultra-low jitter \cite{Pohl2012, Delden2018}. Specifically, in cooperation with the NRAO, we presented a compact LO-module. It is based on a custom-made chip \cite{Braun2022} fulfilling the ngVLA requirements for the upper two-thirds of the E-Band (60--90\,GHz) in \cite{Braun2023}.

\begin{figure}[h!]
	\centering
	\includegraphics[width=0.9\columnwidth]{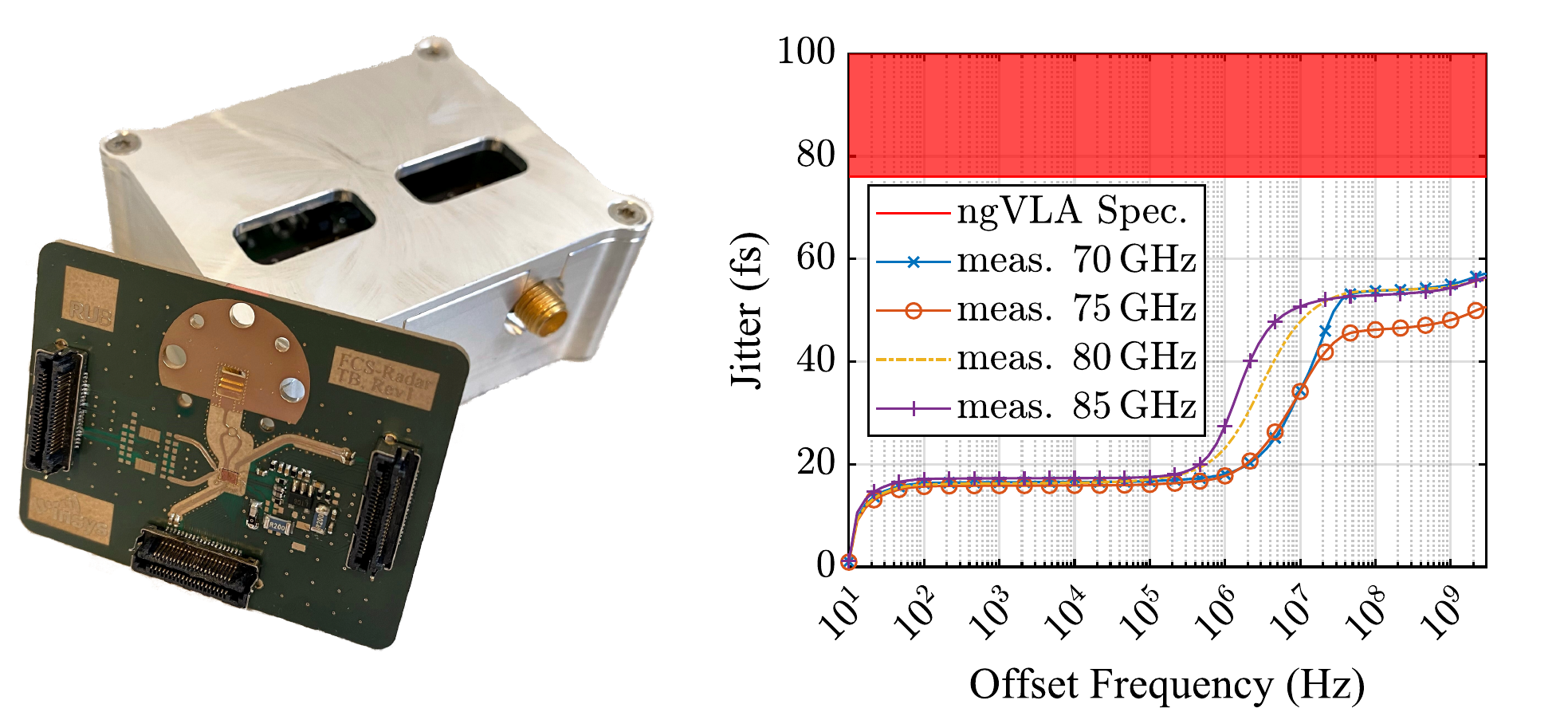}
	\caption{Our developed module (left) is capable of generating LO signals with the stability required for band 6 of the ngVLA. The corresponding jitter measurements (right) show the comfortable fulfillment of the 76\,fs specifications in the frequency range of 70--85\,GHz. 
    %Extending to the full frequency range of 72.5--113.1\,GHz and thereby making measurement equipment more compact, robust, and affordable with integrated synthesizers will be our future goal.
    Extending to the full frequency range of 72.5--113.1\,GHz is ongoing work, which will enable more compact, robust, and affordable LO sources by using integrated synthesizers. 
    }
	\label{Figure}
\end{figure}

At the same time, the module has a size of just 6x5x2\,cm$^3$ and is depicted in Fig. \ref{Figure} (left). The corresponding jitter measurements for LO-frequencies of 70--85\,GHz are shown in Fig. \ref{Figure} (right). With values between 51\,fs and 57\,fs, the specification of the ngVLA is achieved comfortably in the shown frequency range. A lower integration limit of 10\,Hz was chosen instead of the targeted 1\,Hz, as the best reference source available to us did not match what is offered and required in radio telescopes. In combination with the intended reference \cite{Kapathakis2020}, even lower values will be achieved, while starting from 1\,Hz.

Based on those promising results we aim at applying this concept to the specifics of ngVLA’s band 6. With LO-frequencies between 72.5--113.1\,GHz, we intend to address the challenge of covering a frequency range extending beyond the W-band. This will bring the advantages of integrated synthesizers, i.e. high performance, low costs, compactness, and robustness at the same time, into the radio telescopes of the future.
\subsection{Detecting Supermassive Binary Black Holes with the ngVLA}
\RaggedRight\label{britzen01}
\vspace*{\baselineskip}

\noindent \textbf{Thematic Areas:} \linebreak $\square$ Stellar Astrophysics \linebreak $\square$ Solar System, Planetary Systems and Habitability \linebreak
$\square$ Circuit of Cosmic   Matter (incl. star formation) \linebreak $\square$ The Galaxy and the Local Group \linebreak
  $\checked$   Galaxies and AGN \linebreak $\checked$  Cosmology, Large Scale Structure and Early Universe \linebreak
  $\checked$    Extreme conditions in the cosmos, fundamental astrophysics    \linebreak
  $\square$ Interdisciplinary research and technology \linebreak
  
\textbf{Principal Author:}

Name:	Silke Britzen
 \linebreak						
Institution:  Max-Planck-Institut f\"ur Radioastronomie, Auf dem H\"ugel 69, Germany
 \linebreak
Email: sbritzen@mpifr.de
 \linebreak
 
\textbf{Co-authors:} 

E. Kun$^{1,2,3,4,5}$, M. Zaja\v{c}ek$^{6}$, A. Tursunov$^{7}$, I.N. Pashchenko$^{8,9}$
  \linebreak
  
$^{1}$ Theoretical Physics IV: Plasma-Astroparticle Physics, Faculty for Physics \& Astronomy, Ruhr University Bochum, 44780 Bochum, Germany; $^{2}$ Ruhr Astroparticle And Plasma Physics Center (RAPP Center), Ruhr-Universität Bochum 44780 Bochum, Germany; $^{3}$ Astronomical Institute, Faculty for Physics \& Astronomy, Ruhr University Bochum, 44780 Bochum, Germany; $^{4}$ Konkoly Observatory, HUN-REN Research Centre for Astronomy and Earth Sciences, H-1121 Budapest, Konkoly Thege Miklós út 15-17., Hungary; $^{5}$ CSFK, MTA Centre of Excellence, Konkoly Thege Miklós út 15-17., Hungary; $^{6}$ Department of Theoretical Physics and Astrophysics, Faculty of Science, Masaryk University, Kotl\'a\v{r}sk\'a 2, 611 37 Brno, Czech Republic; $^{7}$ Max-Planck-Institut f\"ur Radioastronomie; $^{8}$ Astro Space Center, Lebedev Physical Institute, Russian Academy of Sciences; $^{9}$ Moscow Institute of Physics and Technology, Dolgoprudny, Institutsky per., 9, 141700 Moscow, Russia
\pagebreak

\textbf{Abstract --}
Galaxy collisions are observed throughout the Universe and the natural and final consequences of their evolution are mergers of supermassive black hole pairs detectable in gravitational wave emission with LISA or PTAs. Recently, NANOGrav \cite{nanograv} provided first evidence for gravitational wave background emission from ancient mergers. To detect, image, and dynamically explore the precursors, the closest pairs of supermassive black hole binaries (SMBBHs) with the anticipated and unprecedented superb capabilities of the ngVLA, is the aim of the here presented project.

\textbf{Precessing jets: The smoking gun signature of SMBBHs --}
The formation of SMBBHs in the aftermath of galaxy collisions has been proposed by e.g., \cite{begelman}. While black hole pairs on large scales have been studied across the wavelength regimes e.g., \cite{komossa, krause19}, direct imaging of close pairs with high resolution radio interferometry remains elusive. To the best of our knowledge, only one pair with 7.3 pc separation has so far been imaged \cite{rodriguez06}. 

A mounting number of SMBBH candidates however, has been claimed based on radio interferometric observations of precessing jets (e.g., \cite{caproni04, britzen18, kun23}) . It is the orbital motion of a supermassive black hole binary which leads to the swirling motion of the jet (e.g., \cite{begelman}). Precessing jets can also originate in a misalignment of the black hole spin with the accretion disk angular momentum which leads to the Lense-Thirring effect \cite{lense}.
Precession-induced variability is shown to not only cause pronounced curvature of blazar jets, but also to modulate the variability due to periodically changing viewing angles and Doppler factors (for details see \cite{britzen23} and references therein). Also the Spectral Energy Distribution (SED) correlates with the projected precession phase, as modeled for the most promising SMBBH candidate OJ~287 \cite{britzen23}.

\textbf{Direct detection of absolute motion of the orbiting black hole with the ngVLA --}
Based on the already existing archives of radio interferometric data (MOJAVE, etc.) and our extensive experience in detecting, analyzing, and modeling of precessing jets, we plan to select the best set of the closest pairs of SMBBH candidates for future ngVLA observations.
With the projected 0.1 milliarcsecond resolution and microarcsecond precision astrometry of the ngVLA, we aim at direct imaging of both jets (if present) and cores, as well as dynamically resolving the absolute motion of the orbiting core. As the expected orbital timescale of the two supermassive black holes and hence the radio cores is shorter than the precession timescale \citep{britzen23}, we will also need a temporal coverage of at least a fraction of the year. In Fig. \ref{speed} we show the (circular) orbital speed of the $m_1$ and $m_2$ mass black holes in a binary with a total mass of $m=10^6,~10^7,~10^8,~10^9,~10^{10}~\mathcal{M}_\odot$ (orbital period is $T=1$ yr), where the mass ratio $\nu=m_2/m_1$ ($m_1$$>$$m_2$) is set to $\nu=0.03$ and $\nu=0.3$ (the lower and upper limit of the typical mass ratio range of merging SMBBHs). We show the results for a range of redshifts. To translate the physical scale to angular scale, we adopt the cosmological parameters as follows: $H_0=69.6$~km\,s$^{-1}$\,Mpc$^{-1}$, $\Omega_\mathrm{m}=0.286$ and $\Omega_\Lambda=0.714$. Microarcsecond precision astrometric observations with the ngVLA, conducted on a time-scale of years, might reveal the orbital displacement of at least the faster moving black hole in a SMBBH - the secondary - given it is in active phase.\\ 

With the current Event Horizon Telescope (EHT), the future next generation EHT (ngEHT), and in particular the ngVLA, never before seen radio inteferometric observations will revolutionize direct imaging and astrometric motion studies. These observations will be augmented by multi-wavelength observations by upcoming new instruments (e.g, GRAVITY+, CTA). Together with LISA \cite{lisa} and Pulsar Timing Array observations in the Gravitational Wave regime, SMBBH studies will likely open a new window for the exploration of the fundamentals of gravity.

\begin{figure}
\centering
\includegraphics[clip,width=16cm]{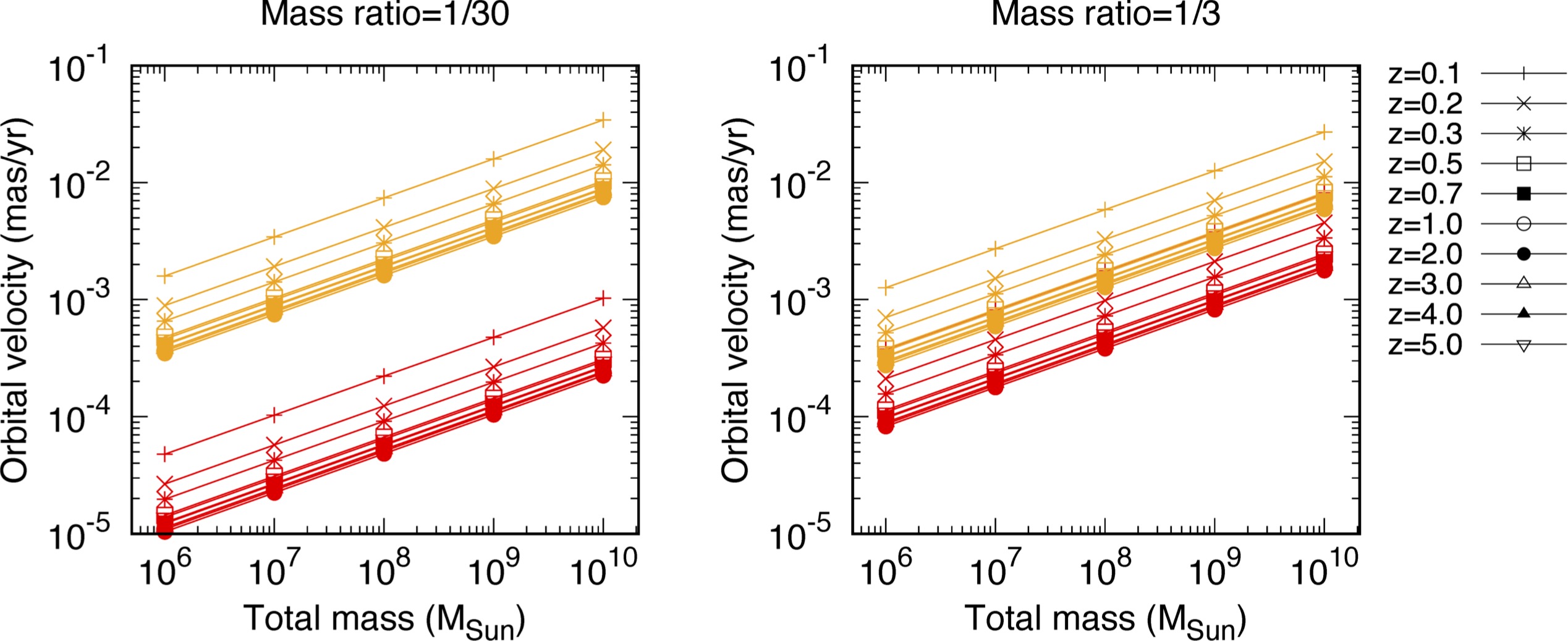}
\caption{Orbital speeds of the black holes for a total mass between $10^6$ and $10^{10}$ M$_{\odot}$ in a redshift range between 0.1 and 5.0, on logarithmic scale. The plot on the left provides this information for a mass ratio of 1/30, the plot on the right for a mass ratio of 1/3.}
\label{speed}
\end{figure}
\subsection{Radio studies of the cosmic web}
\RaggedRight\label{brueggen01}
\vspace*{\baselineskip}

\noindent \textbf{Thematic Areas:} \linebreak $\square$ Stellar Astrophysics \linebreak $\square$ Solar System, Planetary Systems and Habitability \linebreak
$\square$ Circuit of Cosmic   Matter (incl. star formation) \linebreak $\square$ The Galaxy and the Local Group \linebreak
  $\square$   Galaxies and AGN \linebreak $\checked$  Cosmology, Large Scale Structure and Early Universe \linebreak
  $\square$    Extreme conditions in the cosmos, fundamental astrophysics    \linebreak
  $\square$ Interdisciplinary research and technology \linebreak
  
\textbf{Principal Author:}

Name: Marcus Brüggen
 \linebreak						
Institution: Universit\"at Hamburg, Germany
 \linebreak
Email: mbrueggen@hs.uni-hamburg.de
 \linebreak
 
\textbf{Co-authors:} 

M. Hoeft (TLS), D. Hoang (UHH), G. DiGennaro (UHH) 
  \linebreak

The cosmic web usually refers to the structure of galaxies arranged in clusters, filaments, and voids that make up the large-scale structure of the Universe. The synchrotron cosmic web refers to the corresponding synchrotron emission that is produced in intergalactic magnetic fields by electrons accelerated via shocks or turbulence. Studying the synchrotron cosmic web is important to trace the missing baryons in the universe, and constrain magnetogenesis, which refers to the origin and evolution of extragalactic magnetic fields. Detecting magnetic fields in cosmic filaments and voids is key to constrain magnetogenesis scenarios \cite{2016RPPh...79g6901S} because magnetic fields in filaments are more pristine than the magnetic fields in galaxy clusters which are affected by galactic winds and other processes.\\

Simulations predict and model this synchrotron emission \cite{2022A&A...662A..87O}. While these processes are fairly well studied in galaxy clusters, to date, observational confirmation of the synchrotron web has remained elusive \cite{2021Galax...9..109V, 2019Sci...364..981G, 2021A&A...652A..80L}. Stacking of all-sky radio maps has shown the polarization signature of the synchrotron emission with polarization fractions $>$20\% \cite{2023SciA....9E7233V}, but direct detection will only be possible with the ngVLA, owing to its large sensitivity to polarised continuum emission.\\

A related method to probe the magnetic fields in the cosmic web uses linearly polarized radio sources to measure the field strength in thermal magnetized plasma along the line of sight (RM grid). Recent studies have used RM grids and their cross-correlation with large-scale structure to put upper limits on the strength of the intergalactic magnetic field \cite{2020MNRAS.495.2607O}. This has established the RM grid as a powerful method to probe the magnetised cosmic web. The ngVLA can provide a dense RM grid that covers a wide area and a broad range of angular scales, allowing us to measure the strength of the intergalactic magnetic field as a function of spatial scale and redshift.\\

\begin{figure}
    \centering
    \includegraphics[width=0.8\linewidth]{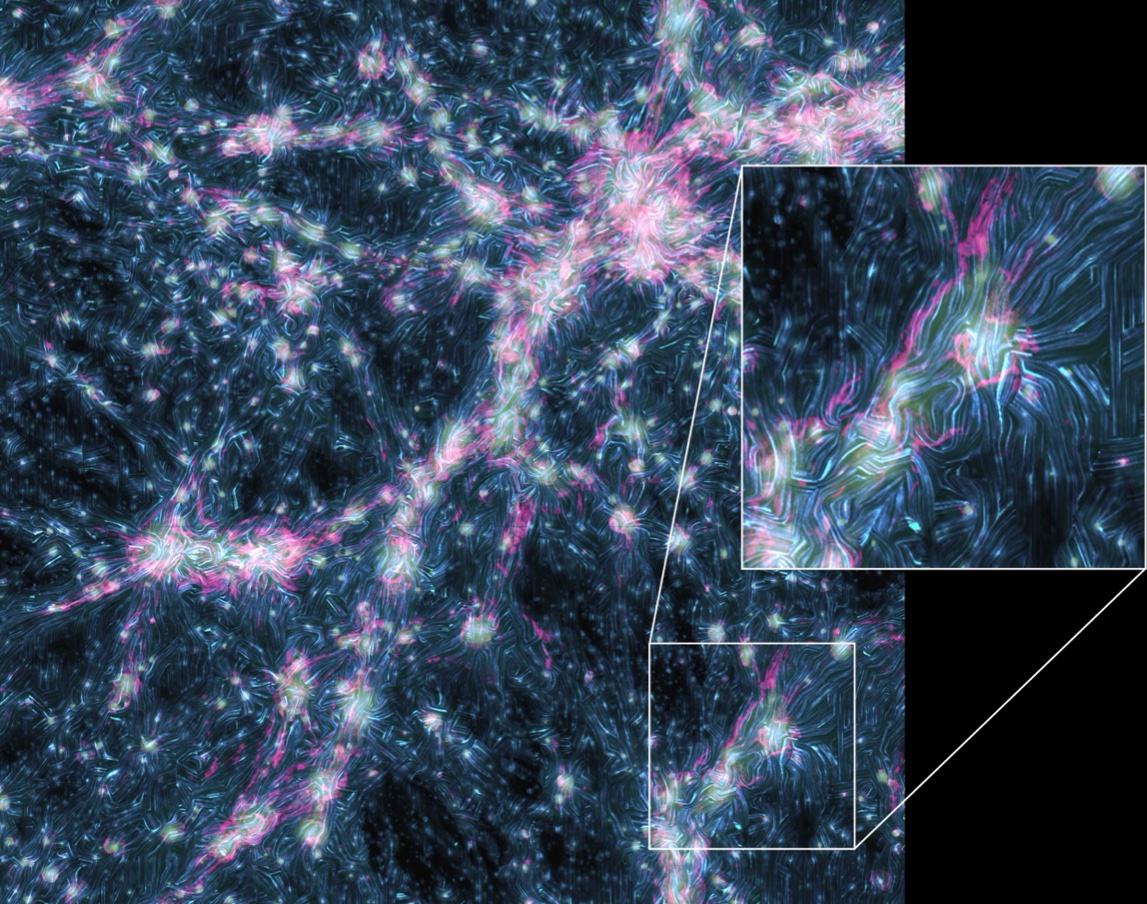}
    \caption{Slice from a cosmological magnetohydrodynamical simulation showing the synchrotron web. The red is the shocked total intensity radio emission. The yellow color shows the temperature and gas density via the X-ray and SZ. The blue lines show the vectors of the magnetic field lines. Simulation by F. Vazza taken from Vernstrom et al. (2023).}
    \label{fig:enter-label}
\end{figure}

Science goals: (i) Magnetic fields in filaments and voids via RM from background sources.\\ 
(ii) Synchrotron emission from the cosmic web to study particle acceleration and magnetic fields.\\
  
\subsection{Galactic Structure}
\RaggedRight\label{brunthaler01}
\vspace*{\baselineskip}

\noindent \textbf{Thematic Areas:} \linebreak $\square$ Stellar Astrophysics \linebreak $\square$ Solar System, Planetary Systems and Habitability \linebreak
$\square$ Circuit of Cosmic   Matter (incl. star formation) \linebreak $\checked$ The Galaxy and the Local Group \linebreak
  $\square$   Galaxies and AGN \linebreak $\square$  Cosmology, Large Scale Structure and Early Universe \linebreak
  $\square$    Extreme conditions in the cosmos, fundamental astrophysics    \linebreak
  $\square$ Interdisciplinary research and technology \linebreak
  
\textbf{Principal Author:}

Name:	Andreas Brunthaler    
 \linebreak						
Institution:  Max-Planck-Institut für Radioastronomie
 \linebreak
Email: brunthal@mpifr-bonn.mpg.de
 \linebreak 
 
\textbf{Co-authors:} Karl M. Menten, Max-Planck-Institut für Radioastronomie
  \linebreak

Determining the structure of the multiple components of the Milky Way has been a topic of active research for decades. Many of the early attempts relied on {\it kinematic distances} that are not very reliable and even unusable in some parts of the Milky Way. In particular the spiral arms are difficult to pinpoint since they are typically obscured by pervasive interstellar dust whose high extinction toward the spiral arms make them not accessible to optical astrometry (i.e. {\it Gaia}) which also makes the use of photometric distances challenging. However, Very Long Baseline Interferometry (VLBI) observations of maser sources (mainly water and methanol masers) in high mass star forming regions have reached accuracies that are comparable or even better than those of {\it Gaia} parallaxes \cite{Brunthaler2011,Sanna2017}. Since these regions are predominantly found in the spiral arms of a galaxy, these measurements enabled for the first time the mapping of the spiral structure of large parts of the Milky Way as well as providing accurate measurements of fundamental parameters of the Milky Way such as its rotation speed and the distance to the Galactic center \cite{Honma2012, Reid2014, Reid2019}.

Recent advancements in the field of VLBI astrometry like MultiView and inverse MultiView \citep{RD2020, Hyland2023}, where multiple calibrators are used to get an improved calibration of the disturbances from the atmosphere, have the potential to improve the accuracy of parallax measurement by up to one order of magnitude to about 1 $\mu$as. The ngVLA, with its multiple antennas at each of the long baseline stations, will be suited perfectly to incorporate these new techniques. Here, one antenna can observe the target source continuously, while the other antennas observe the calibrators. Together with the unprecedented sensitivity, the ngVLA will be able to measure parallaxes of thousands of maser sources in high-mass star forming regions. This will give access to weaker and more distant maser sources to map the spiral arms also on the far side of the Milky Way, a region which will not be accessible to optical surveys. Parallax measurements of masers with the ngVLA will therfore be highly complementary to {\it Gaia}.  
\subsection{Chemical and physical conditions at the dawn of star and planet formation}
\RaggedRight\label{caselli01}
\vspace*{\baselineskip}

\noindent \textbf{Thematic Areas:} \linebreak $\square$ Stellar Astrophysics \linebreak $\checked$ Solar System, Planetary Systems and Habitability \linebreak
$\square$ Circuit of Cosmic   Matter (incl. star formation) \linebreak $\square$ The Galaxy and the Local Group \linebreak
  $\square$   Galaxies and AGN \linebreak $\square$  Cosmology, Large Scale Structure and Early Universe \linebreak
  $\square$    Extreme conditions in the cosmos, fundamental astrophysics    \linebreak
  $\square$ Interdisciplinary research and technology \linebreak
  
\textbf{Principal Author:}

Name: Paola Caselli
 \linebreak						
Institution: Center for Astrochemical Studies, Max Planck Institute for Extraterrestrial Physics
 \linebreak
Email: caselli@mpe.mpg.de
 \linebreak
 
\textbf{Co-authors:} Yuxin Lin, Center for Astrochemical Studies, Max Planck Institute for Extraterrestrial Physics
  \linebreak

The study of the initial conditions in the process of star and planet formation, from cloud to protoplanetary disk scales, will greatly benefit from the high sensitivity, spectral and angular resolution provided by the ngVLA. As shown schematically in Figure\ref{Figure:scheme} (a), ngVLA will allow detailed mapping of the inversion transition of ammonia, to study the connections of dense cores (the units of star formation) to the surrounding cloud. \cite{Choudhury2020} showed that high-sensitivity and high-angular/spectral resolution are needed to disentangle the core emission from that of the core, thus to trace such connection. This is important to understand how dense cores form and evolve during the process of star formation. 

ngVLA will also greatly improve our understanding of the first steps toward protostar and protoplanetary disk formation (see Figure\ref{Figure:scheme} (b)) as contracting prestellar cores can be studied in detail from $\geq$10,000\,au to $\leq$100\,au scales. This is important to compare with predictions from non-ideal MHD simulations of star formation (e.g. \cite{Caselli2019}). Low-J transitions of deuterated molecules, great tracers of cold and dense environments, can be mapped with ngVLA (\cite{Friesen2018}). Ammonia inversion transitions can also be used to reconstruct the physical and chemical structure of prestellar cores (\cite{Pineda2022}). Prestellar cores also host the first steps toward molecular complexity (e.g. \cite{Bacmann2012,JimenezSerra2016,JimenezSerra2022}), with emission typically confined in the outer layers of the cores (e.g. \cite{Vastel2014,Harju2020}). ngVLA will provide great sensitivity to image complex organic molecules, building blocks of pre-biotic molecules, and significantly increase their detection rate. 

Moving from prestellar cores to protostars (see Figure\ref{Figure:scheme} (c)), ngVLA will allow detailed mapping of the recently discovered streamers of material connecting clouds to protoplanetary disks (e.g. \cite{Pineda2020,ValdiviaMena2022}). This will provide a unique way to follow the asymmetric flow of material directly onto the disk and study in detail the effect that such streamers have on the dynamical and chemical evolution of disks, including the dawn of planet formation. Thanks to the lower dust opacities at the ngVLA frequencies, complex organic molecules can also be studied near protostars (e.g. \cite{DeSimone2020}), thus allowing a fresh view of the sublimated ices and more stringent tests on chemical modeling. 

At protoplanetary disk scales, ngVLA will open a new window into the chemical and structural details of disks, where the first steps of planet formation are taking place (see Figure\ref{Figure:scheme} (d) and \cite{SeguraCox2020,Zamponi2021}). Maps of complex organic molecules will become available, thanks to the great ngVLA sensitivity, thus shedding light on the link between disk and exoplanet atmosphere chemistry. Last but not least, ngVLA will provide unique information on the inner zone of protoplanetary disks, where terrestrial planets start to assemble (e.g. \cite{DiFrancesco2019}). Great synergy here is expected between ngVLA, JWST, and GRAVITY+. 

\begin{figure}[h]
\centering
\includegraphics[width=0.9\textwidth]{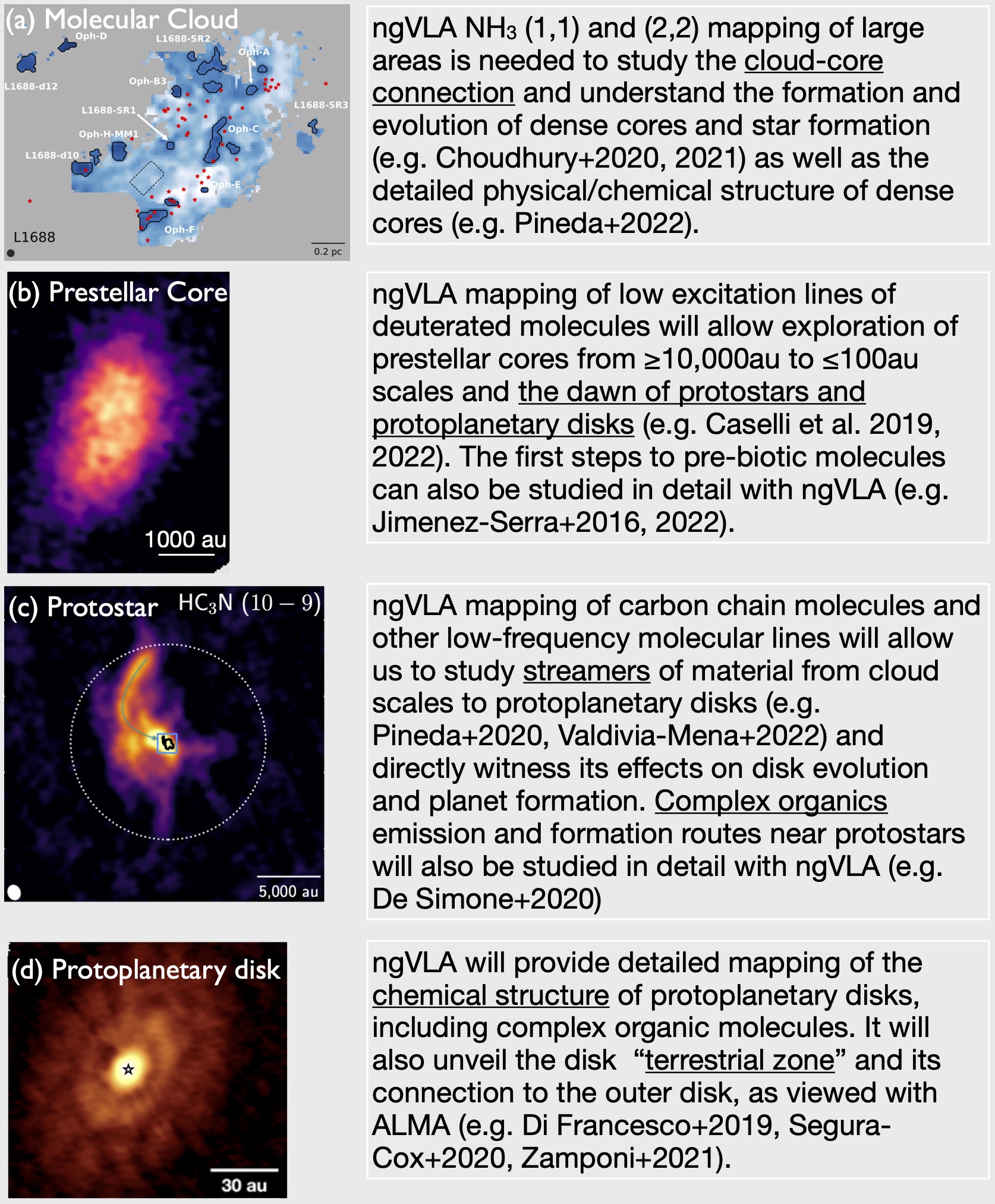}
\caption{Examples of projects that will greatly benefit from ngVLA: (a) the cloud-core connection (the left panel shows the Mach number map obtained using NH$_3$(1,1) and (2,2) mapping of the Ophiuchus molecular cloud (\cite{Choudhury2021}). (b) Detailed physical and chemical structure of prestellar cores and the dawn of protostars and protoplanetary disks (the left panel shows the deuterated ammonia map at 3~mm obtained with ALMA (\cite{Caselli2022}). (c) The streamer connection from cloud to disk scales (the left panel shows the 10,000~au scale streamer discovered by \cite{Pineda2020}). (d) Physical/chemical structure of protoplanetary disks and their "terrestrial zone" (the left panel shows the high-angular resolution image of the youngest protoplanetary disk showing substructure (from \cite{SeguraCox2020}).}
\label{Figure:scheme}
\end{figure}

\subsection{ngVLA studies of gravitational-wave triggers and other energetic transient events}
\RaggedRight\label{elsaesser01}
\vspace*{\baselineskip}

\noindent \textbf{Thematic Areas:} \linebreak $\square$ Stellar Astrophysics \linebreak $\square$ Solar System, Planetary Systems and Habitability \linebreak
$\square$ Circuit of Cosmic   Matter (incl. star formation) \linebreak $\square$ The Galaxy and the Local Group \linebreak
  $\checked$   Galaxies and AGN \linebreak $\square$  Cosmology, Large Scale Structure and Early Universe \linebreak
  $\checked$    Extreme conditions in the cosmos, fundamental astrophysics    \linebreak
  $\square$ Interdisciplinary research and technology \linebreak
  
\textbf{Principal Author:}

Name:	Dominik Elsässer
 \linebreak						
Institution:  TU Dortmund
 \linebreak
Email: dominik.elsaesser@tu-dortmund.de
 \linebreak
 
\textbf{Co-authors:} Kevin Schmidt (TU Dortmund), Matthias Kadler (JMU Würzburg)
 \linebreak

We propose to make energetic transient events, including the early stages of the remnants of binary neutron star mergers, a scientific focus of a German participation to the ngVLA. While these events via the association of  the LIGO-VIRGO event GW170817 with GRB170817A have convincingly been identified as sources of short gamma-ray bursts and as sites of heavy element synthesis, our understanding of the morphology and temporal evolution of the expanding remnants of such collisions is severely lacking. This is especially true for the majority of events where no forward-beamed high-energy emission is observed. Radio-interferometric observations and high frequencies and extreme sensitivities, as offered by the ngVLA, could offer a unique observational window to those processes, and thus – in close concert with gamma-ray observations e.g. with the CTA -- also contribute to elucidating their potential role as cosmic ray acceleration sites.
It is clear, given the large distances to almost all of these sources, that a direct resolution of the morphology, and direct identification of a putative relativistic outflow, may be highly demanding and at times not even feasible. 
Thus, as a first step, we aim to examine the observational data for apparent movements of the emission origin, which would be a strong indicator of relativistic outflows occurring, even for sources where there is negligible forward boosting in the direction of our line of sight, or even de-boosting, and where the source itself is not spatially resolved. This would be a case where, although the source itself would appear to be point-like, the highest possible resolutions are required. For this reason, we propose observations in the cm wavelength range, which corresponds to a frequency range of around 25 GHz.
Identification of an apparent non-stationary and time-variable point source would then indicate movement of the emission site, potentially identified with shocks in a relativistic outflow. 

We aim to embedd our observations in a multi-wavelength context, and to combine them with high-sensitivity observations from gamma-ray observatories like the MAGIC telescopes and the Cherenkov Telescope Array (CTA), where the proposers are long-term collaboration members.
Transients are a central part of the CTA's observing program \cite{cta_science, cta_transient_program}, and high-impact observational data from prototype telescopes is expected already in the next few years \cite{cta_transient}. It is thus timely to think about an ideal approach to include a German participation to the ngVLA into such a strategy.

To analyze the observations, we propose an innovative analysis method based on deep learning techniques. 
The radionets project \cite{radionets_1} focuses on the benefits neural networks can bring to analyzing radio interferometer data.
Radionets cleans up the incompletely calibrated radio interferometer data by reconstructing missing information directly in Fourier space.
This unique approach not only delivers high-resolution images faster, but also tends to produce fewer artifacts in the cleaned images compared to conventional analysis approaches \cite{radionets_2}. This can prove especially beneficial in light of the scientific question posed above, a search for apparent motion in faint point-like, or near point-like, sources.
This deep learning approach is presently being further developed to incorporate information from other frequency bands in a meaningful way to create the imaging product.

Furthermore, we will research machine learning methods as developed e.g. in the Lamarr center for outlier detection to facilitate rapid and reliable transient detection in the observational data. 
\subsection{Hyper Compact Radio sources in the Galactic Center}
\RaggedRight\label{vonFellenberg01}
\vspace*{\baselineskip}

\noindent \textbf{Thematic Areas:} \linebreak $\checked$ Stellar Astrophysics \linebreak $\square$ Solar System, Planetary Systems and Habitability \linebreak
$\square$ Circuit of Cosmic   Matter (incl. star formation) \linebreak $\checked$ The Galaxy and the Local Group \linebreak
  $\square$   Galaxies and AGN \linebreak $\square$  Cosmology, Large Scale Structure and Early Universe \linebreak
  $\checked$    Extreme conditions in the cosmos, fundamental astrophysics    \linebreak
  $\square$ Interdisciplinary research and technology \linebreak
  
\textbf{Principal Author:}

Name: Sebastiano von Fellenberg
 \linebreak						
Institution: Max Planck Institute for Radio Astronomy, Bonn
 \linebreak
Email: sfellenberg@mpifr-bonn.mpg.de
 \linebreak
 
\textbf{Co-authors:} Arman Tursunov, Andrei Lobanov, Gunther Witzel 
  \linebreak

\textbf{Stellar Remenants in the Galactic Center --}
{A large number of stellar remnants in expected to be found in the Galactic Center} \citep[e.g.,][]{Morris1993,Hailey2018,Generozov2018}, which is trivially explained by the large number of massive stars currently observable in the Galactic Center today (e.g., \cite{vonFellenberg2022}). 
Of these stellar remants, {at least one bonafide stellar remnant has been detected}, that is the Magnetar SGR J174-2900 located just 3'' from Sgr~A* \cite{Kennea2013,MoriKe2013,Rea2013}.

\textbf{A census of Hyper Compact Objects --}
\cite{Zhao2022} conducted {a deep $33~\mathrm{GHz}$ and $44.6~\mathrm{GHz}$ JVLA survey of the central arcseconds} and identified 68 hyper-compact objects in the Galactic Center (see \autoref{fig:hypercompact_objects}). The nature of these objects is not finally determined, however, the authors categorized the objects into three categories based on their spectral properties.
\begin{compactitem}
\item{\bf Flat-spectrum HCRs:}
\cite{Zhao2022} speculated that these sources may be unresolved peaks in HII regions because their characteristic emission profile matches that of the free-free emission.

\item{\bf Inverted-spectrum HCRs:} 
\cite{Zhao2022} argue that two scenarios are likely: The spatial correlation indicates these sources are due to strong stellar winds. Nevertheless, the spectral similarity of these sources with the spectrum of the GC magnetar indicates that a fraction of the sources may be other magnetars.

\item{\bf Steep-spectrum HCRs:} These sources have spectral indices that indicate none-thermal emission. \cite{Zhao2022} argue that this implies that {{these sources are most likely associated with massive stellar remnants}}. 
\end{compactitem}

The ngVLA  will drastically improve the sensitivity compared to JVLA, and {we thus expect a significant increase} the number of detectable HCRs. If these sources can be robustly identified as the radio counterparts of compact stellar remnants such as black holes or neutron stars, their measurement will allow to {derive the first direct constraints of the dark mass distribution close to a super massive black hole}.\\ 

\textbf{Astrometry of HCR-sources --}
Due the extreme senstivity of the ngVLA, we expect to achieve nano-arcsecond astrometry, implying that these objects can be followed with extreme astrometric precssion. 

Due to the likely absence of emission lines for these sources, astrometric information alone is insufficient to determine orbital solutions. Thus these objects cannot directly be used as gravitional probes. They still offer the chance of being used as for parallax studies, which is on the order of $1/8200pc = x [''] \sim 125~\mathrm{\mu as}$. The parallax may thus yield an independent distance measurement to the Galactic Center, similar to the maser stars used traditionally (e.g., Reid \& Brunthaler 2020). \\

\textbf{Binary-systems --}
Due to graviational in-spiral and the Hills mechanism, binary systems are not stable close to Sgr A*. {At larger distances, binaries can survive for a long time and are thus expected to be found}. Two binary systems, consisting of at least one star have been identified so far \citep{Genzel2010}. HCR sources may represent the {radio-detectable counterpart of NS or BH-binary system}, {ngVLA observations offer the chance for an astrometric detection of such binaries}.
Depending on the separation distance, binary systems can be divided into close and wide binaries. The former are often associated with transient phenomena and as gravitational wave (GW) emitting sources, allowing one to test gravity in a strong regime. %On the other hand, the wide binaries are expected to interact more frequently with neighboring objects (Erez et al. 2023) including stars, influencing dynamics, thus changing their eccentricity. For the Galactic center, it is especially important for the measurement of the gravitational redshifts during pericentre passages of S-stars.  %[\href{https://ui.adsabs.harvard.edu/abs/2023arXiv231002558M/}{REFERENCE}], 

Direct observation of  binary systems in the Galactic center could  be important in the context of future GW telescopes, such as LISA, providing tighter constraints on the expected rate of GW events and observing the GW emitting binaries prior to merger.  In this sense, in addition to the NS-BH binary studies, mentioned above, the systems without NS, but involving black holes and white dwarfs are of a special interest. According to \citep{Wang2021} BH-BH  binaries and those involving WD provide the most plausible systems observable by future LISA GW experiment.

% \paragraph{The rosseta measurement}
% A Binary consisting of a ngVLA detectable HCO + a detectable star, which would allow high sensitivity radial velocity measurements. For instance, if the star is a giant, NIR K-band spectroscopy will give m/s precision radial velocities (Davies et al. 2018). Thus NIR + ngVLA observations will allow for direct measurement of the both dynamical components, with 6D constraints on the infrared star, and 4D constraints on the motion of the BH or NS, allowing for unprecedented accuracy in tests of the gravitational potential that both stars see. 

\begin{figure}
    \centering
    \includegraphics[width=0.95\linewidth]{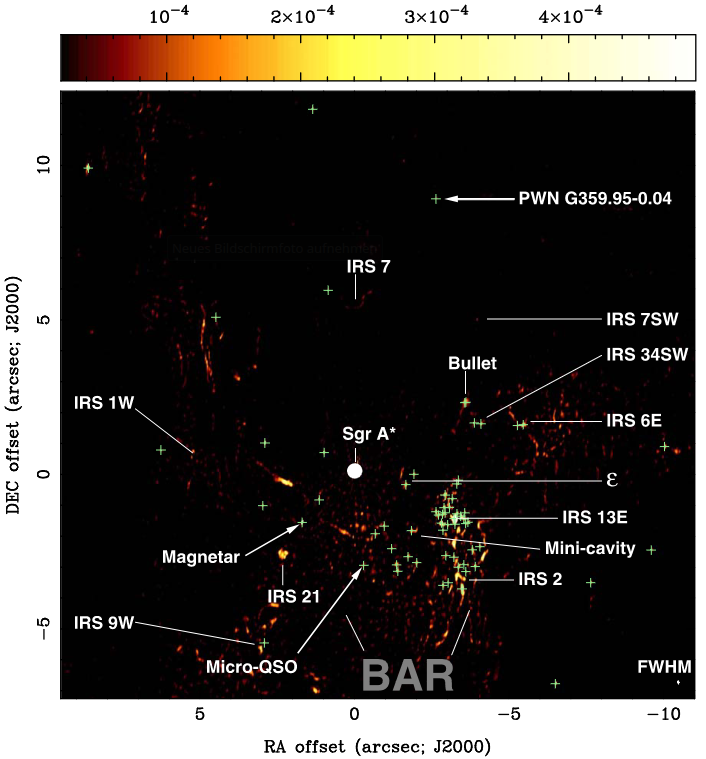}
    \caption{JVLA map of the Galatic Center. Green crosses show the location of hyper-compact objects as classified by \cite{Zhao2022}.}
    \label{fig:hypercompact_objects}
\end{figure}

\subsection{Substructures of dust rings in protoplanetary disks}
\RaggedRight\label{flock01}
\vspace*{\baselineskip}

\noindent \textbf{Thematic Areas:} \linebreak $\square$ Stellar Astrophysics \linebreak $\checked$ Solar System, Planetary Systems and Habitability \linebreak
$\square$ Circuit of Cosmic   Matter (incl. star formation) \linebreak $\square$ The Galaxy and the Local Group \linebreak
  $\square$   Galaxies and AGN \linebreak $\square$  Cosmology, Large Scale Structure and Early Universe \linebreak
  $\square$    Extreme conditions in the cosmos, fundamental astrophysics    \linebreak
  $\square$ Interdisciplinary research and technology \linebreak
  
\textbf{Principal Author:}

Name: Mario Flock
 \linebreak						
Institution:  Max-Planck Institute for Astronomy
 \linebreak
Email: flock@mpia.de
 \linebreak
 
\textbf{Co-authors:} Takahiro Ueda, MPIA, Til Birnstiel, LMU, Sebastian Wolf, Uni Kiel, Myriam Benisty, IPAG Grenoble
\linebreak

Since the revolution of high-angular resolution observations at submillimeter/millimeter wavelengths of protoplanetary disks we have learned, their morphology is made of azimuthally symmetric or point-symmetric substructures, in some cases with spiral arms or localized spur- or crescent-shaped features. 

Even the majority of theoretical studies to explain the observational results have focused on disk models with embedded planets, only in very few cases have exoplanets been detected in these systems. 

Furthermore, the concentration of solid material at pressure traps are expected to appear before planets form, as they are necessary to drive the concentration of small solids which can lead to the formation of planetesimals. 

In the work series \citep{flo20,bla21} we present observational predictions from high-resolution 3D radiative hydrodynamical models that follow the evolution of gas and solids in a prototoplanetary disk. We focus on substructures in the distribution of millimeter-sized and smaller solid particles produced by the vertical shear instability.

\begin{figure*}[h]
\centering
\includegraphics[width=0.49\textwidth]{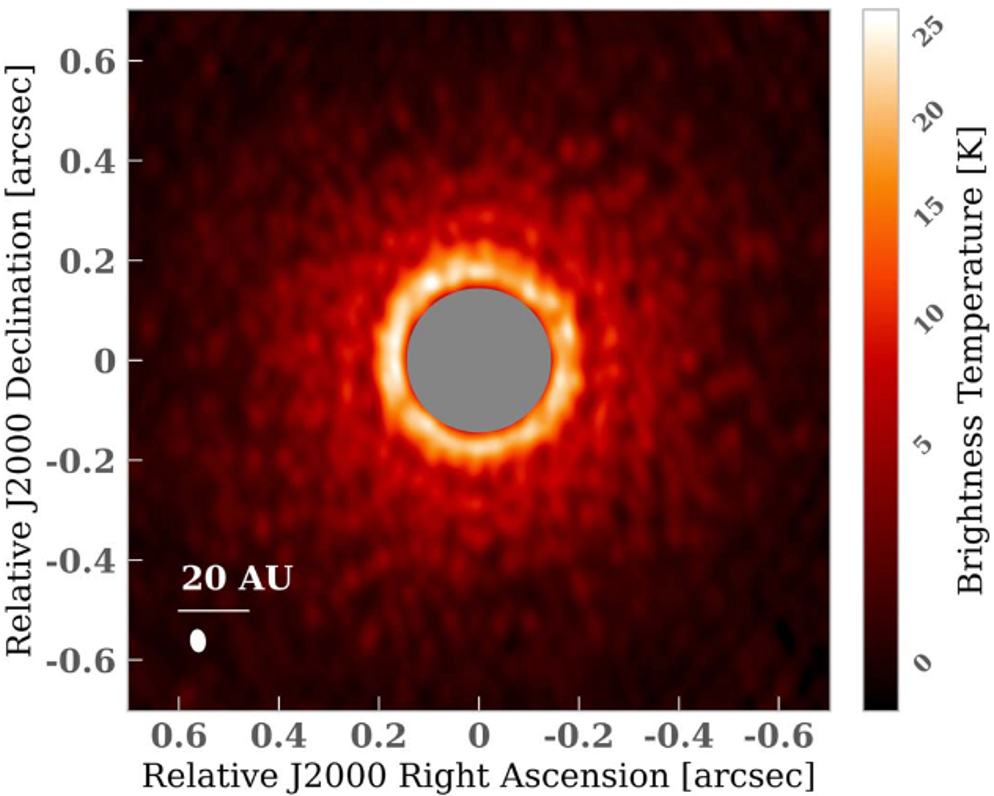}
\includegraphics[width=0.49\textwidth]{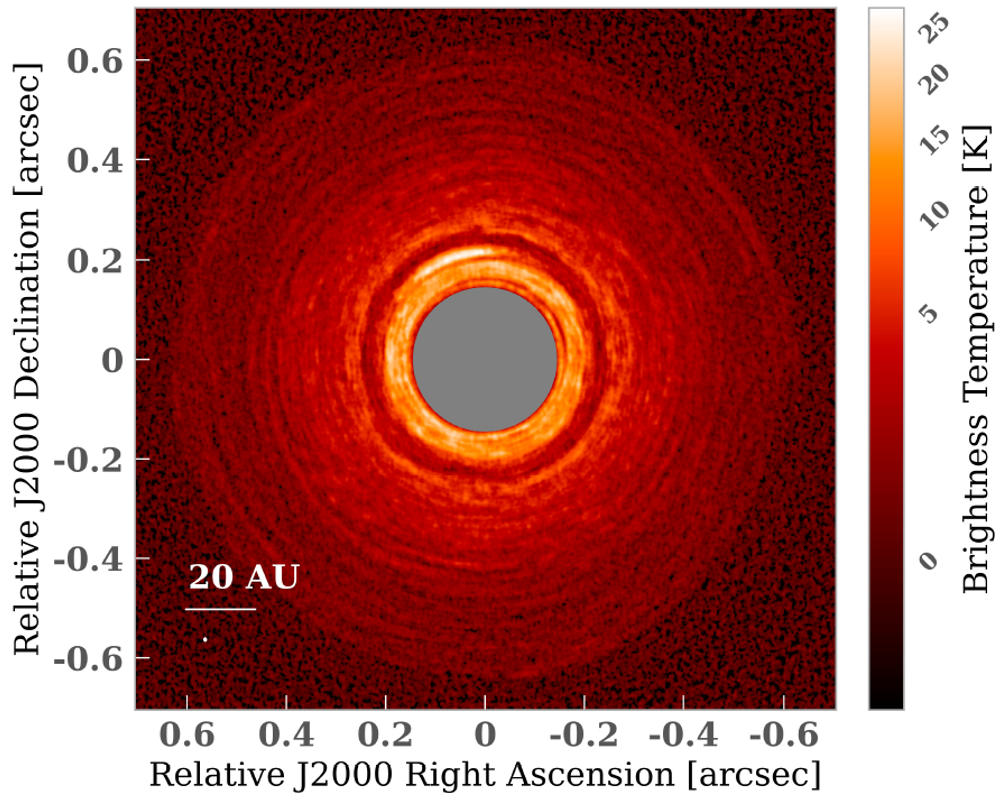}
\caption{
Left: Image of the ALMA dust-continuum emission at 1.3 mm obtained with observational characteristics that resemble those obtained for the ALMA DSHARP project. The resultant rms noise on the map and synthesized beam, shown in the lower left corner, are about 15.0 $\mu$Jy beam$^{-1}$ and 43 mas $\times$ 28 mas, respectively. Right: Simulated long ngVLA observations for the dust emission at 3 mm for our face-on disk model. The resultant rms noise and the synthesized beam, correspond to about 0.1 $\mu$Jy beam$^{-1}$ and 8 $\times$ 6 mas, respectively.}
\label{fig1}
\end{figure*}

We show that their characteristics are compatible with some of the shallow gaps detected in recent observations at sub-mm/mm wavelengths and present predictions for future observations with better sensitivity and angular resolution with ALMA and a Next Generation Very Large Array. 
\subsection{Probing SgrA$^\star$ flares with the ngVLA}
\RaggedRight\label{fromm01}
\vspace*{\baselineskip}

\noindent \textbf{Thematic Areas:} \linebreak $\square$ Stellar Astrophysics \linebreak $\square$ Solar System, Planetary Systems and Habitability \linebreak
$\square$ Circuit of Cosmic   Matter (incl. star formation) \linebreak $\checked$ The Galaxy and the Local Group \linebreak
  $\square$   Galaxies and AGN \linebreak $\square$  Cosmology, Large Scale Structure and Early Universe \linebreak
  $\checked$    Extreme conditions in the cosmos, fundamental astrophysics    \linebreak
  $\square$ Interdisciplinary research and technology \linebreak
  
\textbf{Principal Authors:}
%Name: Erandi Chavez \linebreak
%Institution: Center for Astrophysics Harvard \& Smithsonian, 60 Garden St, Cambridge, MA 02138, USA \linebreak
%Email: erandi.chavez@cfa.harvard.edu
%\linebreak
Christian M. Fromm	
 \linebreak						
Institution:  JMU W\"urzburg, W\"urzburg, Germany
 \linebreak
Email: christian.fromm@uni-wuerzburg.de
\linebreak

\textbf{Co-authors:}  Luciano Rezzolla (Inst. f\"ur Theor. Physik, Goethe Universit\"at, Frankfurt, Germany),
Michael D. Johnson (CfA Harvard \& Smithsonian, USA)
 \linebreak

The supermassive black hole (SMBH) in the centre of our Milky Way, SgrA$^\star$, represents a perfect laboratory for fundamental and plasma physics \cite{BHC2017}. Observations with the GRAVITY experiment provided a mass value of $4.14\times 10^6\,M_\odot$ \cite{Gravity2019}. However, the spin of the black hole, $a^\star$, and its inclination, $\vartheta$, are less constrained. The recent Event Horizon Telescope (EHT) observation shows a ring like structure and the best bet models are characterised by $a^\star\geq0.5$ and an inclination $\vartheta\leq 30^\circ$ and favour a jet-dominated emission model \cite{EHTSgrAP1}. In addition, SgrA$^\star$ is well known for its flaring behaviour in the near infrared and in the x-ray regime \cite{Gravity2018,Witzel2018}. However, the detailed mechanism for these flares are still on debate.\\
A promising model for the flares in the galactic centre is related to magnetic re-connection, i.e., the re-configuration of the magnetic field topology while energizing the radiative particles. Recent Particle in Cell (PIC) simulations have shown that during the re-connection events electrons can be accelerated from a thermal Maxwell-J\"uttner distribution into a non-thermal distribution \cite{Ball2016,Meringolo2023} which could be connected to the SgrA$^\star$ flares. Global general relativistic magneto-hydro-dynamical (GRMHD) simulations of accreting black holes have identified the equatorial current sheet generated during the accretion process as possible location for magnetic  re-connection events. This current sheet is in-stable against the tearing instabilities and lead to the formation of plasmoids, i.e. magnetic islands, while heating the particles \cite{Nathanail2020,Ripperda2022,Jiang2023}. The formed plasmoids orbit as coherent structure the black holes and are sheared out over time in the larger accretion flow \cite{Nathanail2022}. In addition to the plasmoids generated in the equatorial current sheet plasmoids can also be generated in the shear layer between the jet and wind region due to Kelvin-Helmholtz (KH) instabilities. Some of those KH-driven plasmoids grow in size while gaining energy and finally become unbound by the black hole \cite{Nathanail2020}. Based on our simulations we could track large-scale plasmoids out to a radius of $r\sim100$ gravitational radii \cite{Nathanail2020}.\\

Indeed, the observed GRAVITY flares in the infrared could be associated with motion of plasmoids \cite{Gravity2020}. Thus the question arises whether we can detect the formation of plasmoids and further investigate the physical mechanism of the SgrA$^\star$ flares with future observations. The ngVLA provides an unprecedented coverage of the u-v plane, high frequency observing capabilities from 70\,GHz to 116\,GHz together with improved sensitivity\footnote{https://ngvla.nrao.edu/page/performance}. Given that the characteristic time scale of the galactic centre is $\sim 20\,s$ standard imaging of several hours long VLBI observations will merge several orbital periods of the plasmoids and their signature will most likely smeared out. However, the impressive u-v coverage of ngVLA will allow for snapshot-imaging\footnote{see for example https://library.nrao.edu/public/memos/ngvla/NGVLA105.pdf}, i.e, the splitting of the observations into several short time bins during the imaging process, and thus mitigate the variability of SgrA$^\star$ during the imaging. Given the common flaring activity of the galactic centre observing campaigns of several consecutive days with the ngVLA will most likely catch a flare and thus allow us test the hypothesis of magnetic re-connection driven outburst.

\begin{figure}[!htb]
        \centering
        \includegraphics[width=1.0\linewidth]{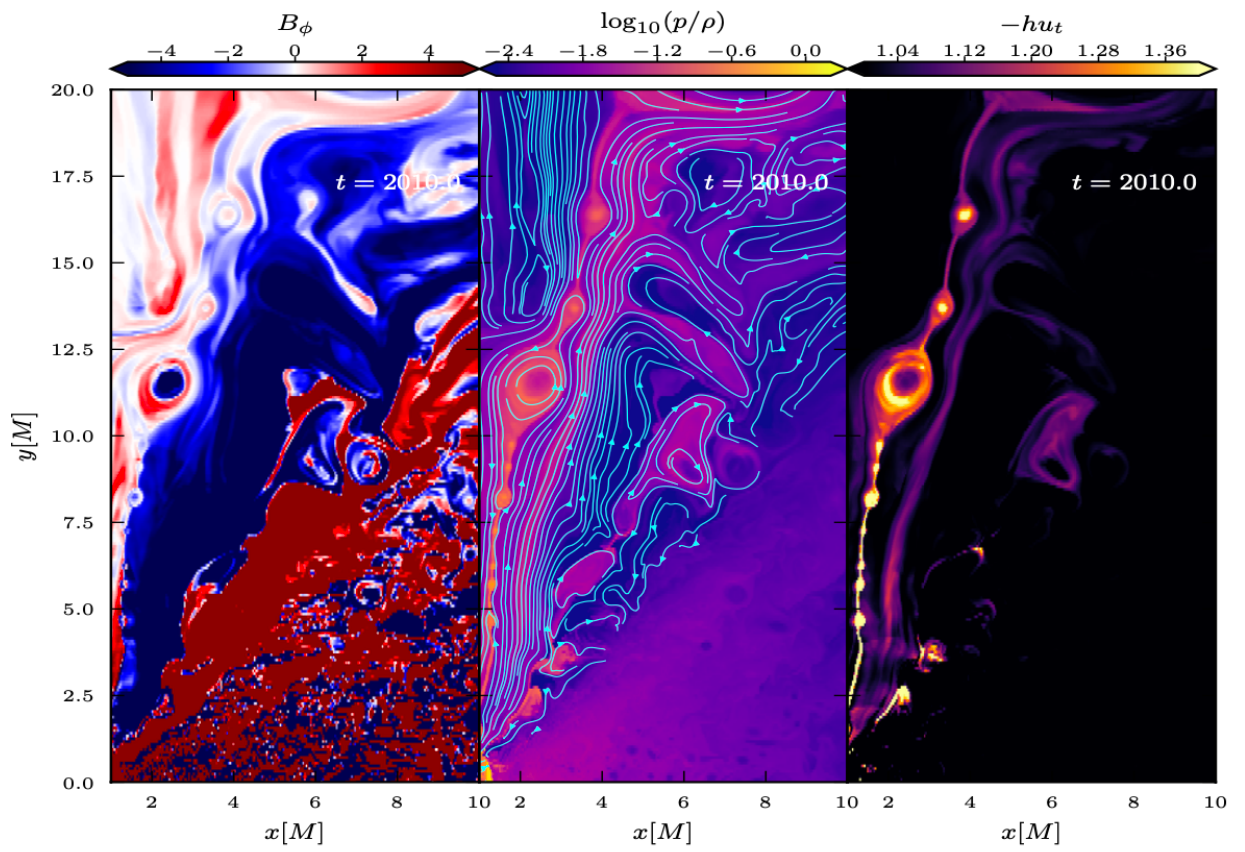}
        \caption{Properties of a plasmoid chain formed in GRMHD simulations of a black hole with $a^\star=0.94$. The panels show from left to right the azimuthal component of the magnetic field, $B_\phi$, the ratio of pressure to density and the Bernoulli parameter, $-hu_t$, indicating bound and unbound plasma. The formed plasmoids are clearly visible in the left most plot \cite{Nathanail2020}.} 
\end{figure}

To summarize, the ngVLA with its distinguishing features mentioned above provides the unique opportunity to test the formation of plasmoids during magnetic-reconnection events and their possible connection to the flaring behaviour of SgrA$^\star$. Therefore, future ngVLA observations of SgrA$^\star$ in close combination with state-of-the art numerical modelling will allow us to provide answers to one of the most pending questions regarding the black hole in our Milky Way.
\subsection{Jet Launching and Radiation Microphysics}
\RaggedRight\label{fromm02}
\vspace*{\baselineskip}

\noindent \textbf{Thematic Areas:} \linebreak $\square$ Stellar Astrophysics \linebreak $\square$ Solar System, Planetary Systems and Habitability \linebreak
$\square$ Circuit of Cosmic   Matter (incl. star formation) \linebreak $\square$ The Galaxy and the Local Group \linebreak
  $\checked$   Galaxies and AGN \linebreak $\square$  Cosmology, Large Scale Structure and Early Universe \linebreak
  $\checked$    Extreme conditions in the cosmos, fundamental astrophysics    \linebreak
  $\square$ Interdisciplinary research and technology \linebreak
  
\textbf{Principal Author:}

Name: Christian M. Fromm	
 \linebreak						
Institution: Institute for Theoretical Physics and Astrophysics, JMU W\"urzburg, Emil-Fischer Str. 31, 97074 W\"urzburg, Germany
 \linebreak
Email: christian.fromm@uni-wuerzburg.de
 \linebreak
 
\textbf{Co-authors:} Anne-Kathrin Baczko (Chalmers Univ. Gothenburg, Sweden), J. Anton Zensus (MPIfR), Luciano Rezzolla (Inst. f\"ur Theor. Physik, Goethe Universit\"at, Frankfurt, Germany)
 \linebreak

Relativistic jets are the most prominent and fascinating feature in Active Galactic Nuclei (AGN). 
They are anchored in the nucleus of AGNs, propagating as a collimated outflow several thousands of 
light years through their host galaxy before finally encountering the interstellar medium. 
According to our current understanding, relativistic jets are powered either by the potential energy
released during the accretion of matter from the surrounding disk onto supermassive black holes (SMBHs)
or by directly tapping the rotational energy of the spinning SMBH \cite{Bla77,Bla82}.
In both scenarios magnetic fields play a crucial role. Yet, its origin and geometry is unclear.\\

Most of our knowledge on relativistic jets is inferred from Very Long Baseline Interferometric (VLBI) 
observations which can resolve their structure from their outskirts down to the horizon scales \cite{2019ApJ...875L...1E,Algaba2021}. Given the advances in numerical methods and available computational resources the launching of relativistic jets from black hole accretion disk systems can be modelled using General Relativistic Magneto-Hydrodynamic (GRMHD) simulations \cite[see][for a review]{Mizuno2022}. Despite the substantial observational and theoretical progress in understanding jets many fundamental questions regarding their formation, acceleration and collimation remain unanswered \citep[see][for a review]{Blandford2019}.
\newline The recent 86\,GHz observations of M\,87 using the Global Millimetre VLBI array detected both, a ring like inner structure and a launched large scale jet providing the opportunity to observe and model both main features simultaneously \cite{Lu2023}.\\
The ngVLA will provide both, an unprecedented u-v coverage, a wide frequency coverage (1\,GHz-116\,GHz) and improved sensitivity. These unique capabilities will allow us to improve not only the observations of \citet{Lu2023} but also observations of relativistic jets in AGNs in general. We can further investigate i) the magnitude and geometry of the magnetic field in the direct vicinity of SMBHs and ii) possible particle acceleration mechanisms.
Performing ngVLA multi-frequency observations of AGNs and using modern image reconstruction algorithms \cite{Chael2023} the distribution of the turnover flux density, $S_\mathrm{t}$, and frequency, $\nu_\mathrm{t}$, can be obtained. From the turnover values the magnitude of the magnetic field and the particle density can be computed \cite{Fromm2022,Osorio2022,Fromm2023}. The geometry of the magnetic field can be extracted from possible polarisation observations which will also benifit from the improved observing capabilties of the ngVLA. In addition to the turnover values, the spectral slope will be obtained from the multi-frequency observations. Comparing the distribution of the spectral index along the jet with advanced GRMHD simulations \cite{Mizuno2021} will provide details on the particle acceleration mechanism and side. Figure 1 shows the distribution of the turnover frequency, $\nu_\mathrm{t}$, turnover flux density, $S_\mathrm{t}$, and optically thin spectral index, $\alpha$ from a 3D GRMHD simulations of M\,87 for a thermal particle distributions (top) and for a kappa distribution (bottom) \cite{Fromm2023}.\\

\begin{figure}[h]
        \centering
        \includegraphics[width=1.0\linewidth]{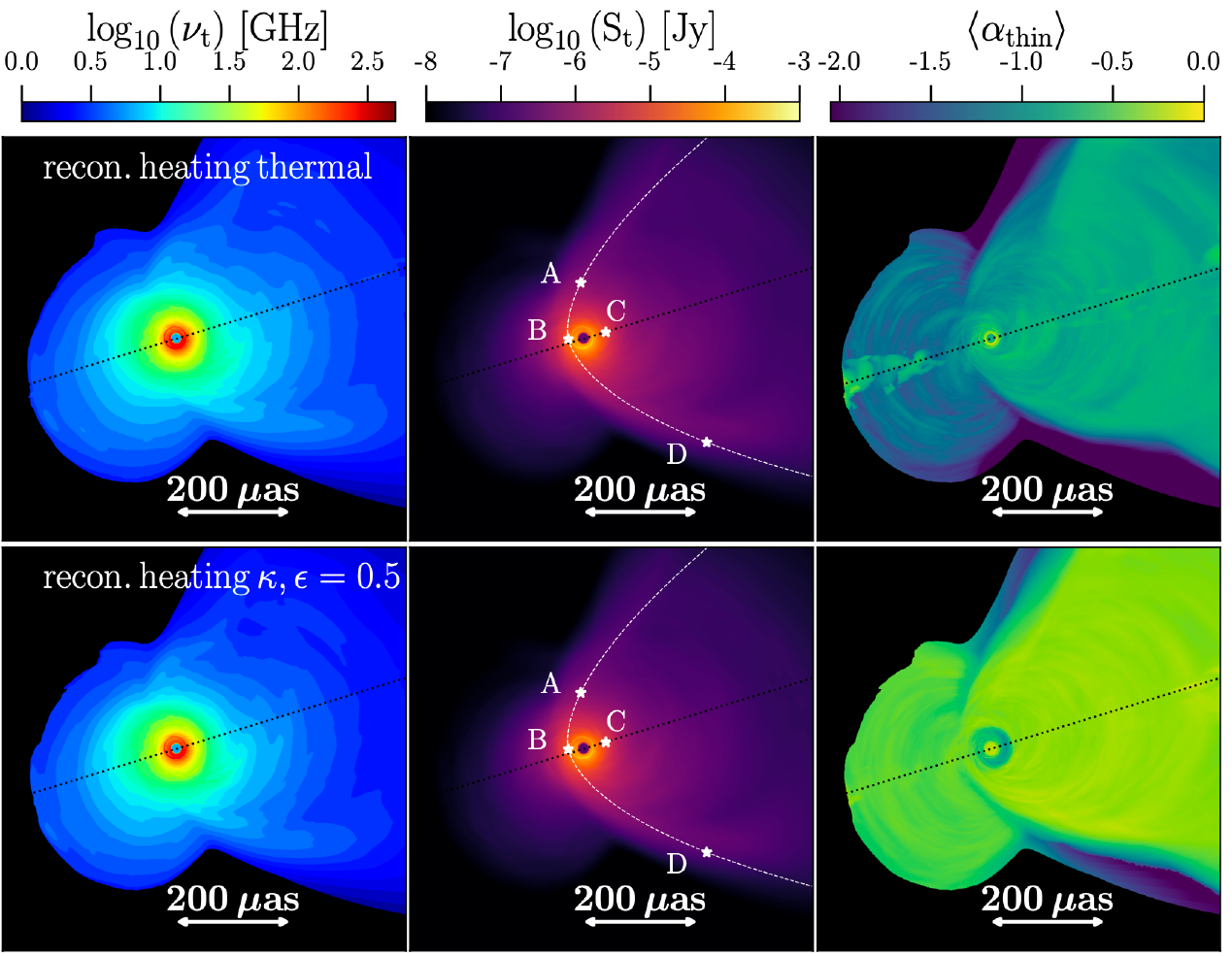}
        \caption{Turnover values from a 3D GRMHD simulation for M87 including magnetic reconnection heating.  Distribution of the turnover frequency, $\nu_\mathrm{t}$, the turnover flux density, $S_\mathrm{t}$, and optically thin spectral index, $\alpha$, for a thermal particle distribution (top) and non-thermal (kappa) particle distribution (bottom) \cite{Fromm2023}}. 
\end{figure}

To summarize, future ngVLA observations in close combination with GRMHD simulations will shield light on the disk-jet connection in AGNs and on the acceleration of particles in relativistic jets and will provide answers to the pending questions of AGN and jet physics.
\subsection{Bi-/Multistatic Radar for Space Situational Awareness and Near-Earth Object Studies}
\RaggedRight\label{fuhrmann01}
\vspace*{\baselineskip}

\noindent \textbf{Thematic Areas:} \linebreak $\square$ Stellar Astrophysics \linebreak $\checked$ Solar System, Planetary Systems and Habitability \linebreak
$\square$ Circuit of Cosmic   Matter (incl. star formation) \linebreak $\square$ The Galaxy and the Local Group \linebreak
  $\square$   Galaxies and AGN \linebreak $\square$  Cosmology, Large Scale Structure and Early Universe \linebreak
  $\square$    Extreme conditions in the cosmos, fundamental astrophysics    \linebreak
    $\checked$ Interdisciplinary research and technology \linebreak
  
\textbf{Principal Author:}

Name:	
Lars Fuhrmann \linebreak						
Institution:  
Fraunhofer-Institut für Hochfrequenzphysik und Radartechnik (FHR) \linebreak
Email: 
lars.fuhrmann@fhr.fraunhofer.de \linebreak
 
\textbf{Co-authors:}
Marcel Laubach,   
Robert Perkuhn,  
Stephan Stanko,  
Marcus Albrecht (FHR) \linebreak

In contrast to the decades following the beginning of the Space Age with a fairly constant number of rocket launches per year, the latter has increased dramatically over the last 5--10 years. The number of payloads placed into Earth orbit increased accordingly. Compared to the 221 objects launched into orbit in 2016, this number changed by a factor 12 in 2022 \cite{ESAreport2022}. With this upward trend and pace, the density of Earth orbiting space objects, their collision probability and the (large-/small-scale) space debris population is also increasing continuously \cite[e.g.][]{Kessler1978,undseth2021}. 

Among questioning the future general access to Space due to the Kessler syndrome (at least for certain orbit regimes), these developments pose severe challenges to current and future Space Situational Awareness (SSA) and Space Traffic Management (STM) capabilities in e.g. detecting, tracking, cataloging and characterizing these growing number of Earth-orbiting objects (operational satellites and space debris). The currently most relevant orbit regime for a dynamic understanding of activities is the Low-Earth Orbit (LEO, $\lesssim$ 2000\,km), however, activities are increasing up to Geosynchronous/Geostationary Orbits (GEO belt: $\sim$\,36.000\,km). Beyond these classical regimes, cislunar space out to the Earth-Moon Lagrange Point 2 is becoming increasingly important. The growth of current and planned space activities with e.g. sending spacecraft into and through cislunar space for future robotic and human Moon/solar system exploration requires an enhancement of SSA capabilities also towards these distances \cite[e.g.][]{frueh2021}.

In addition, the proliferation of smaller space assets with decreased radar cross section and more compact mountings requires better detection as well as higher resolution imaging capabilities for detailed object analysis in the future. Finally, the vast majority of space debris ($>$\,99\,\%) is currently not tracked at all giving no possibility to accurately predict potential collisions and provide in-time collision avoidance. The small-scale space debris population at size-scales of about 1\,mm to 1\,cm with about 130 million objects in Earth orbit is of particular concern given its growing collision hazard \cite{ESAreport2022}. Objects of this size are dominating the collision risk profile of operational spacecraft in LEO \cite{maclay2021}. Orbital debris models are important to statistically describe the current and future debris environment and to perform impact risk assessment for spacecraft and satellites. Here, sensitive radar measurements of the orbital debris flux as well as single fragmentation events down to size scales as small as possible and towards large distances are required \cite{mehrholz2004,letsch2009,vierinen20192018,cerutti2023}.    

While ground-based radar is state-of-the-art for SSA in LEO, the GEO and cislunar regimes are particularly challenging due to the strong distance dependence ($\sim R^{-4}$) of radar sensitivity \cite[e.g.][]{leushacke1993,mehrholz1997}. One possibility to improve on the sensitivity of a monostatic radar system (co-located transmit (TX) and receive (RX) paths) is to increase the collecting area by extending the monostatic system with a high gain/large aperture RX system in a bistatic configuration. A multistatic configuration with an array of large aperture RX and powerful TX systems would further improve the detection, tracking and e.g. orbit determination and collision avoidance capabilities. In this framework, the future ngVLA will be extremely valuable as a large array (or selected stations of multiple elements) of low-noise receivers when paired with high transmit-power radars on different baseline scales. Long (up to intercontinental) baseline bi-/multistatic configurations would mostly address the MEO/GEO and cislunar distance regime due to common geometric visibility constraints. Medium (several hundred kilometer) baselines would also allow observations in LEO, including the possibility of simultaneously obtaining estimates of the 3D position and velocity vector of an object given the delay and Doppler measurements obtained at different TX/RX pair geometries (multi-lateration). In addition, bi-/multistatic high-resolution Inverse Synthetic Aperture Radar (ISAR) imaging \cite{yousfi2024} (see Fig. \ref{fig:isar}), as well as interferometric 3D image reconstruction of LEO objects becomes possible given the large number of simultaneous TX/RX pairs at different viewing angles \cite{albrecht2023}. This will strongly enhance the capabilities in object identification and characterization including e.g. damage analysis, status assessment for re-entries and Active Debris Removal (ADR) missions, as well as improved spin axis and intrinsic rotation analysis of LEO objects \cite[e.g.][]{rosebrock2011,svenja2017,karamanavis2023}. The enhanced sensitivity would also allow to further push the object size limit for small-scale space debris population studies and debris cloud analysis to the lower end of the mm-regime ($\lesssim$\,5\,mm).   

Given the ngVLA baseline receiver configuration with its broad frequency coverage and large bandwidths including the typical radar operational frequency bands (L-Band, S-Band, X-Band, Ku-Band, Ka-Band), several existing radar facilities would be suitable as powerful illuminator. Among e.g. the future radar capability at the Green Bank Telescope (GBT) in the U.S., the ongoing system upgrades (including a new broadband Ka-Band radar) will make the 34\,m Tracking and Imaging Radar (TIRA) of Fraunhofer FHR in Germany a suitable transmitter and receiver in such bi-/multistatic configuration \cite{mehrholz1993,mehrholz1997radar,klare2024}). Furthermore, new phased-array radar systems and distributed networks for space surveillance, such as the German Experimental Surveillance and Tracking Radar and its extensions (GESTRA, \cite{reising2022,albrecht2023}) could be integrated into such a configuration to further enhance baseline and sky coverage, transmit power and other capabilities relevant for space surveillance and tracking.
\vspace*{\baselineskip}

As natural objects relevant for SSA, planetary defense and planetary science, Near-Earth Objects (NEOs) denote asteroids and comets having their perihelion distance less than 1.3 au. A fraction of this population is posing a possible future impact hazard to Earth. In identifying these Potentially Hazardous Asteroids (PHAs), dedicated NEO surveys are currently conducted by ground- and space-based optical/IR facilities \cite{milam2019}. These surveys are still far from being complete for object sizes $\ge$\,140\,m and, in particular, for Tunguska/Chelyabinsk impactor sizes of the order $\sim$\,20--50\,m. Consequently, the number of new survey detections will largely increase over the next decade. These optical/IR detections with plane-of-sky astrometry require post-discovery rapid follow-up radar tracking to obtain precise orbit determination by adding high-precision line-of-sight range and Doppler measurements. 
 
In addition, physical characterization is important for e.g. population studies and possible mitigation strategies for objects with high impact probability. Deep space radar capabilities are crucial in providing these necessary physical/dynamical characterization allowing to study size, shape, bulk density, surface characteristics, rotational period, spin axis and possible companions \cite[e.g.][]{ostro2002}. Given the large distances involved, large antenna apertures and high-power transmitters are required. In the past, the now defunct Arecibo Observatory have been a key facility, while NASA's Goldstone Solar System Radar at X-Band (GSSR with DSS-14/DSS-13 antennas in bistatic configuration, occasionally including GBT/VLBA/VLA) is currently the primary planetary radar facility though with limited observing time \cite[e.g.][]{naidu2016}. While first feasibility studies on European level were performed \cite{pupillo2023}, no comparable TX/RX system currently exists in Germany/Europe which could adequately support these activities with radar tracking/characterization and significant observing time for a subset of new NEO detections. A future bi-/multistatic configuration of Fraunhofer's TIRA system as transmitter with a larger number of RX elements of the future ngVLA (and other European facilities) could change this situation. The large enhancement in sensitivity would not only improve data quality, parameter estimation and the access to smaller object sizes, but also push further the observable volume in which new NEO detections will be accessible for radar follow-up observations.       

\begin{figure}[h!]
	\vspace{2cm}
	\centering
	\includegraphics[width=1.0\linewidth]{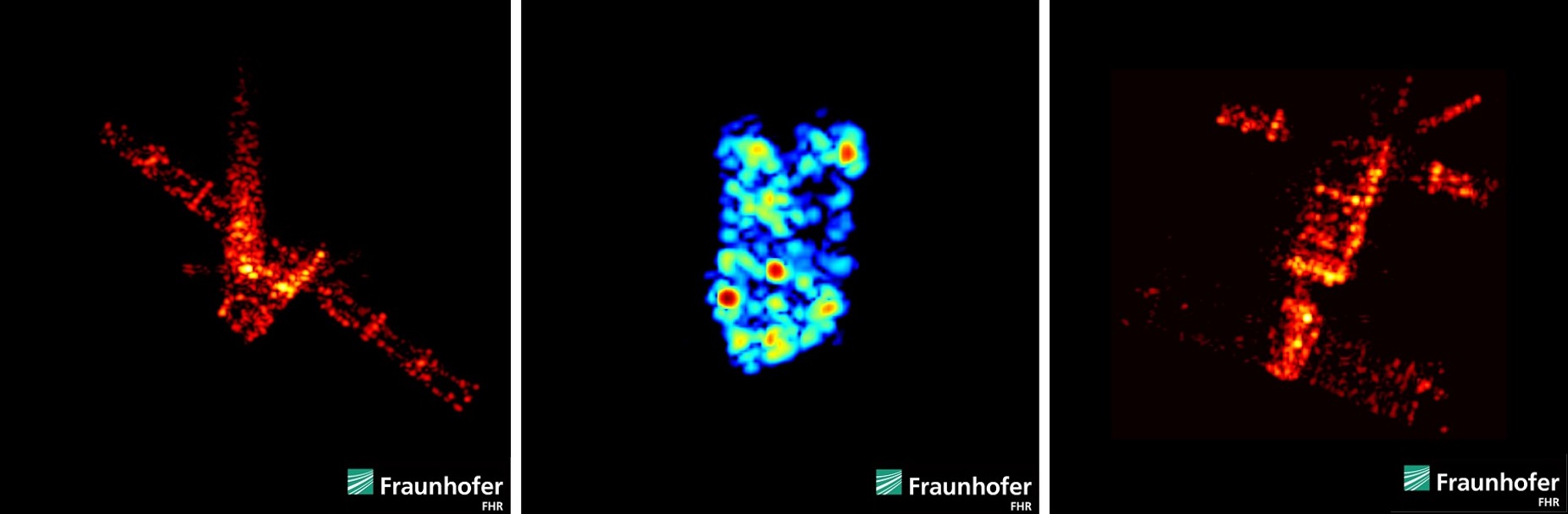}
	\caption{Examples of monostatic ISAR imaging obtained during recent re-entry campaigns (FHR for the joint German Space Situational Awareness Center, GSSAC) with the TIRA broadband Ku-Band system: last images of (i) ESA satellite Aeolus during the assisted re-entry in July 2023 (left), (ii) the International Space Station's battery pack in March 2024 (middle) and (iii) ESA satellite ERS-2 in February 2024 (right). Future bi-/multistatic high-resolution ISAR imaging will further enhance the capabilities in space object identification and characterization.}
	\label{fig:isar}
\end{figure}

\subsection{Constraining the general circulation of the terrestrial planet's atmospheres by Doppler wind measurements with the  ngVLA}
\RaggedRight\label{hartogh01}
\vspace*{\baselineskip}

\noindent \textbf{Thematic Areas:} \linebreak $\square$ Stellar Astrophysics \linebreak $\checked$ Solar System, Planetary Systems and Habitability \linebreak
$\square$ Circuit of Cosmic   Matter (incl. star formation) \linebreak $\square$ The Galaxy and the Local Group \linebreak
  $\square$   Galaxies and AGN \linebreak $\square$  Cosmology, Large Scale Structure and Early Universe \linebreak
  $\square$    Extreme conditions in the cosmos, fundamental astrophysics    \linebreak
  $\square$ Interdisciplinary research and technology \linebreak

\textbf{Principal Author:}

Name: Paul Hartogh
 \linebreak						
Institution: MPI for Solar System Research
 \linebreak
Email: hartogh@mps.mpg.de
 \linebreak

General circulation models \citep[e.g.][]{Hartogh2005} of Mars and Venus are usually constrained by temperature measurements either from ground or from satellites. Since the forcing mechanisms are not restricted to thermal contrasts, but include waves and eddies, e.g. generated by flow over topography, the winds predicted by the models may be rather wrong. In Figure~\ref{fig:hartogh_fig1} \citep{Medvedev2011}, the difference in zonal wind speeds for model calculations with and without gravity waves are displayed. \\

\begin{figure}[h!]
    \centering
    \includegraphics[width=0.65\linewidth]{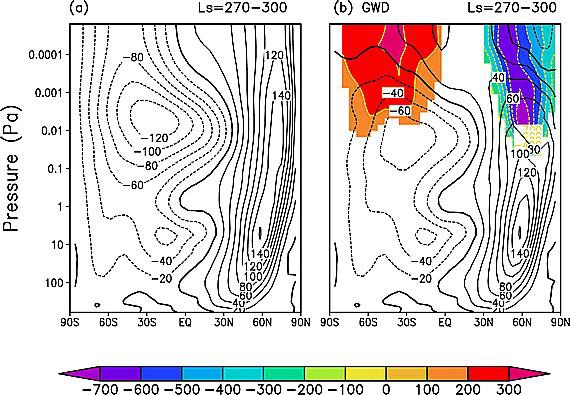}\hfill
        \caption{Simulation of the MAOAM MGCM for the same season: a) without gravity waves and b) with wave drag (shaded).}
    \label{fig:hartogh_fig1}
\end{figure}

Not only the amplitude of the wind speed, but also its direction differs for some altitudes/latitudes. Discrepancies between different models may be even larger (Fig.~\ref{fig:hartogh_fig2}).\\

\begin{figure}[h!]
    \centering
    \includegraphics[width=\linewidth]{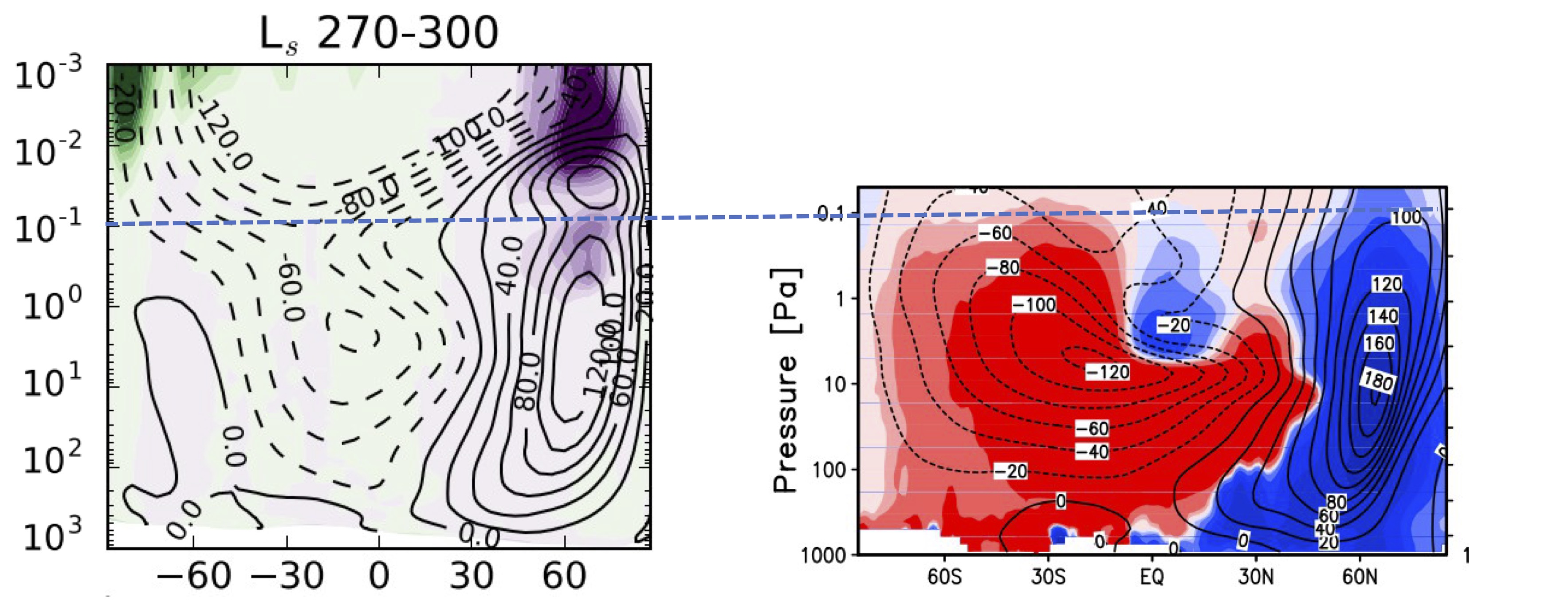}
    \caption{Comparison of winds in two Mars GCMs \citep{Neary2018} and \citep{Kuroda2015}.}
    \label{fig:hartogh_fig2}
\end{figure}

Doppler wind measurements of Mars and Venus from satellites in the millimeter and submillimeter regime of the electromagnetic spectrum were proposed for a long time, but were not implemented thus far. \cite{Shah1991} demonstrated this technique for the first time from ground using the 115 GHz CO transition with a single dish telescope. More precise measurements were performed later at higher transitions \citep{Rengel2008}. \cite{Moreno2009} presented interferometric wind measurements derived from 115\,GHz CO observations of the PdB interferometer with baselines ranging from 15 to 75\,m, corresponding to a spatial resolution of 3--7\,arcsec, while Mars had an apparent diameter between 9.5 and 23$^{\prime\prime}$. Figure~\ref{fig:hartogh_fig3} shows that the spatial resolution of these measurements is not high enough to resolve the martian atmosphere in limb (only absorption lines are detected). Limb resolved observations require a spatial resolution of 0.05 to 0.1\,arcsec. They will allow the retrieval of several vertically resolved layers of wind. The capabilities of ngVLA will allow us to perform such highly resolved observations at 115\,GHz. Regular wind and temperature measurements will provide a strong constraint to MGCMs.  The assimilation of these observations into MGCMs are essential for future weather predictions required for lander or balloon missions in the terrestrial planets.

\begin{figure}[h!]
    \centering
    \includegraphics[width=0.95\linewidth]{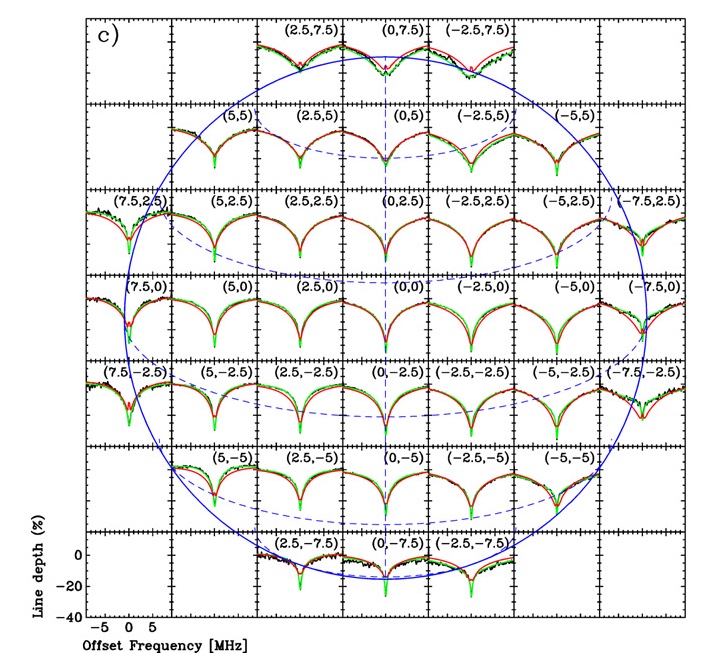}
    \caption{PdB map of the 115 GHz CO transition. From these spectra Doppler winds were retrieved.}
    \label{fig:hartogh_fig3}
\end{figure}
\subsection{Galaxy clusters with ngVLA}
\RaggedRight\label{hoeft01}
\vspace*{\baselineskip}

\noindent \textbf{Thematic Areas:} \linebreak $\square$ Stellar Astrophysics \linebreak $\square$ Solar System, Planetary Systems and Habitability \linebreak
$\square$ Circuit of Cosmic   Matter (incl. star formation) \linebreak $\square$ The Galaxy and the Local Group \linebreak
  $\checked$   Galaxies and AGN \linebreak $\checked$  Cosmology, Large Scale Structure and Early Universe \linebreak
  $\square$    Extreme conditions in the cosmos, fundamental astrophysics    \linebreak
  $\square$ Interdisciplinary research and technology \linebreak
  
\textbf{Principal Author:}

Name: Matthias Hoeft
 \linebreak						
Institution: Thüringer Landessternwarte
 \linebreak
Email: hoeft@tls-tautenburg.de
 \linebreak
 
\textbf{Co-authors:} M. Brüggen (UHH), Aritra Basu (TLS) 
  \linebreak

Radio emission from galaxy clusters is an important probe of cosmic magnetogenesis, plasma physics, and the interaction of active galaxies with their environment \cite{2018SSRv..214..122D,2018Galax...6..142V,2019SSRv..215...16V,2019AJ....157..126G,2021MNRAS.501.3332S,2022hxga.book...56K}. Knowing the properties of the intra-cluster medium and studying its evolution is crucial for determining how magnetic fields are amplified during cosmic evolution, for exploring the physics of particle acceleration in the very low density environment, and for solving the puzzle of how the feedback from the AGN in the central galaxy in a relaxed cluster regulates its accretion from the intra-cluster medium \cite{2022hxga.book....5H}.
\\

Radio emission probes in particular the non-thermal component in clusters, namely the magnetic field and cosmic ray electrons. Clusters show a rich variety of outbursts from the central galaxies, which are thought to compensate for the cooling of the central intra-cluster medium, self-regulation of accretion and feedback, perturbed AGN, stripped (jellyfish) star-forming galaxies, and diffuse emission features, relics and halos, which are thought to trace merger shocks and merger-induced turbulence. Neither the acceleration of electrons to relativistic energies nor the strength, structure and evolution of cluster magnetic fields are well understood. 
\\

Radio observations in L-, S- and C-band have been shown to shed light on CRe and magnetic fields, e.g. when using background sources to study the RM of the ICM, the sources could be largely depolarised at low frequencies. The same is true for emission AGN, radio galaxies, relics embedded in the ICM, depending on the amount of ICM in front of these sources, they may be depolarised at low frequencies. Another crucial aspect is the broad-band study of the morphology of the sources, which is necessary to understand to what extent the filamentary structure may directly reflect the magnetic field structure of the ICM. Very broadband, sensitive, spatially resolved observations are essential to determine the complex structure of the ICM in Faraday space.
\\

\begin{figure}
    \centering
    \includegraphics[width=\linewidth]{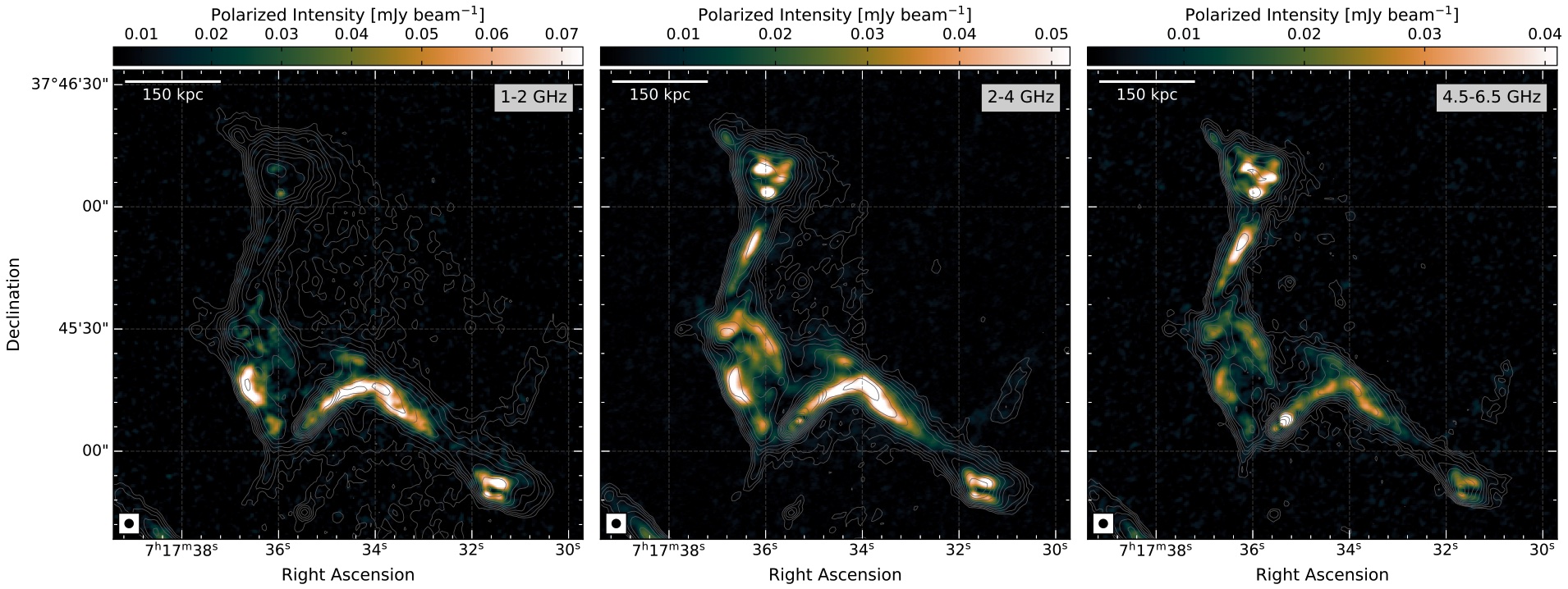}
    \caption{Polarization intensity images of the radio relic and AGN in the galaxy cluster MACS J0717.5+3745 at 2 arcsecond resolution, showing that the polarization emission is distributed in a clumpy manner. The image also revels fine, small-scale filaments visible in the total power emission. Contour levels are drawn at $[1, 2, 4, 8, . . . ] \times \sigma_{\rm rms}$ and are from the VLA L-, S-, and C-band Stokes I images. The beam sizes are indicated in the bottom left corner of the each image. The image has been published in Rajpurohit et al. (2022) \cite{2022A&A...657A...2R}.}
    \label{fig:enter-label}
\end{figure}

{\it The scientific objectives} of the ngVLA would be: (i) to use very broadband spectro-polarimetric observations to study the magnetic field structures. (ii) measure the spectral properties up to the highest frequencies to constrain the acceleration processes in the intra-cluster medium. 
\\[1ex]
\subsection{Resolving the Doppler Crisis}
\RaggedRight\label{kadler01}
\vspace*{\baselineskip}

\noindent \textbf{Thematic Areas:} \linebreak $\square$ Stellar Astrophysics \linebreak $\square$ Solar System, Planetary Systems and Habitability \linebreak
$\square$ Circuit of Cosmic   Matter (incl. star formation) \linebreak $\square$ The Galaxy and the Local Group \linebreak
  $\checked$   Galaxies and AGN \linebreak $\square$  Cosmology, Large Scale Structure and Early Universe \linebreak
  $\checked$    Extreme conditions in the cosmos, fundamental astrophysics    \linebreak
  $\square$ Interdisciplinary research and technology \linebreak
  
\textbf{Principal Author:}

Name:	Matthias Kadler
 \linebreak						
Institution:  JMU Würzburg
 \linebreak
Email: matthias.kadler@astro.uni-wuerzburg.de
 \linebreak
 
\textbf{Co-authors:} Eduardo Ros (MPIfR)
  \linebreak

Blazars are a subclass of AGN that emit 
violently 
variable broadband emission from radio to $\gamma$-ray energies. 
Understanding their physics is of very high interest for astroparticle physics as these objects are possible sources of ultra-high-energy cosmic rays and neutrinos \citep[e.g.,][]{Hillas1984,Mannheim1995}.
Since the 1990s, Imaging Air Cherenkov Telescopes have been highly successful in detecting gamma-ray emission up to TeV photon energies from an increasing number of blazars. 
Already today, about 80 TeV blazars are known and
it has been demonstrated that TeV observations can efficiently be combined with radio single-dish and VLBI results  to probe source emission models \citep{Kadler2012,blackholelightning}. However,
the majority of all known TeV blazars are high-synchrotron-peaked (HSP) objects of the BL\,Lac type which can be detected at TeV energies typically only during flaring states and which are faint radio sources.
A major improvement will be achieved in the next decade with the Cherenkov Telescope Array \citep[CTA; ][]{CTAScienceCase} thanks to its sensitivity and low energy threshold \citep{diPierro2019}.
With the improved flux sensitivity of CTA, it will be possible to study a large population of blazars both in high and low states and measure their high-energy spectral variability. In combination with the next generation Very Large Array (ngVLA; \cite{murphy18}), it will become possible to solve current major questions regarding the high-energy emission of blazars.

One of the most pressing questions in this context is the so-called "Doppler-crisis" \citep{Henri2006}.
Usually, the highly variable gamma-ray emission of blazars is explained by very high Doppler factors up to $\delta>40$ \citep[e.g.,][]{Tavecchio2010}. In stark contrast to this, radio observations have found that HSP jets typically move at comparatively low speeds and show relatively low brightness temperatures, indicative of low Doppler factors. 
One possible explanation is that the radio emission and the variable $\gamma$-ray emission may not originate from the same emitting region.
This could be explained by the existence of a "spine-sheath" structure \citep{Ghisellini2005}  where the outer layers of the relativistic jet (the sheath) have a slower bulk velocity along the jet axis than the inner layers (the spine) and can serve as a seed-photon field for inverse-Compton processes within the latter. 
Other scenarios involve the natural presence of stationary recollimation shocks in jets \citep[e.g.,][]{Hervet2019}. 
Such models can be constrained by high-resolution VLBI observations 
but previous observations have only yielded rather low signal-to-noise data \cite{Piner2018} or studies of the very brightest individual sources at low frequencies \citep{Giroletti2008}. The vast majority of the HSP blazar population is much too faint for current-generation VLBI instruments. With the ngVLA, it will become possible to resolve different emission regions in structured blazar jets both in total intensity and in polarized emission at unprecedented angular resolution and sensitivity. This will yield the key data to test multi-zone broadband emission models towards a solution of the Doppler crisis. 

\begin{figure}
\centering
\includegraphics[width=0.465\textwidth]{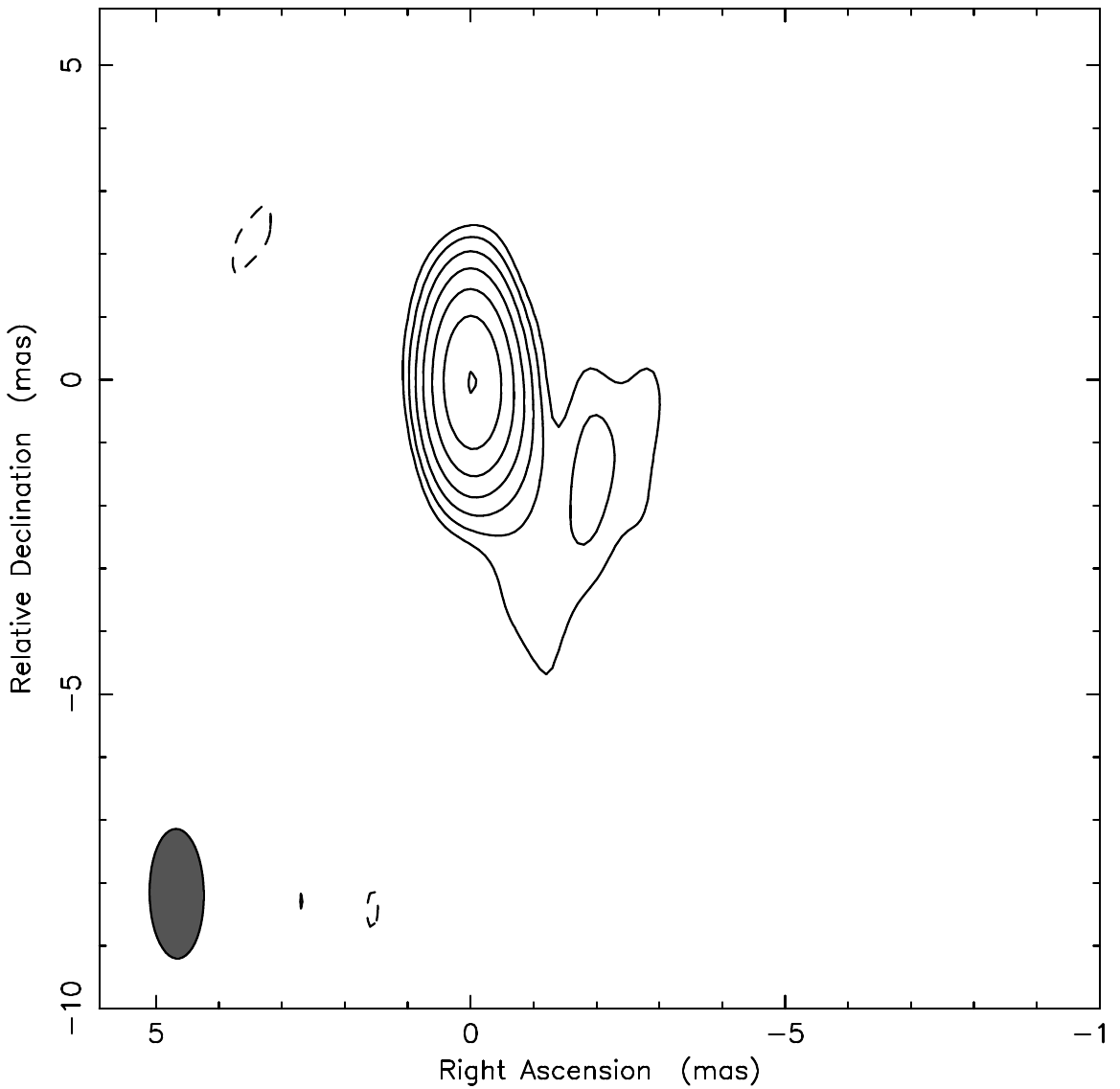}
\includegraphics[width=0.5\textwidth]{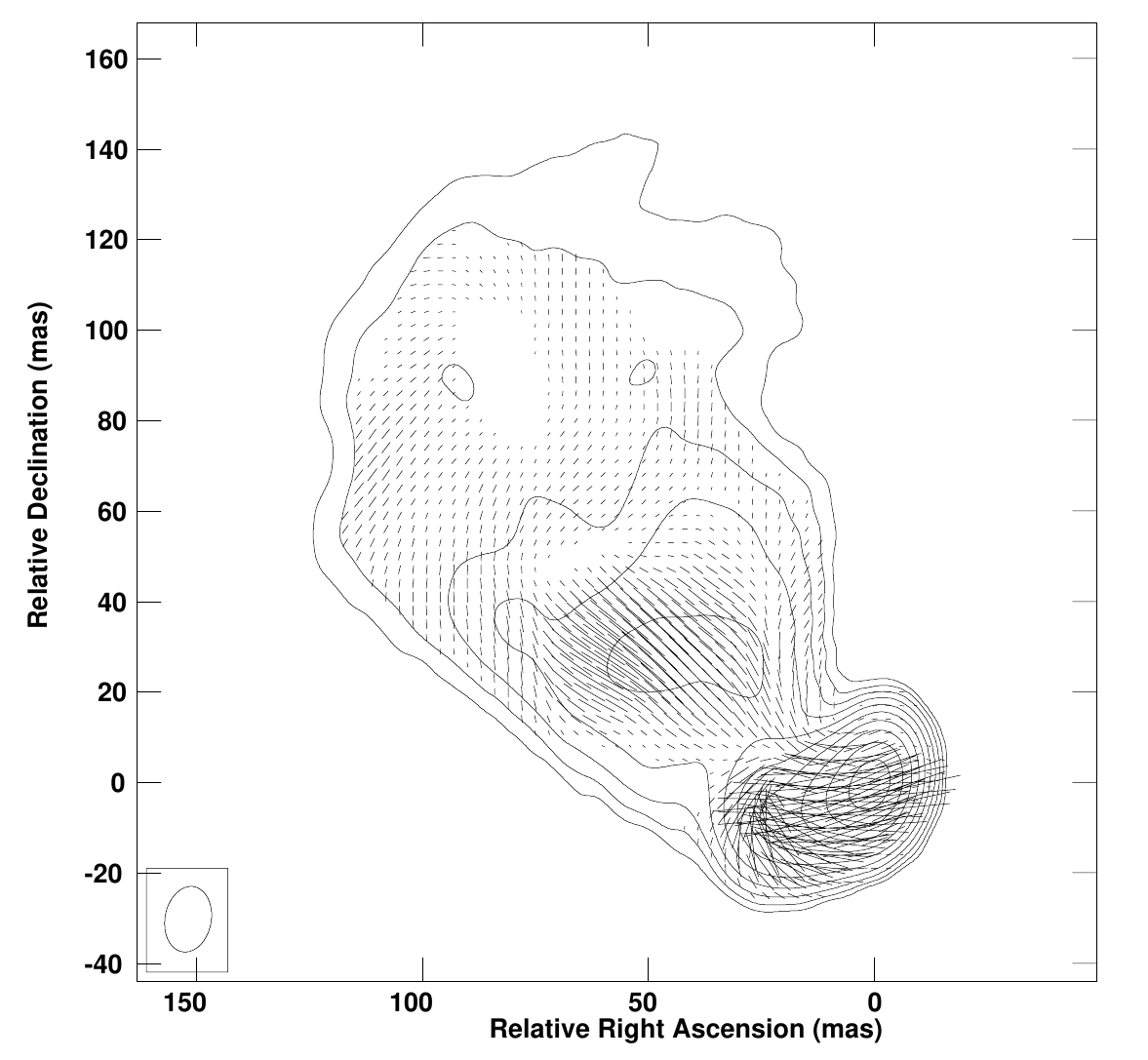}
\caption{VLBA image at 8\,GHz of a typical faint TeV blazar (1ES 0347$-$121; \cite{Piner2016}; left) and High-Sensitivity-Array image at 1.4\,GHz of the brightest HSP blazar (Mrk\,501; \cite{Giroletti2008}; right). The ngVLA will provide the image sensitivity needed to detect regions of low surface brightness and measure their polarization structure for the bulk HSP blazar population. This is currently possible only for the very brightest sources and only at low frequencies in long experiments.
  }
\end{figure}    
\subsection{VLBI Probes of Neutrino Emission in Blazars}
\RaggedRight\label{kadler02}
\vspace*{\baselineskip}

\noindent \textbf{Thematic Areas:} \linebreak $\square$ Stellar Astrophysics \linebreak $\square$ Solar System, Planetary Systems and Habitability \linebreak
$\square$ Circuit of Cosmic   Matter (incl. star formation) \linebreak $\square$ The Galaxy and the Local Group \linebreak
  $\checked$   Galaxies and AGN \linebreak $\square$  Cosmology, Large Scale Structure and Early Universe \linebreak
  $\checked$    Extreme conditions in the cosmos, fundamental astrophysics    \linebreak
  $\square$ Interdisciplinary research and technology \linebreak
  
\textbf{Principal Author:} 

Name:	Matthias Kadler
 \linebreak						
Institution:  JMU Würzburg
 \linebreak
Email: matthias.kadler@astro.uni-wuerzburg.de
 \linebreak
 
\textbf{Co-authors:} Yuri Kovalev (MPIfR), Eduardo Ros (MPIfR)
  \linebreak

The origin of the high-energy cosmic neutrinos detected by the IceCube observatory is a hotly debated topic in astroparticle physics with growing evidence that some of these neutrinos may be associated with active galactic nuclei (AGN), in particular blazars.
In 2017, an association of the track-like muon neutrino event IC\,170922A with the $\gamma$-ray blazar TXS\,0506+056 has been found at a significance of $\sim3\,\sigma$ \cite{IceCube2018}. An independent second analysis found a $3.5\,\sigma$ significant association between the TXS\,0506+056 blazar and a so-called 'flare' of lower-energy neutrinos in 2014/15 \cite{IceCube2018b} during a low $\gamma$-ray emission state. This surprising apparent lack of $\gamma$-ray emission during at least some episodes of enhanced neutrino production has been widely discussed in the literature \citep[e.g.,][]{Reimer2019,Rodrigues2021,Kun2021,Mastichiadis2021,Kun2023}. A leading idea is that particle cascades could lead to a redistribution of the calorimetric output toward lower energies. This can eventually lead to strong shocks on parsec scales and explain associated radio variability. 
The radio band is thus of utmost importance for a better understanding of neutrino emission in blazars and for the related major puzzle regarding the origin of high-energy cosmic rays \citep{Hillas1984}. 

%Recently, the IceCube Collaboration \cite{IceCube2022a} reported a statistical correlation of neutrinos with  known $\gamma$-ray emitters at a significance of $3.3\,\sigma$ with the largest contributions from the nearby active galaxy NGC\,1068 and the three blazars TXS\,0506+056, PKS\,1424+240 and GB6\,J1542+6129. 
Statistical correlations have been claimed by multiple groups 
%both between IceCube neutrinos and blazars from the Roma-BZCat catalog 
\cite{Plavin2020,Hovatta2021,Plavin2021,Plavin2023,Buson2022,IceCube2022a,Buson2023,Bellenghi2023,Suray2024} but recent studies also revealed the difficulties and limitations of matching large source samples with neutrino events \citep[e.g.,][]{IceCube2023}.  
Here, we focus on the complementary approach to investigate individual sources with novel high-quality observational data with the aim to understand the neutrino emission mechanisms of blazars.

%There are two different explanations for the gamma-ray emission of blazars: inverse Compton scattering and hadronic emission models. 
In hadronic emission models, the high-energy $\gamma$-ray emission of blazars is produced by interactions of relativistic protons in the jet with soft ambient seed photons \citep[e.g.][]{Mannheim1993}  and  neutrino emission is naturally expected. The origin and properties of the seed photon field are a major open question.
An attractive model \citep{Tavecchio2014,Tavecchio2015} assumes that 
the high-energy protons of the fast inner spine of the blazar jet interact with soft target photons of the sheath, a slower jet-layer surrounding the spine, leading to high-energy $\gamma$-ray and neutrino emission. 
Other models suggest that neutrino production could happen closer to the jet base, e.g., in standing recollimation shocks \citep[e.g.,][]{Kalashev2022}. VLBI observations can probe such models if sufficient angular resolution, sensitivity and image fidelity are provided. The putative  substructures are likely to have different magnetic-field configurations so that polarization information is of high relevance.

Previous VLBI observations of neutrino-candidate blazars have indeed found indications of limb-brightened structures, as predicted in spine-sheath models \cite{Ojha2010,Karamanavis2016,Ros2020} albeit at limited image fidelity so that the candidate regions for the photopion-process seed photon fields cannot  be fully resolved and characterized.
Fig.~\ref{fig:spine-sheath-pol}
shows 15\,GHz MOJAVE images of three neutrino-candidate blazars, which demonstrate the need for higher resolution. This is generally achieved at high frequencies and since a good $(u,v)$ coverage is driving image fidelity, the next generation Very Large Array (ngVLA; \cite{murphy18}) will be superior to other VLBI instruments in this science context. 

\begin{figure}[b!]
    \centering
    \includegraphics[width=\textwidth]{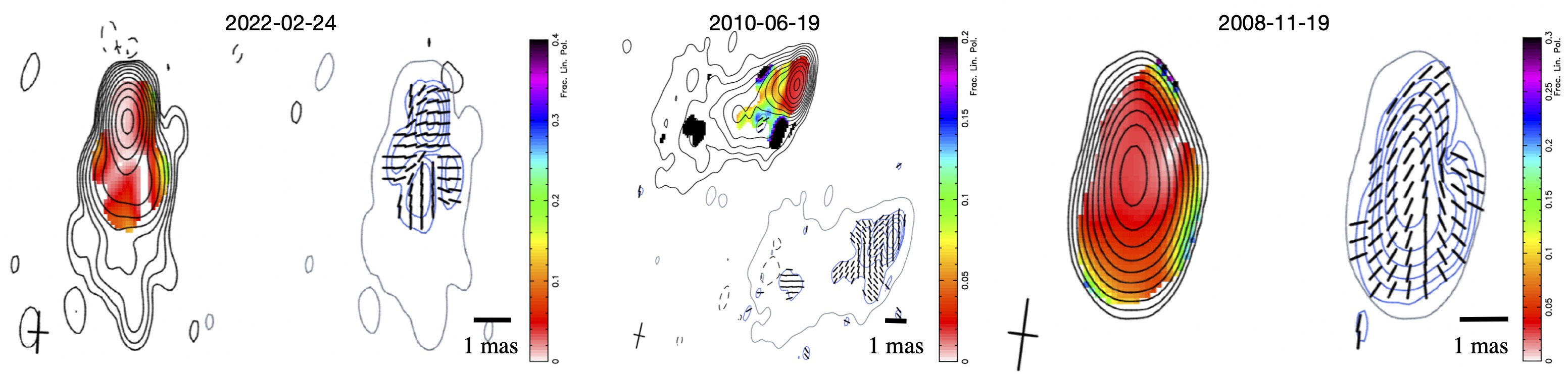}
    \caption{Spine-sheath structures seen at 15\,GHz (VLBA, MOJAVE): TXS\,0506+056 (left), PKS1502+106 (middle) and PKS\,1741$-$038 (right).  EVPAs at the outer edges of the jets (the sheaths) are near-perpendicular to the inner part (the spine). For each source, the left sub-panel shows  total intensity contours and fractional polarization in color. The right sub-panel shows the $3\, \sigma$ total-intensity contour and inner contours of linear-polarized intensity with EVPAs superimposed. Beam sizes and scale bars are shown in the bottom left and right corner. 
    %{Observations with current instruments are limited in sensitivity and show these structures only in selected epochs. Their limited angular resolution is insufficient to separate spine and sheath in total intensity or probe for stationary recollimation shocks close to the jet base. }
    }
    \label{fig:spine-sheath-pol}
\end{figure}  
\subsection{Low-$J$ CO Line Excitation}
\RaggedRight\label{keenan01}
\vspace*{\baselineskip}

\noindent \textbf{Thematic Areas:} \linebreak $\square$ Stellar Astrophysics \linebreak $\square$ Solar System, Planetary Systems and Habitability \linebreak
$\checked$ Circuit of Cosmic   Matter (incl. star formation) \linebreak $\square$ The Galaxy and the Local Group \linebreak
  $\checked$   Galaxies and AGN \linebreak $\square$  Cosmology, Large Scale Structure and Early Universe \linebreak
  $\square$    Extreme conditions in the cosmos, fundamental astrophysics    \linebreak
  $\square$ Interdisciplinary research and technology \linebreak
  
\textbf{Principal Author:}

Name: Ryan P. Keenan
 \linebreak						
Institution: Max-Planck-Institut f\"ur Astronomie
 \linebreak
Email: keenan@mpia.de
 \linebreak
 
\textbf{Co-authors:} Leindert A. Boogaard (MPIA), Fabian Walter (MPIA), Melanie Kaasinen (ESO), Dominik Riechers (Uni K\"oln), Nikolaus Sulzenauer (MPIfR)
  \linebreak

Molecular hydrogen (H$_2$) is the fuel for star formation. Because the abundance of H$_2$ is difficult to measure directly, rotational emission from carbon monoxide (CO), the second most abundant interstellar molecule, is the preferred tracer of molecular gas abundance. The fundamental $J=1\rightarrow0$ transition of CO (hereafter CO(1--0)) is easily excited under conditions prevalent in molecular clouds and has been carefully calibrated as a molecular gas tracer \cite{bolatto2013}. Transitions amongst the higher energy rotational states of CO form a ladder of emission lines which trace conditions of increasing gas density and temperature. The relative intensities of multiple CO lines can therefore be used to derive the density and temperature molecular gas clouds \cite{daddi+15,kamenetzky+17,boogaard+20,valentino+20,liu+21}, which may provide new insight into how gas abundance, ISM conditions and stellar feedback couple to regulate star formation and drive the evolution of the galaxy main sequence.

The lowest energy transitions are critical to such measurements, as they are sensitive to cold, diffuse gas components, which may dominate the total molecular gas mass but do not contribute significantly to luminosity the $J_u \gtrsim 3$ lines \cite{kamenetzky+17}. Figure~\ref{fig:rpk-sled} shows a compilation of low-$J$ CO spectral line energy distributions (SLEDs) from \cite{daddi+15}, \cite{lamperti+20} and \cite{keenan23thesis}. Galaxies exhibit a wide range of line intensity ratios, and a significant fraction have SLEDs which cannot easily be reconciled with simple models of molecular clouds, motivating the need for more sophisticated models. Interestingly, while $z\sim0$ line ratios show evidence for correlation with star formation rate surface density, not all high-$z$ galaxies match this behavior, suggesting that the coupling between molecular cloud properties and star formation rate is evolving over time.

The current generation of sub/millimeter facilities are capable of measuring the low-$J$ CO lines only in small samples of high-$z$ galaxies, typically by combining VLA observations of CO(1--0) with ALMA or NOEMA observations of the higher-$J$ lines \cite{daddi+15,riechers20}. Small sample sizes limit the power of current studies to map out statistical trends in $z>1$ CO SLEDs and ISM properties. The ngVLA will be $\sim5$ times more sensitive than the VLA between 10 and 50 GHz and ALMA between 70 and 115 GHz. This results in a factor of 25 reduction in observing time for CO(1--0) at $10>z>0$ as well as CO(2--1) at $z>1$ and CO(3--2) at $z>3$, and will enable measurements of the lowest $J$ transitions for representative samples of high redshift galaxies. 

The resolution of the ngVLA will also enable resolved mapping of CO line ratios at $\sim$kpc scales. This will make it possible to study the evolving effects of galactic environments like bars and active nuclei, which are known to affect gas conditions in $z\sim0$ galaxies \cite{koda+12,denbrok+21,yajima+21} and likely play an important role at high redshift as well \cite{sulzenauer+21}.

\begin{figure}
    \centering
    \includegraphics[width=\textwidth]{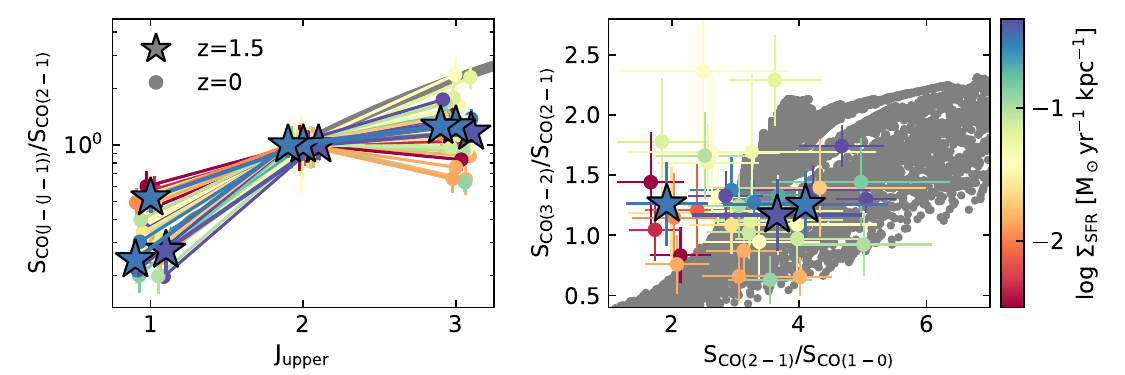}
    \caption{Left: The $J_u=1$--$3$ SLEDs, normalized by CO(2--1) flux, for a samples of $z\sim0$ (circles) and $z\sim1.5$ (stars) galaxies are shown, color coded by star formation rate surface density. Right: CO(2--1)/CO(1--0) and CO(3--2)/CO(2--1) flux ratios for the same sample are shown, compared to a grid of cloud models from \cite{leroy+22}. Galaxies show a diverse array of SLEDs. For local galaxies, the line ratios, which are sensitive to the density and temperature of the emitting gas, correlate with properties such as SFR surface density, pointing to a coupling between the bulk conditions in molecular clouds and their ability to form stars. However the small sample of $z\sim1.5$ observations do not uniformly obey these trends, providing tentative evidence that this connection is evolving over time. Larger samples of $z>1.0$ CO SLEDs, only possible with the ngVLA, are necessary to robustly explore this possibility.}
    \label{fig:rpk-sled}
\end{figure}
  
\subsection{Radio-optical link: AGN-based VLBI and \Gaia fundamental reference systems and their application for astrometry and astrophysics} \RaggedRight\label{kovalev01}
\vspace*{\baselineskip}

\noindent \textbf{Thematic Areas:} \linebreak $\square$ Stellar Astrophysics \linebreak $\square$ Solar System, Planetary Systems and Habitability \linebreak
$\square$ Circuit of Cosmic   Matter (incl. star formation) \linebreak $\square$ The Galaxy and the Local Group \linebreak
  $\checked$   Galaxies and AGN \linebreak $\square$  Cosmology, Large Scale Structure and Early Universe \linebreak
  $\square$    Extreme conditions in the cosmos, fundamental astrophysics    \linebreak
  $\square$ Interdisciplinary research and technology \linebreak
  
\textbf{Principal Author:}

Name:	Yuri Kovalev
 \linebreak						
Institution:  MPIfR, Bonn
 \linebreak
Email: yykovalev@gmail.com
 \linebreak
 
\textbf{Co-authors:} Eduardo Ros (MPIfR), Andrei Lobanov (MPIfR)
  \linebreak

The most recent results of the \Gaia astrometry and photometry information is summarized in the \Gaia DR3 catalog \cite{2023A&A...674A...1G}.
%The complete set of \textit{Gaia} data, spanning every epoch of observation for all targets since 2014, is anticipated to be available by 2026. 
Through cross-identification with radio-bright active galactic nuclei (AGN) hosting parsec-scale jets\footnote{Radio Fundamental Catalog (RFC): http://astrogeo.smce.nasa.gov/rfc}, several thousand matches have been discovered. Approximately 10\% of these matches exhibit significant positional VLBI-\Gaia shifts along their jets, primarily due to the extension of optical jets while VLBI pinpoints the compact parsec-scale core (Figure~\ref{f:VLBI-Gaia}). Opacity-driven frequency-dependent core-shift and extended radio structure also plays some role in this effect. Additionally, it has been deduced that \textit{Gaia} positions significantly fluctuate due to optical flares occurring in their cores.
Details can be found in \cite{2017A&A...598L...1K,2017MNRAS.467L..71P,2017MNRAS.471.3775P,2019MNRAS.482.3023P,2019ApJ...871..143P,2020MNRAS.493L..54K}. 
Optical and VLBI AGN data applied together can help to differentiate between flares occurring in accretion disks (which are expected to make the spectrum more blue since the accretion disk thermal emission  peaks in UV) and synchrotron flares in the jet. This joint analysis they will allow us to dissect the disk-jet system in AGN.

ngVLA will provide continuous frequency coverage and significantly improved fringe detection sensitivity, image dynamic range and image fidelity. As a result, the sample of AGN with detectable jets will increase by more than an order of magnitude with significantly enhanced accuracy of the absolute astrometry measurements.
This will deliver the next generation of the radio fundamental reference frame and provide grounds for a transformational change in our ability to study the accretion disk -- relativistic jet interaction in active galaxies addressing questions of jet formation and energy extraction.

\begin{figure}
    \centering
    \includegraphics[width=0.5\columnwidth,trim=1cm -0.7cm 0cm 0cm]{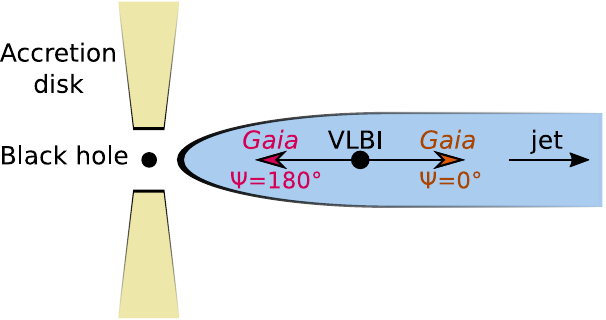}
    \includegraphics[width=0.35\columnwidth,trim=15cm 0cm 0cm 0cm,clip]{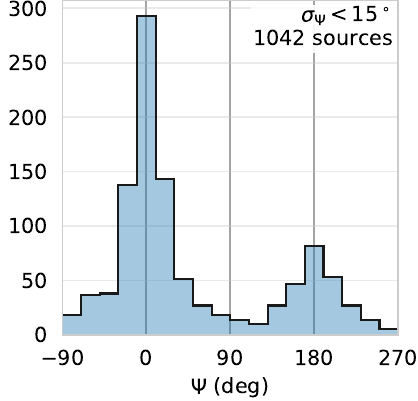}
    \caption{\textit{Left:} Diagram explaining the definition taken for the two preferred VLBI-\Gaia offset directions with respect to the parsec-scale jet: downstream $\Psi= 0^\circ$ and upstream $\Psi=180^\circ$.
    \textit{Right:} A distribution of the angle $\Psi$ between the VLBI-\Gaia positional shift direction and direction of VLBI jet filtered on its error value. Gaia DR3 optical and RFC radio positions are used, jet position angles are taken from \cite{2022ApJS..260....4P}.
    \label{f:VLBI-Gaia}}
\end{figure}

\subsection{Blazars as neutrino candidates: complete sample studies}
\RaggedRight\label{kovalev02}
\vspace*{\baselineskip}

\noindent \textbf{Thematic Areas:} \linebreak $\square$ Stellar Astrophysics \linebreak $\square$ Solar System, Planetary Systems and Habitability \linebreak
$\square$ Circuit of Cosmic   Matter (incl. star formation) \linebreak $\square$ The Galaxy and the Local Group \linebreak
  $\checked$   Galaxies and AGN \linebreak $\square$  Cosmology, Large Scale Structure and Early Universe \linebreak
  $\checked$    Extreme conditions in the cosmos, fundamental astrophysics    \linebreak
  $\square$ Interdisciplinary research and technology \linebreak
  
\textbf{Principal Author:}

Name:	Yuri Kovalev
 \linebreak						
Institution:  MPIfR, Bonn
 \linebreak
Email: yykovalev@gmail.com
 \linebreak
 
\textbf{Co-authors:} Matthias Kadler (JMU Würzburg)
  \linebreak

Currently, studies of high-energy astrophysical neutrinos and their sources are limited by the sensitivity and resolution of neutrino observatories. A new era is coming, as their capabilities are increasingly improving. The next-generation IceCube-Gen2 will grow the telescope volume tenfold, from 1 to 10 cubic kilometers, aiming at a corresponding increase in detection rates by 2033 \citep{IC_Gen2}. 
At the same time, it is expected that water-based detectors will provide a better angular resolution than the current IceCube, both for muons and for cascades.
The Baikal-GVD detector has already reached the instrumental volume of 0.5 cubic kilometer and continues to grow and improve event reconstruction algorithms \citep{2023PhRvD.107d2005A}. 
KM3NeT, a neutrino observatory in the Mediterranean, is being constructed and has already started yielding its first results \citep{2019APh...111..100A}. 
Together, these instruments will provide a qualitative leap forward in both the number of detected astrophysical neutrinos and in the precision of their localization.

While neutrino telescopes continue to gather more neutrino detections, radio monitoring of AGN before, during, and after these events is crucial for identifying the potentially neutrino-associated changes in their compact jet structure. Given the current small number statistics, one cannot predict which AGN will be positionally associated with future neutrinos. Thus observing a large, complete candidate sample is necessary to conduct a wide range of statistical analyses \citep[e.g.,][]{2022A&A...666A..36L,Plavin2023,Hovatta2021}.
We note that studying a complete sample of AGN with well-defined properties will allow us not only to relate the observed changes to detected neutrinos but also to set a robust significance on that association.
Brighter blazars are more likely counterparts to detected neutrinos,
as has been specifically demonstrated by \citet{Plavin2023}. This motivates selecting AGN by their parsec-scale flux density in order to maximize the probability of obtaining associations.

While the original claims of the association are confirmed when the same data are used, newer reconstructions of IceCube data or different selection procedures may result in the apparent reduction of the association significance \citep{IceCube2023}. This trend was reported for the original TXS~0506$+$056 association by \citet{IceCube-TXS0506-2023} as well as
for various blazar statistical associations by \citet{Bellenghi2023}. 
This additionally supports continuous VLBI monitoring of complete samples to provide grounds for future analysis against evolving neutrino data.

ngVLA with its enhanced sensitivity and observing efficiency will allow for such monitoring to study conditions in and around AGN cores before during and after high energy neutrino arrival. Its highest observing frequency of about 100 GHz will allow us to study regions close to the central engines mitigating synchrotron opacity.
This opens exciting opportunities to understand particle acceleration and neutrino production.  
\subsection{Dense gas in nearby galaxies}
\RaggedRight\label{kreckel01}
\vspace*{\baselineskip}

\noindent \textbf{Thematic Areas:} \linebreak $\square$ Stellar Astrophysics \linebreak $\square$ Solar System, Planetary Systems and Habitability \linebreak
$\checked$ Circuit of Cosmic   Matter (incl. star formation) \linebreak $\square$ The Galaxy and the Local Group \linebreak
  $\checked$   Galaxies and AGN \linebreak $\square$  Cosmology, Large Scale Structure and Early Universe \linebreak
  $\square$    Extreme conditions in the cosmos, fundamental astrophysics    \linebreak
  $\square$ Interdisciplinary research and technology \linebreak
  
\textbf{Principal Author: }

Name:	Kathryn Kreckel
 \linebreak						
Institution:  Heidelberg University
 \linebreak
Email: kathryn.kreckel@uni-heidelberg.de
 \linebreak
 
\textbf{Co-authors:} Frank Bigiel (Univ. Bonn), Eva Schinnerer (MPIA)
  \linebreak

The physical processes that drive the formation of stars inside galaxies is regulated by the cosmic matter cycle, and operates on the sub-100~pc scales defined by the sizes of individual giant molecular clouds and the radiative and mechanical reach of stellar feedback.   
%The emergence of molecular gas from the neutral gas phase, its collapse to form new stars as well as its subsequent disruption and potential reformation can be traced directly via radio observations at the necessary physical scales. 
In recent years, the old idea of a universal population of Giant Molecular Clouds (GMCs) has been replaced by a more diverse, dynamical view (e.g., \cite{Hughes2013, Meidt2015}) based on clear evidence that cloud properties (e.g., gas mass surface density, line width, internal turbulence) are shaped by their environment (e.g., \cite{Colombo2014, Sun2022}). Pinpointing these correlations requires an external view of complete galaxies, but the distance to which the relevant scales are accessible by current facilities is limited. %In particular, our physical understanding of how dense gas responds to extreme star formation and stellar feedback conditions is limited by the physical conditions we can observe in relatively local (D$<$ 10 Mpc) galaxies. 
The ngVLA has the potential to directly map molecular gas across a much larger sample of nearby galaxies, achieving 50~pc scales out to 100 Mpc distances. 

Tracing dense gas in extragalactic targets is challenging with today's instruments as lines (e.g N$_2$H$^+$ \cite{Jimenez2023}) and line ratios (e.g. HCN/CO \cite{Jimenez2019} or N2H$^{+}$/HCN \cite{Jimenez2023}) are simply too faint to be mapped over large parts of nearby galaxy disks and remain challenging even with ALMA. Extragalactic observations must use the brightest higher-critical density lines, such as HCN, HCO$^+$, and HNC where ratios between lower and higher critical density lines, e.g., between CO~(1-0) and HCN~(1-0), yield an approximate, first-order gauge of the gas density \cite[e.g.][]{Gao2004,Bigiel2016,Jimenez2019,Beslic2021}. 
The ngVLA would enable Milky Way science across a broad spectrum of galaxies, at or below the cloud scale in both CO and molecular gas tracers. 
The enormous sensitivity gain with ngVLA will open up direct routine mapping of N$_2$H$^+$, the most robust tracer of dense star-forming gas in the Milky Way, across entire galaxies, with high spatial resolution for a diverse galaxy population. Lines such as HCN and HCO$^+$ will be mapped in fainter galaxies, including dwarfs and more distant galaxies galaxies with more extreme starburst conditions. 
These direct constraints of multiple dense gas tracers would provide a statistical view towards understanding the exact role gas density plays in regulating star formation. 

%Sharpening our view of CO emission maps the bulk molecular gas mass, with isotopologues and its physical properties like excitation, optical depth and abundances \cite[e.g.,][]{Cormier2018, Denbrok2022}. 
Stellar feedback is expected to contribute to preventing gas within a GMC from undergoing gravitational collapse \cite[e.g.][]{Walch2015, Rathjen2021}, setting the efficiency of gas to form stars. 
Early feedback from both stellar winds and ionizing radiation is crucial in setting the local conditions to maximize the impact of supernovae events \cite{Peters2017}, %which occur only after $\sim$5 Myr when the most massive stars reach the end of their lifetime, 
and suppress the accretion of fresh gas soon after the cluster has formed \cite{Gatto2017}. 
Recent studies have shown how supernova feedback plays a critical role in shaping the structure of the ISM by driving large $>$100~pc superbubbles into the molecular gas \cite{Watkins2023}. Without higher angular resolution views of the molecular gas, we are unable to directly constrain the point at which pre-supernova feedback processes dominate the early evolution of these bubbles (Figure \ref{fig:watkins}), or the efficiency with which these processes couple with the gas \cite{Egorov2023}, critical parameters with strong metallicity dependencies when simulating star formation in galaxies. ngVLA observations combined with new JWST data constraining the young stellar populations can place direct quantitative constraints on these processes, even for the earliest embedded phases of the cosmic matter cycle. 

\begin{figure}[h]
\centering
  \begin{minipage}[c]{0.46\textwidth}
\includegraphics[width=\textwidth]{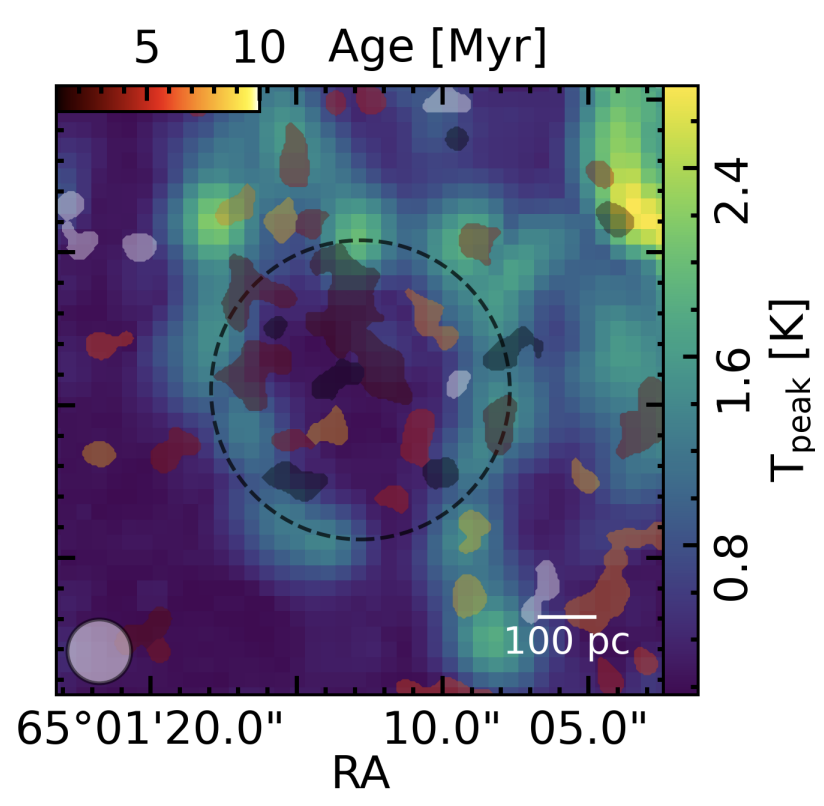}
\end{minipage}\hfill
    \begin{minipage}[c]{0.53\textwidth}
\caption{A superbubble in molecular gas in NGC 1566. The molecular ring (blue-green colorscale, dashed line) traced in CO(2-1) is driven by the stellar feedback processes from young star clusters (red-yellow colorscale). By modeling the energetics of this system, it is possible to determine that the bubble expansion must be driven by supernova explosions, as might be expected for such large $>$100pc diameter features. Direct calculation of the energetics due to stellar winds and radiation require a sharper view to identify the smaller (earlier) bubbles within nearby galaxies. Figure adapted from \cite{Watkins2023}.}
\label{fig:watkins}
  \end{minipage}
\end{figure}  
\subsection{Wide-band polarization AGN studies, covering scales from parsecs to kiloparsecs, measuring extreme Faraday rotation and complex Faraday effects}
\RaggedRight\label{livingston01}
\vspace*{\baselineskip}

\noindent \textbf{Thematic Areas:} \linebreak $\square$ Stellar Astrophysics \linebreak $\square$ Solar System, Planetary Systems and Habitability \linebreak
$\square$ Circuit of Cosmic   Matter (incl. star formation) \linebreak $\square$ The Galaxy and the Local Group \linebreak
  $\checked$   Galaxies and AGN \linebreak $\square$  Cosmology, Large Scale Structure and Early Universe \linebreak
  $\checked$    Extreme conditions in the cosmos, fundamental astrophysics    \linebreak
  $\square$ Interdisciplinary research and technology \linebreak
  
\textbf{Principal Author:}

Name:	Jack Livingston       
 \linebreak						
Institution:  Max Planck Institute for Radio Astronomy (MPIfR)
 \linebreak
Email: jack.david.livingston+academic@gmail.com
 \linebreak
 
\textbf{Co-authors:} Yuri Kovalev (MPIfR) and Andrei Lobanov (MPIfR)
  \linebreak

%\noindent\textbf{ngVLA Background:} The ngVLA is a ground breaking extension to the existing VLA and will provide may new opportunities in Active Galactic Nuclei (AGN) research, in particular advancements in the study of the polarization of AGN. The ngVLA will have 244 antennas with diameters of 18 meters, with 19 supplemental 6 meter diameter antennas to cover the sampling of short baselines. These antennas will have six band receivers that span a total coverage of 1.2 -- 116 GHz. The longest baselines of the ngVLA array will be at parody with the current VLBA, while providing high-fidelity and high surface brightness sensitivity and more complete UV-coverage than the VLBA. The resolution of the maximum baselines at 93 GHz will be $\sim 0.08$ mas; for a nearby AGN such as 3C120 ($z \sim 0.033$) with a central mass of $\sim 6 \times 10^7 M_{\odot}$, this means we could probe to around 10 times the Schwarzschild radius of the center engine. 
\noindent\textbf{ngVLA Background:} The ngVLA is a ground breaking extension to the existing VLA and will provide may new opportunities in Active Galactic Nuclei (AGN) research, in particular advancements in the study of the polarization of AGN. The ngVLA will have 244 antennas with diameters of 18 meters, with 19 supplemental 6 meter diameter antennas to cover the sampling of short baselines. These antennas will have six band receivers that span a total coverage of 1.2 -- 116 GHz. The longest baselines of the ngVLA array will be at par with the current VLBA, while providing high-fidelity and high surface brightness sensitivity and more complete UV-coverage than the VLBA. The resolution of the maximum baselines at 93 GHz will be $\sim 0.08$ mas; for a nearby AGN such as 3C120 ($z \sim 0.033$) with a central mass of $\sim 6 \times 10^7 M_{\odot}$, this means we could probe to around 10000 times the Schwarzschild radius of the center engine.
\\
\noindent\textbf{Science Case for AGN Polarization Studies:} The central engines of AGN generate extreme luminosities, and are believed to be powered by supermassive black-holes (SMBHs). These SMBH should have strong magnetic fields and high electron densities in, and above, the inner regions of their accretion disks which play a major role in the formation of relativistic jets that are visible at radio frequencies. Magnetic fields of up to $B_\mathrm{jet} \sim 1$ G are typically found near the base of the jet at around $r \sim 1$ pc \citep{Lobanov1998,Pushkarev2012}. We can measure these magnetic fields via the Faraday rotation of the polarization angle of synchrotron radiation. Faraday rotation near the base of the jet, assuming a mean electron volume density of $10^2 \sim 10^4\,\mathrm{cm^{-3}}$, should be on the order of $10^{8} \sim 10^{10}\mathrm{rad\,m^{-2}}$. Using ALMA, \cite{MartiVidal2015} found a Faraday rotation value for PKS 1830-211 of $10^8\mathrm{rad\,m^{-2}}$. Multiple Faraday rotating media can be present along the line-of-sight as well as complex Faraday effects like co-spatial Faraday rotation and polarized emission; RM-synthesis using the ngVLA will allow us to measure extreme Faraday rotation and other complex Faraday effects.
\\
\noindent\textbf{RM Synthesis with the ngVLA:} With polarization data over a large range of frequencies, the distribution of Faraday rotation along a sight-line can be associated with the complex polarized surface brightness of emission, via RM-synthesis \citep{Brentjens2005}, bypassing extreme Faraday depolarization and the n--$\pi$ ambiguity. This process can disentangle multiple Faraday rotating regions and probe for the presence of other complex Faraday effects. The accuracy and precision of RM-synthesis are highly dependent on the properties of observed frequencies \protect\citep{Brentjens2005,Dickey2019}. With the capabilities of the ngVLA, we are able to resolve Faraday rotation structures to $\frac{69}{\mathrm{SNR}}\mathrm{rad\,m^{-2}}$ for the lowest frequency band and probe Faraday rotation up to $\sim 10^9\mathrm{rad\,m^{-2}}$ with the $70 - 116$ GHz band (channel width of $2$ MHz). With the current VLA, for the VLA Q band ($40 - 50$ GHz, channel width of $2$ MHz), we are limited to $\sim 10^8\mathrm{rad\,m^{-2}}$. Unlike the VLBA, the superior sensitivity and frequency coverage of the ngVLA mean we are not limited by extreme Faraday rotation depolarization and the n--$\pi$ ambiguity, and the ngVLA will not be as limited by opacity and resolution based effects which normally affect our ability to observe extreme Faraday rotation with the VLA.   
\subsection{Astrophysical searches for physics beyond the Standard Model}
\RaggedRight\label{lobanov01}
\vspace*{\baselineskip}

\noindent \textbf{Thematic Areas:} \linebreak $\square$ Stellar Astrophysics \linebreak $\square$ Solar System, Planetary Systems and Habitability \linebreak
$\square$ Circuit of Cosmic   Matter (incl. star formation) \linebreak $\square$ The Galaxy and the Local Group \linebreak
  $\square$   Galaxies and AGN \linebreak $\square$  Cosmology, Large Scale Structure and Early Universe \linebreak
  $\checked$    Extreme conditions in the cosmos, fundamental astrophysics    \linebreak
  $\square$ Interdisciplinary research and technology \linebreak
  
\textbf{Principal Author:}

Name: Andrei Lobanov
 \linebreak						
Institution: Max-Planck-Institut f\"ur Radioastronomie, Bonn
 \linebreak
Email: alobanov@mpifr-bonn.mpg.de
 \linebreak
 
\textbf{Co-authors:}  Aritra Basu (TLS Tautenburg), Simona Vegetti (MPA)
 \linebreak

Finding experimental evidence for physics beyond the standard model
(SM) of particle physics is one of the pinnacles of the present day
physical research. A number of the SM extensions feature a new class of 
ultra-light, weakly interacting sub-eV particles (WISP), including axions, axion-like particles, dark photons, millicharged particles, and chamaeleons \citep[see][for a review]{2013arXiv1311.0029E}, most of which
are also strong candidates for the dark matter and dark energy "carrier" particles. \\

The ngVLA frequency range of 1.4--116\,GHz will allow probing the existence of such particles via different avenues, including signals from dark matter conversion \cite{2009JETP..108..384P}, astrophysical generation \cite{2022PhRvL.129y1102F,2023PhRvD.108f3001B}, propagation through space \cite{apl2013}, and other specific effects such as birefringence \cite{2020PhRvL.124f1102C,2021PhRvL.126s1102B}. \\

The impact of ngVLA in this areas can be exemplified by its potential for detection of dark photons through their kinetic mixing with ordinary photons which should result in oscillations  between the normal photon state ($\gamma$) and a "dark" or "hidden" photon state ($\gamma_\mathrm{s}$) in which the photons acquire a non-zero mass and propagate along time-like geodesics but do not interact with any normal matter \cite{1982JETP...56..502O,1986PhLB..166..196H}. The properties of such dark photon can be completely described by its mass $m_{\gamma_\mathrm{s}}$ and the kinetic mixing parameter, or mixing angle, $\chi$ ($\ll 1$) describing the probability of the oscillation. These oscillations will induce a specific pattern \cite{2008PhRvL.101m1801J} in broadband spectra of astrophysical objects, which can be used for detecting dark photon and characterising their mass and kinetic mixing parameter \cite{apl2013}.

\begin{figure}[h!] 
\begin{center}
\includegraphics[width=0.8\textwidth]{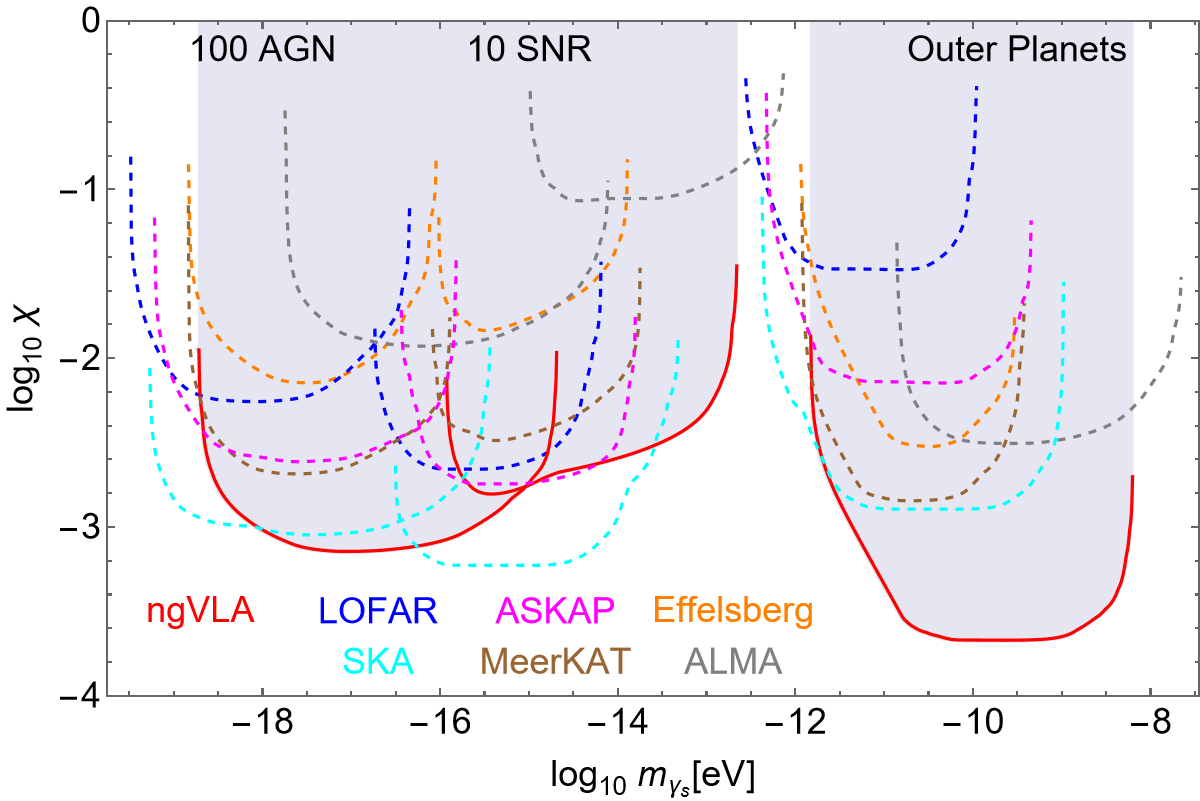} 
\end{center}
\caption{Expected sensitivity of ngVLA observations for
detecting the dark photon oscillations compared to the respective sensitivities of  other instruments. The sensitivity is calculated
assuming observations of 100 AGN, 10 SNR, and 100 measurements of
radio emission from outer planets. Dark photon searches with the ngVLA would cover over ten orders of magnitude in the particle mass in which they would provide unique
constraints on the dark photon coupling, $\chi$ to the ordinary photons. Calculations and figure are adapted from \cite{apl2013}.}  
\label{fig:Lobanov01-fig1} 
\end{figure}

The expected sensitivity of the ngVLA for detecting dark photon oscillations is shown in Fig.~\ref{fig:Lobanov01-fig1} which compares the results that could be obtained by combining together observations of 100 typical AGN, or 10 Galactic, or 100 measurements of the outer Solar system planets performed at different separations from the Earth. \\

One can see that such measurements will cover  $>10$ orders of magnitude of the particle mass and will probe well into the couplings of $\chi\leq 0.01$ in which dark photons can constitute the dark matter \cite{2009JHEP...11..027G}. In all of this mass range, the ngVLA would be one of the most competitive instruments, often delivering the most sensitive measurements.\\

\begin{figure}[h!] 
\begin{center}
\includegraphics[width=0.8\textwidth]{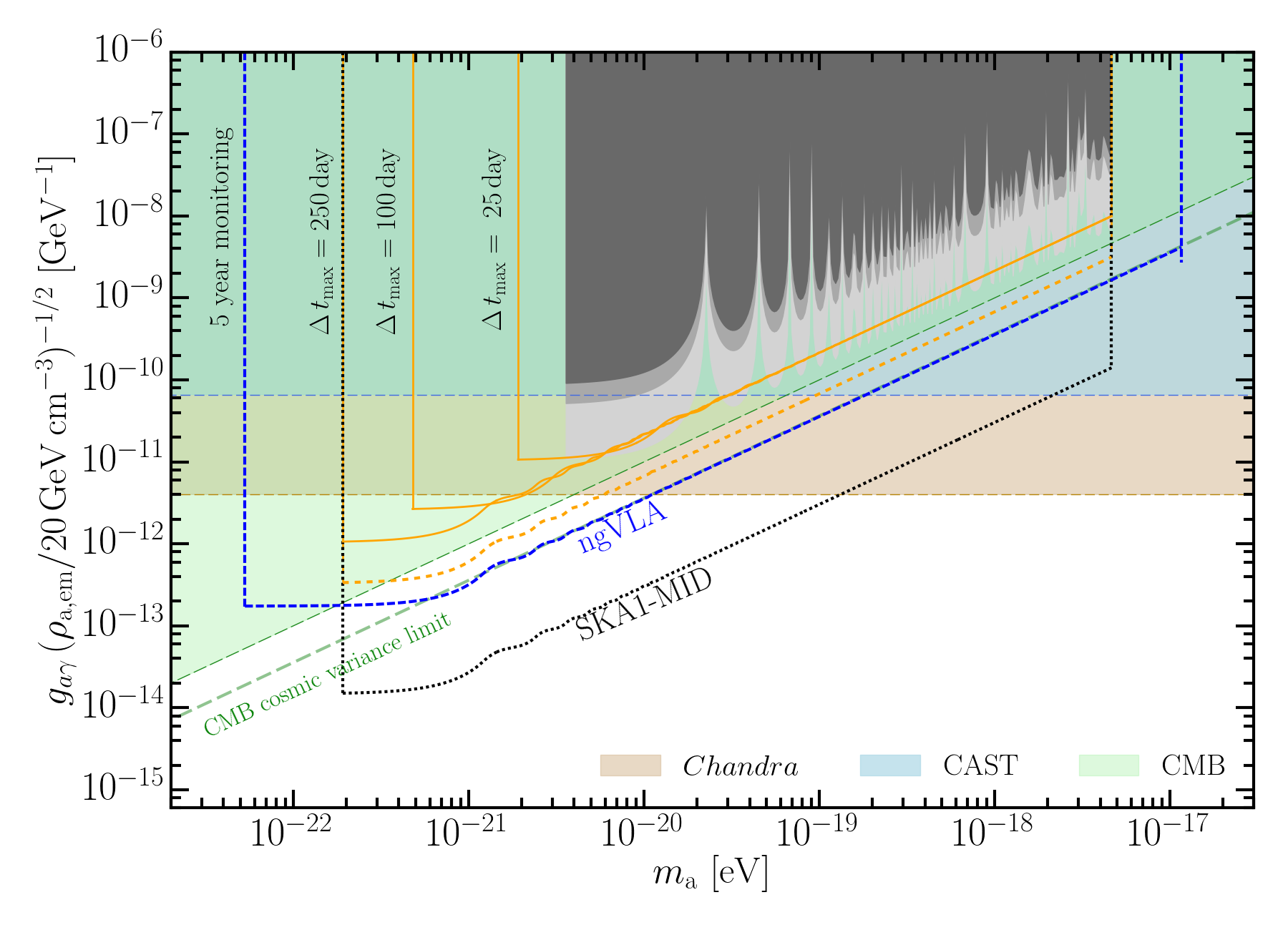} 
\end{center}
\caption{Expected sensitivity of the ngVLA to differential birefringence from ultra-light ALPs.}  
\label{fig:ulalps} 
\end{figure}

\noindent{}\textbf{Ultra-light axion-like particles:} Light boson degrees of freedom, as for example, axions and axion-like particles represent a potential and well-motivated candidate for dark matter. One of the most interesting aspects of this type of dark matter is that of achromatic birefringence \cite{HARARI199267, PhysRevLett.81.3067}. When electromagnetic radiation propagates through axionic fields, a tiny interaction ($g_{a\gamma}$) between each other leads to different phase velocities of the right- and left-handed polarisation states of the photons. As a consequence, the plane of a linearly polarized light undergoes rotation by $\Delta\theta_a$ which directly depends on the properties of the axionic field. Observations of astrophysical and cosmological sources that emit polarised light have been used to detect this effect and set constraints on the properties of the axionic field \cite{2019JCAP...02..059I,2019PhRvL.122s1101F}. However, due to instrumental and calibration related offsets, and often insufficient knowledge of the intrinsic polarization of the source, $\Delta\theta_a$ is difficult to measure. Strong gravitational lensing provides a measure of  differential birefringence between the multiple images of the same lensed source. Hence it represents a unique opportunity to constrain the axion field in a way that alleviates all astrophysical and instrumental systematics \cite{2021PhRvL.126s1102B}.

The high angular resolution, and sensitivity of the ngVLA  will play a pivotal role in advancing the search for ultra-light axionic fields with mass $\lesssim10^{-17}$\,eV. In particular, broadband polarimetry will allow for searching for differential birefringence signals from strong gravitational lenses that produce multiple images. As can be seen from Fig.~\ref{fig:ulalps}, the expected parameter space a sample of gravitational lenses monitored for 5\,years with the ngVLA will pessimistically probe is already comparable to the cosmic variance limit of birefringence measurements via the polarized CMB \cite{PhysRevD.100.015040}. 
\subsection{Physics and astrometry with photon rings}
\RaggedRight\label{lobanov02}
\vspace*{\baselineskip}

\noindent \textbf{Thematic Areas:} \linebreak $\square$ Stellar Astrophysics \linebreak $\square$ Solar System, Planetary Systems and Habitability \linebreak
$\square$ Circuit of Cosmic   Matter (incl. star formation) \linebreak $\square$ The Galaxy and the Local Group \linebreak
  $\square$   Galaxies and AGN \linebreak $\square$  Cosmology, Large Scale Structure and Early Universe \linebreak
  $\checked$    Extreme conditions in the cosmos, fundamental astrophysics    \linebreak
  $\square$ Interdisciplinary research and technology \linebreak
  
\textbf{Principal Author:}

Name: Andrei Lobanov
 \linebreak						
Institution: Max-Planck-Institut f\"ur Radioastronomie, Bonn
 \linebreak
Email: alobanov@mpifr-bonn.mpg.de
 \linebreak
 
\textbf{Co-authors:}  Eduardo Ros (MPIfR), Anton Zensus (MPIfR)
 \linebreak

The remarkable success of the Event Horizon Telescope (EHT) using very long baseline interferometry (VLBI) at 230\,GHz (wavelength $\lambda=1.3$\,mm) for imaging ring like structures in Sgr~A$^\star$ \cite{2022ApJ...930L..12E} and M87 \cite{2019ApJ...875L...1E} has opened up new perspectives in detecting and characterising relativistic photon ring structures in AGN and using these measurements for addressing a range of fundamental questions \cite{2023Galax..11...61J}. 

Studies of photon rings are one of the key science drivers for upgrading the EHT, including the planned {\em next generation}~EHT (ngEHT, \cite{2023Galax..11...61J}), and potentially also for future space VLBI missions \cite{2020SciA....6.1310J}. These approaches largely rely on increasing the instrumental resolution, $\theta_\mathrm{inst} \approx 1.22 \lambda_\mathrm{obs}/B_\mathrm{max}$, by either shortening the observing wavelength, $\lambda_\mathrm{obs}$, or increasing the maximum baseline, $B_\mathrm{max}$, of interferometric measurements.  

However, the effective resolution, $\theta_\mathrm{im}$, of interferometric observations also depends on  signal-to-noise ratio of visibility measurements \cite{2005astro.ph..3225L,2015A&A...574A..84L} and, respectively, on the dynamic range, $D_\mathrm{im}$, of images reconstructed from these visibilities, so that $\theta_\mathrm{im} \approx \theta_\mathrm{inst}/\sqrt{\ln 3\, D_\mathrm{im}}$ . This effect, commonly referred to as "superresolution" of CLEAN deconvolution of interferometric images, has now been fully exploited in a range of new imaging algorithms and successfully demonstrated by recovering a ring like structure in M87 from VLBI observations at 86\,GHz ($\lambda= 3.6$\,mm) \cite{Lu2023}.

Observations with the ngVLA are expected to provide an unprecedented improvement of visibility and image noise \cite{Sel18}, and at 93\,GHz they should reach down to $\sigma_\mathrm{rms}=4.5\,\mu\mathrm{Jy}$ on angular scales below 100\,$\mu$as within one hour of observing time. This should allow for imaging the most compact emission at dynamic ranges $D_\mathrm{im}\gtrsim 200\,000\,[S/1\,\mathrm{Jy}]$, which should result in considerable improvements of both effective resolution and accuracy of positional location of compact features such as photon rings.

The resulting positional accuracy would approach microarcsecond level, similar to that expected from frequency phase transfer mode of VLBI at 86\,GHz \cite{2023arXiv230604516D}, which would provide a firm foundation for performing state of the art astrometric measurements of annual parallaxes at distances up to $\approx 100$\,kpc, galactic proper motions up to distances of $\sim 10$\,Mpc, and secular parallaxes at distances up to $\sim 100$\,Mpc.

Prospects of ngVLA observations to be used for photon ring detection and characterisation are presented in Fig.~\ref{fig:Lobanov02-fig1} which outlines dynamic ranges required to distinguish between a ring of finite thickness and other types of compact structure. The figure demonstrates that ngVLA would provide excellent capabilities for imaging both M87 and Sgr~A$^\star$ at 116\,GHz (overcoming the $0.76\times 0.40$\,mas scatter broadening expected at this frequency \cite{2019ApJ...871...30I}) and may also enable robust photon ring characterisation in a good subset of objects considered for ngEHT observations \cite{Rama2023}. The latter objects will particularly benefit from operating the ngVLA (with $B_\mathrm{max}=8680$\,km) in the VLBI mode, with external antennas on transcontinental baselines including potential new locations in Germany and Europe, increasing the maximum baseline length to over 11,000\,km. In this capacity, ngVLA will have an exceptionally strong impact on studies of photon rings.

\begin{figure}[h!] 
\begin{center}
\includegraphics[width=1.0\textwidth]{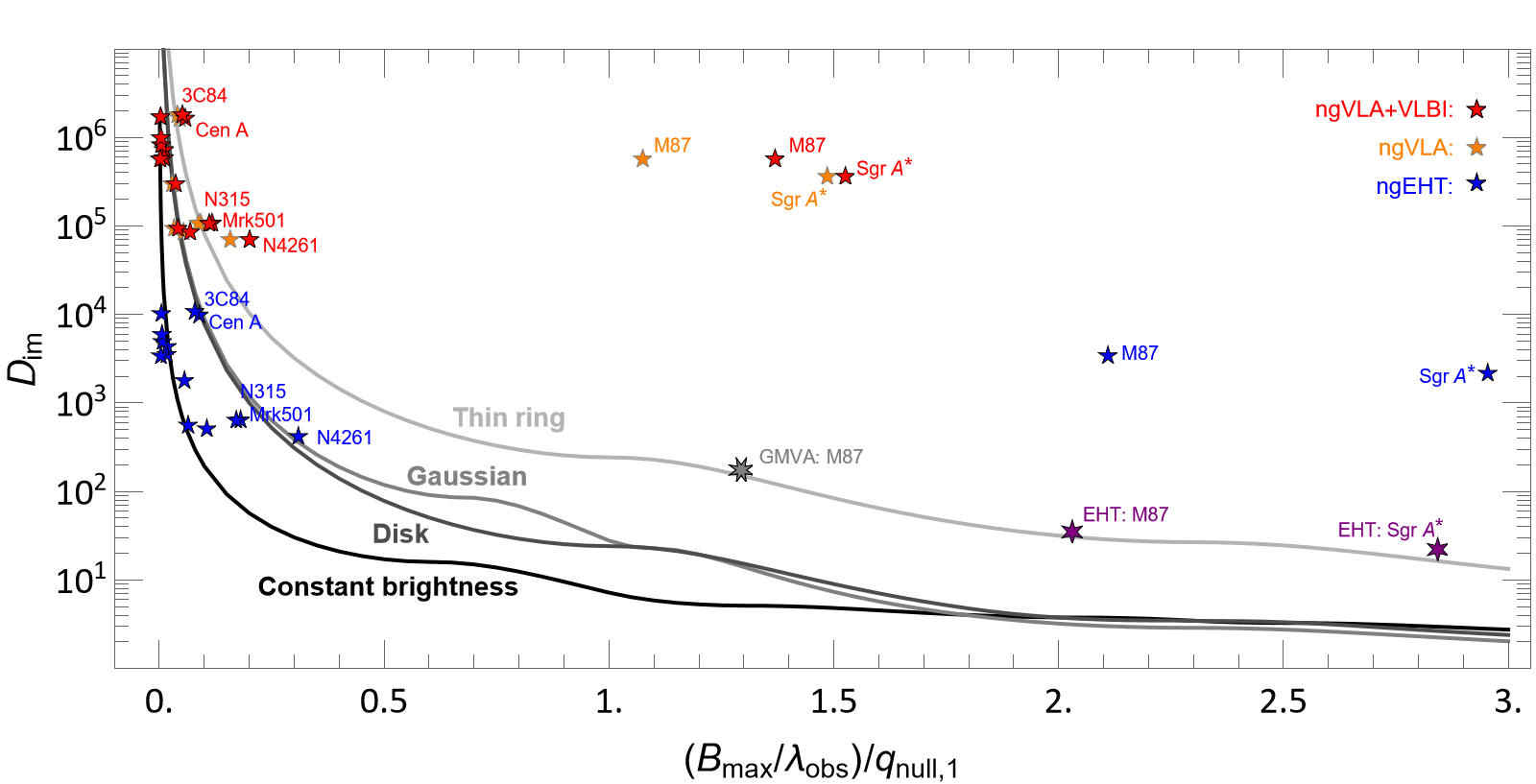} 
\end{center}
\caption{Estimated capability for detecting and characterising photon rings in M87, Sgr~A$^\star$, and selected AGN from the ETHER sample \cite{Rama2023}, compared for the existing EHT (purple stars; \cite{2019ApJ...875L...1E,2022ApJ...930L..12E}) and GMVA (grey star; \cite{Lu2023}) observations and at dynamic ranges, $D_\mathrm{im}$, projected for the ngEHT \cite{Rama2023} (blue stars) at 230 GHz and for 116~GHz observations with the ngVLA \cite{Sel18} (orange stars) and ngVLA operating in VLBI mode (red stars). For all arrays, the maximum sampled spatial frequency, $B_\mathrm{max}/\lambda_\mathrm{obs}$, is normalized by the spatial frequency, $q_\mathrm{null,1}$, of the first null in the visibility response of a thin ring. Scatter broadening \cite{2019ApJ...871...30I} is taken into account in the ngVLA estimates for Sgr~A$^\star$. The four curves show the minimum dynamic ranges which would allow distinguishing a ring with finite thickness from the respective types of structure observed on the most compact scale accessible for a given array. The remarkable sensitivity improvements of the ngVLA should make it highly competitive for detecting and characterising potential photon ring structures in AGN, particularly when operating in VLBI mode with external antennas on transcontinental baselines.}  
\label{fig:Lobanov02-fig1} 
\end{figure}
 
\subsection{Scattering mitigation of the Galactic Center using ngVLA}
\RaggedRight\label{mueller_mus01}
\vspace*{\baselineskip}

\noindent \textbf{Thematic Areas:} \linebreak $\square$ Stellar Astrophysics \linebreak $\square$ Solar System, Planetary Systems and Habitability \linebreak
$\square$ Circuit of Cosmic   Matter (incl. star formation) \linebreak $\checked$ The Galaxy and the Local Group \linebreak
  $\checked$   Galaxies and AGN \linebreak $\square$  Cosmology, Large Scale Structure and Early Universe \linebreak
  $\checked$    Extreme conditions in the cosmos, fundamental astrophysics    \linebreak
  $\square$ Interdisciplinary research and technology \linebreak
  
\textbf{Principal Author:}

Name: Hendrik M\"uller \& Alejandro Mus\footnote{both authors have and will equally contribute}
 \linebreak						
Institution: MPIfR, Bonn \& Univ. de Val\`encia, Val\'encia
 \linebreak
Email: hmueller@mpifr-bonn.mpg.de, alejandro.mus@uv.es
 \linebreak
 
\textbf{Co-authors:} Andrei Lobanov, Eduardo Ros \& Guang-Yao Zhao
  \linebreak

\textbf{Motivation --}
The Galactic Center (GC) represents an invaluable astrophysical laboratory, hosting a diverse assemblage of celestial remnants, including supernova remnants, pulsars, magnetars, and radio transients, among other sources~\cite{Heywood2022}. A focal point of paramount scientific interest is the super-massive black hole, Sagittarius A* (SgrA*) \cite[e.g.][]{Gravity2020, 2022ApJ...930L..12E}. However, the GC is susceptible to the perturbing influence of the interstellar medium (ISM), giving rise to refractive and diffractive scattering phenomena that significantly perturb visibility measurements, particularly at frequencies below approximately 120 GHz \cite{Johnson2015, 2019ApJ...871...30I}. % \gy{below 86 GHz?}.

The forthcoming capabilities of the ngVLA hold the promise of enabling observations of SgrA* even at lower frequencies than attainable with the Event Horizon Telescope (EHT), thereby expanding the dynamic range of observations. Nonetheless, an effective strategy for the mitigation of scattering effects is imperative.

To address this challenge, the German research team is poised to develop innovative mathematical tools aimed at ameliorating the influence of scattering, thus affording astronomers the ability to attain high-quality imaging results. The degree of scattering is intrinsically linked to the square of the observation wavelength ($\lambda^2$), whereby higher frequencies experience proportionally reduced scattering effects. The extensive frequency bandwidth envisioned for the ngVLA is poised to capitalize on this wavelength-dependent characteristic, albeit with stochastic perturbations affecting visibilities unevenly. The research team advocates for the introduction of a novel mathematical function, termed a "regularizer," which leverages high-frequency visibilities to counterbalance the adverse impacts of scattering upon their low-frequency counterparts.

\textbf{Methodology --}
To tackle the problem of the stochastic scattering noise affecting the visibilities, we propose a mathematical function that will use the full wide bandwidth to mitigate the corruptions.
A first regularizer was developed in~\cite{johnson2016, Johnson2018, 2019ApJ...871...30I}. This regularizer will be included in a forward-modeling approach. Due to a number of degeneracies between the intrinsic resolution and variability of SgrA*, and the properties of the scattering screen itself, a multimodal and multiobjective strategy is needed as recently developed in  ~\cite{moeadI, moeadII}.
In this way, even a dynamical multifrequency reconstruction of the screen will be able to be done.

% Forward modeling, RML solution
% Extension MJ18
% MOEAD + scattering + multifrequency + new regularizer

\textbf{Science cases --}
Once this method is successfully implemented, the advantages for the astronomical community will be profound, with implications for various aspects of astrophysical research:
\begin{compactitem}
    \item \textbf{Constrain the Scattering Screen Mapping:} The ability to constrain the scattering screen during one long observation holds significant scientific value. This technique would enable astronomers to investigate the temporal evolution of interstellar scattering, to better deal with the turbulence and structures in the interstellar medium. Exploiting the wide bandwidth on high frequencies to constrain and mitigate the variability on lower frequencies, even doing so in a dynamic way, can provide a better calibration techniques that will improve our understanding of celestial objects affected by these screens, such as pulsars and distant galaxies.

    \item \textbf{Improved Transient Detection and Astrometry:} The mitigation of scattering effects is essential for the precise detection and characterization of transient astrophysical phenomena, including Fast Radio Bursts (FRBs), gamma-ray bursts, and transients. In particular, recent ``super-slow'' pulsars have been observed~\cite{caleb2022, beniamini2023evidence}, but none near the GC. By reducing scattering-induced distortions and signal loss, we can obtain more accurate information about these events, even increasing the chances to find exotic sources such as `super-slow'' neutron stars.

    \item \textbf{Observing SgrA* at Lower Frequencies and Frequency-Dependent Ring Shadow Analysis:} Lower-frequency observations of SgrA* offer the potential for more detailed investigations of the black hole's immediate vicinity. It can facilitate the study of the shadow cast by SgrA*, the dynamics of hotspots in its vicinity, and determine whether the appearance of the ring shadow is frequency-dependent. This analysis is crucial for testing General Relativity, validating or refining models of black hole accretion, and assessing potential deviations from its predictions. In essence, lower-frequency observations not only probe a larger area in the emission region but also provide opportunities to advance our understanding of the fundamental physics governing black hole dynamics and gravity in strong gravitational fields.
\end{compactitem}
\vspace*{\baselineskip}

\noindent
In summary, the successful implementation of our method holds great promise for advancing the frontiers of astrophysical research, enabling scientists to probe the underlying physics of scattering screens, transient phenomena, and the fundamental nature of black holes and gravity in a more nuanced and precise manner. These scientific advancements would deepen our understanding of the universe and its intricate mechanisms.

% If we are able to successfully apply our method, the astronomical community will be benefecit from:
% - We could construct a dynamic map of the scattering screen.
% - Mittigating scattering would help in search of transients, FRB, etc.
% - See SgrA* shadow at lower freq. and trace hotspot evolution, test general relativity, determine if ring shadow depends on the freq.
\subsection{The Cosmic Density and Excitation of Cold Molecular Gas}\RaggedRight\label{riechers01}
\vspace*{\baselineskip}

\noindent \textbf{Thematic Areas:} \linebreak $\square$ Stellar Astrophysics \linebreak $\square$ Solar System, Planetary Systems and Habitability \linebreak
$\square$ Circuit of Cosmic   Matter (incl. star formation) \linebreak $\square$ The Galaxy and the Local Group \linebreak
  $\checked$   Galaxies and AGN \linebreak $\checked$  Cosmology, Large Scale Structure and Early Universe \linebreak
  $\square$    Extreme conditions in the cosmos, fundamental astrophysics    \linebreak
  $\square$ Interdisciplinary research and technology \linebreak
  
\textbf{Principal Author:}

Name:	Dominik A. Riechers
 \linebreak						
Institution:  Universit\"at zu K\"oln
 \linebreak
Email: riechers@ph1.uni-koeln.de
 \linebreak
 
\textbf{Co-authors:} Fabian Walter (MPIA), Leindert Boogaard (MPIA), Ryan Keenan (MPIA), Pierre Cox (IAP)
  \linebreak

Substantial efforts including multiple large programs on the Atacama Large Millimeter/submillimeter Array (ALMA), the Karl G. Jansky Very Large Array (VLA) and the NOrthern Extended Millimeter Array (NOEMA) over the past decade have now resulted in the first delineation of the evolution of the cold molecular gas content of galaxies per unit volume as a function of redshift (e.g., \cite{riechers19,riechers20,decarli20,boogaard2023}, and references therein). Combined with studies of stellar mass, atomic gas, and the star formation activity, this has now led to a consensus picture of the evolution of baryons associated with galaxies over cosmic time and space (e.g., \cite{walter20}). While the current constraints already required a major revision of simulations of galaxy evolution, substantial uncertainties remain due to the limited volumes ($\sim$700,000\,Mpc$^3$) and depth possible with current observatories, as well as the fact that studies were carried out in different rotational lines of CO, which leaves residual uncertainties due to gas excitation. With its vastly increased survey speed and bandwidth, providing simultaneous coverage of multiple CO lines including continuous coverage of the fundamental CO $J$=1$\to$0 transition (toward which all cold gas masses are calibrated) at all redshifts and the typically brightest CO $J$$\geq$3--5 lines out to $z$$>$6, the next generation Very Large Array (ngVLA; \cite{murphy18}) is expected to push this area of research from a consensus picture to a precision measurement, both in terms of accuracy of individual detections and survey volume.

%Use the \cite command for references, e.g., \cite{Kadler2012}
%Add entries to bibliography.bib

\begin{figure}[h]
    \centering
	\includegraphics[angle=0.0,width=0.63\textwidth]{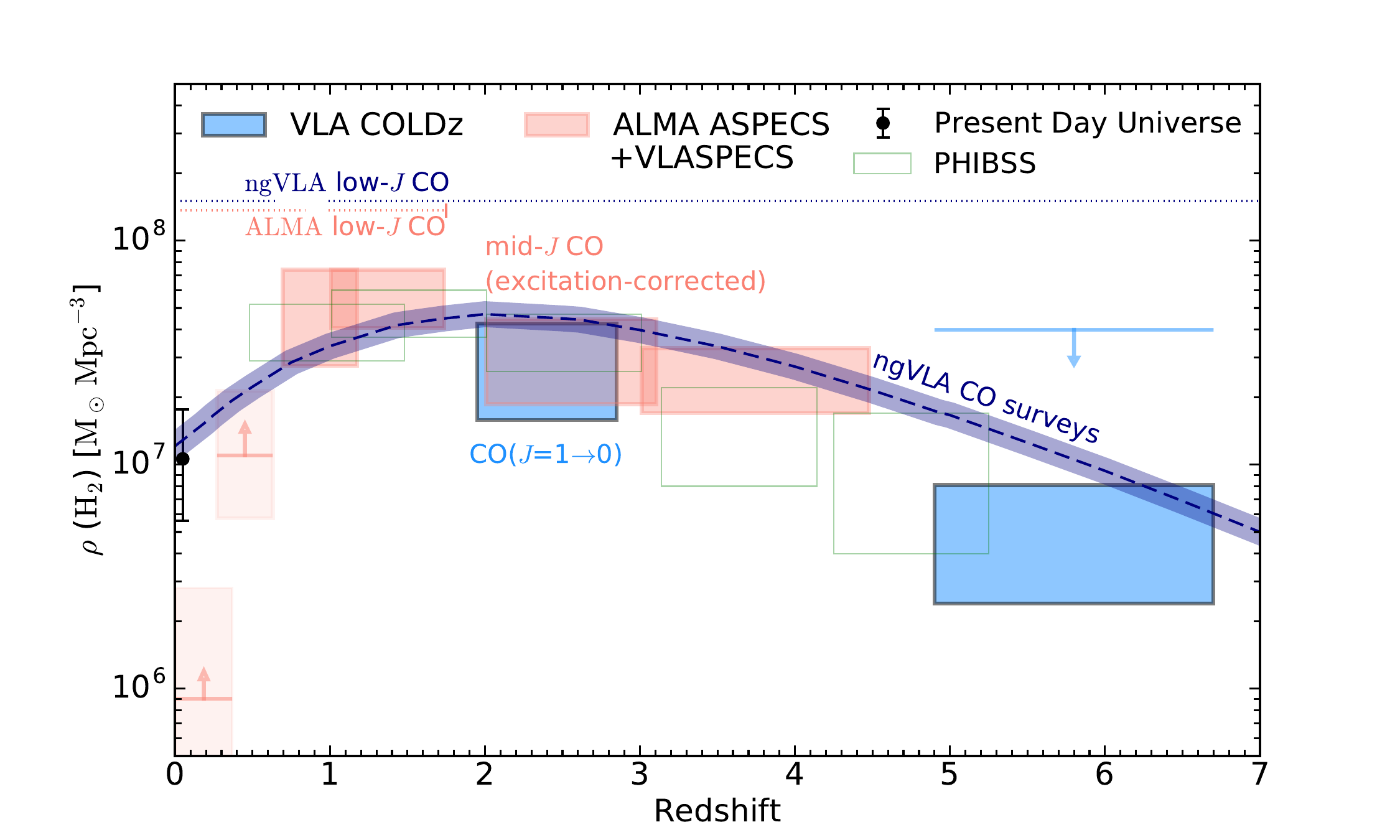}
 	\includegraphics[angle=0.0,width=0.36\textwidth]{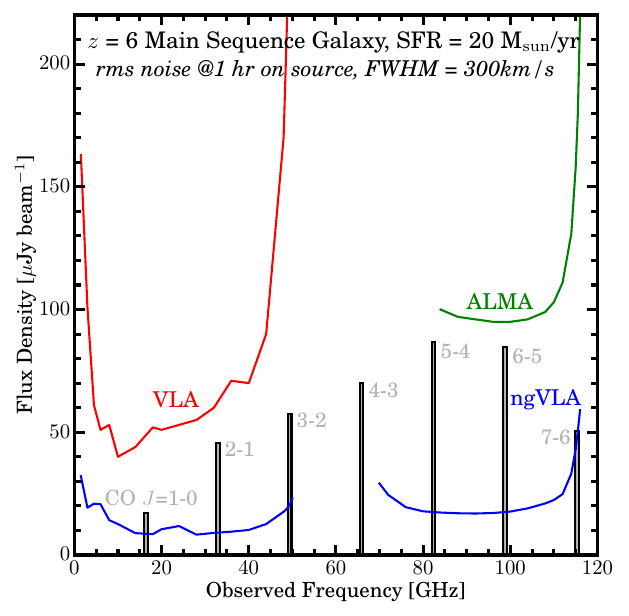}
    \vspace{-6mm}
	\caption{Left:\ Legacy CO line scan surveys based on $>$1000\,hr of dedicated observations with ALMA, NOEMA, and the VLA over the span of a decade (boxes; e.g., \cite{riechers19,riechers20,decarli20,walter20}, and references therein), and improvements in the uncertainties expected due to the vastly improved survey speed of the ngVLA (blue dashed line and shaded curve). Right:\ Accessibility of the CO line ladder for a $z$=6 main sequence galaxy, showing that the ngVLA will have both the sensitivity and bandwidth to cover six of the seven lowest-$J$ rotational transitions for studies of CO excitation. Clearly, the ngVLA is required for the next major breakthrough in this key field of research.}
	\label{f1}
\end{figure}  

\subsection{Molecular Line Absorption Against the CMB as a Probe of Cosmological Parameters}
\RaggedRight\label{riechers02}
\vspace*{\baselineskip}

\noindent \textbf{Thematic Areas:} \linebreak $\square$ Stellar Astrophysics \linebreak $\square$ Solar System, Planetary Systems and Habitability \linebreak
$\square$ Circuit of Cosmic   Matter (incl. star formation) \linebreak $\square$ The Galaxy and the Local Group \linebreak
  $\square$   Galaxies and AGN \linebreak $\checked$  Cosmology, Large Scale Structure and Early Universe \linebreak
  $\checked$    Extreme conditions in the cosmos, fundamental astrophysics    \linebreak
  $\square$ Interdisciplinary research and technology \linebreak
  
\textbf{Principal Author:}

Name:	Dominik A. Riechers
 \linebreak						
Institution:  Universit\"at zu K\"oln
 \linebreak
Email: riechers@ph1.uni-koeln.de
 \linebreak
 
\textbf{Co-authors:} Axel Wei\ss\ (MPIfR), Fabian Walter (MPIA), Pierre Cox (IAP)
  \linebreak

With the ever-growing precision in measurements of fundamental cosmological parameters, standard $\Lambda$CDM cosmology has so far stood the test of time. However, in light of potential complications like the "Hubble Tension" (e.g., \cite{riess23} and references therein), deviations from this simple model are still important to investigate, since they may provide a pathway toward the discovery of new fundamental physical phenomena, such as different forms of dark matter, exotic dark energy, revised prescriptions of gravity, or the presence of currently unknown particles or coupling processes. One particularly promising method to study deviations from the standard model are distortions imprinted on the cosmic microwave background (CMB; e.g., \cite{planck20}) by astrophysical sources across a broad range in redshifts. While galaxies have now been detected out to $z$$>$13 thanks to the James Webb Space Telescope (JWST; e.g., \cite{fujimoto23}), powerful methods to study CMB distortions like the Sunvaey-Zel'dovich effect associated with galaxy clusters remain
%% remained largely 
limited to $z$$\leq$1 
%% in the past
(e.g., \cite{hurier14}).
It has recently been established that the fundamental 1$_{10}$--1$_{01}$ rotational transition of the ortho-H$_2$O molecule can be used to measure 
%% fundamental 
key parameters like the CMB temperature $T_{\rm CMB}$ toward massive starburst galaxies 
% out to 
at redshifts $z$$>$6, where the $T_{\rm CMB}$$>$20\,K is sufficient to excite this transition (Figs.\ \ref{f1} and \ref{f2}; see \cite{riechers22}). While detections are possible with the best current millimeter-wave interferometers like the Atacama Large Millimeter/submillimeter Array (ALMA) and the NOrthern Extended Millimeter Array (NOEMA), the signal is too weak to be spatially resolved in reasonable integration times (Fig.~\ref{f1}). The lack of spatial resolution is the currently largest source of uncertainty of the $T_{\rm CMB}$ measurements. The next generation Very Large Array (ngVLA; \cite{murphy18}) will offer the boost in sensitivity and resolution to provide detailed images of the absorption signal against the CMB, providing detailed column density distributions of the H$_2$O vapour. Moreover, the ngVLA will have the sensitivity to detect the signal in the most massive star-forming galaxies at $z$$>$10 (see Fig.~\ref{f2}), which are 
%% likely 
expected to be an order of magnitude fainter than those seen at $z$=4--6.

%Use the \cite command for references, e.g., \cite{Kadler2012}
%Add entries to bibliography.bib

\begin{figure}[h]
    \centering
	\includegraphics[angle=0.0,width=0.67\textwidth]{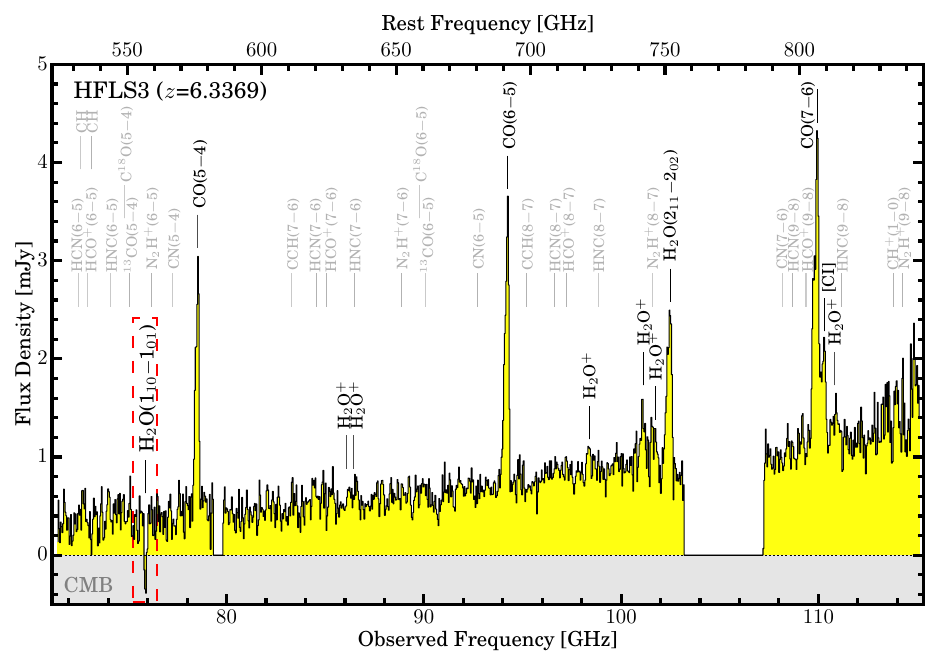}
    \vspace{-6mm}
	\caption{NOEMA detection of H$_2$O 1$_{10}$--1$_{01}$ absorption toward the $z$=6.34 starburst galaxy HFLS3 (red dashed box; \cite{riechers22}). The line is about twice as strong as the galaxy's continuum emission, and thus, absorbs into the CMB. When the dust spectral energy distribution and size of the galaxy are well measured, the strength of the absorption only depends on the CMB properties, in particular its temperature -- providing a direct measurement at the redshift of the galaxy (see Fig.~\ref{f2}). The ngVLA will have 
 % provide 
   the sensitivity and spatial resolution to resolve the absorption signal, which will 
   %% provide 
   result in substantially more accurate measurements.}
	\label{f1}
\end{figure}

\begin{figure}[h]
    \centering
	\includegraphics[angle=0.0,width=0.44\textwidth]{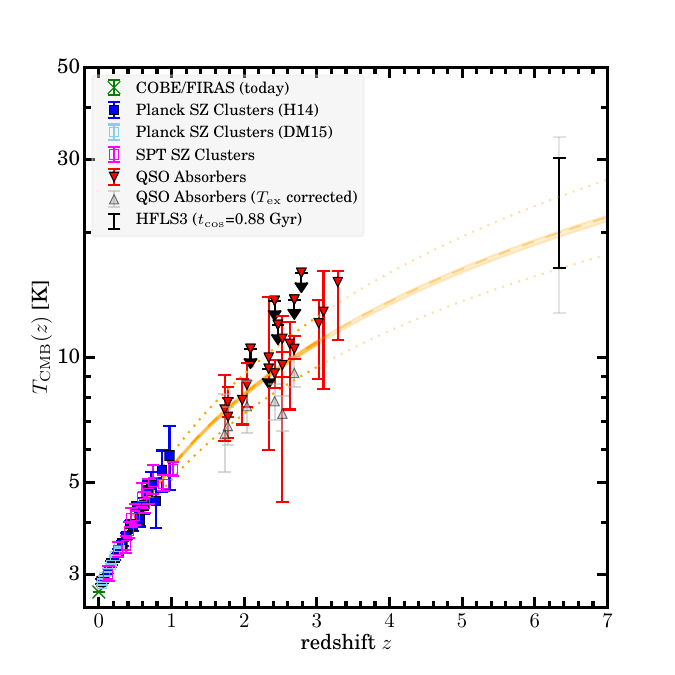}
 	\includegraphics[angle=0.0,width=0.47\textwidth]{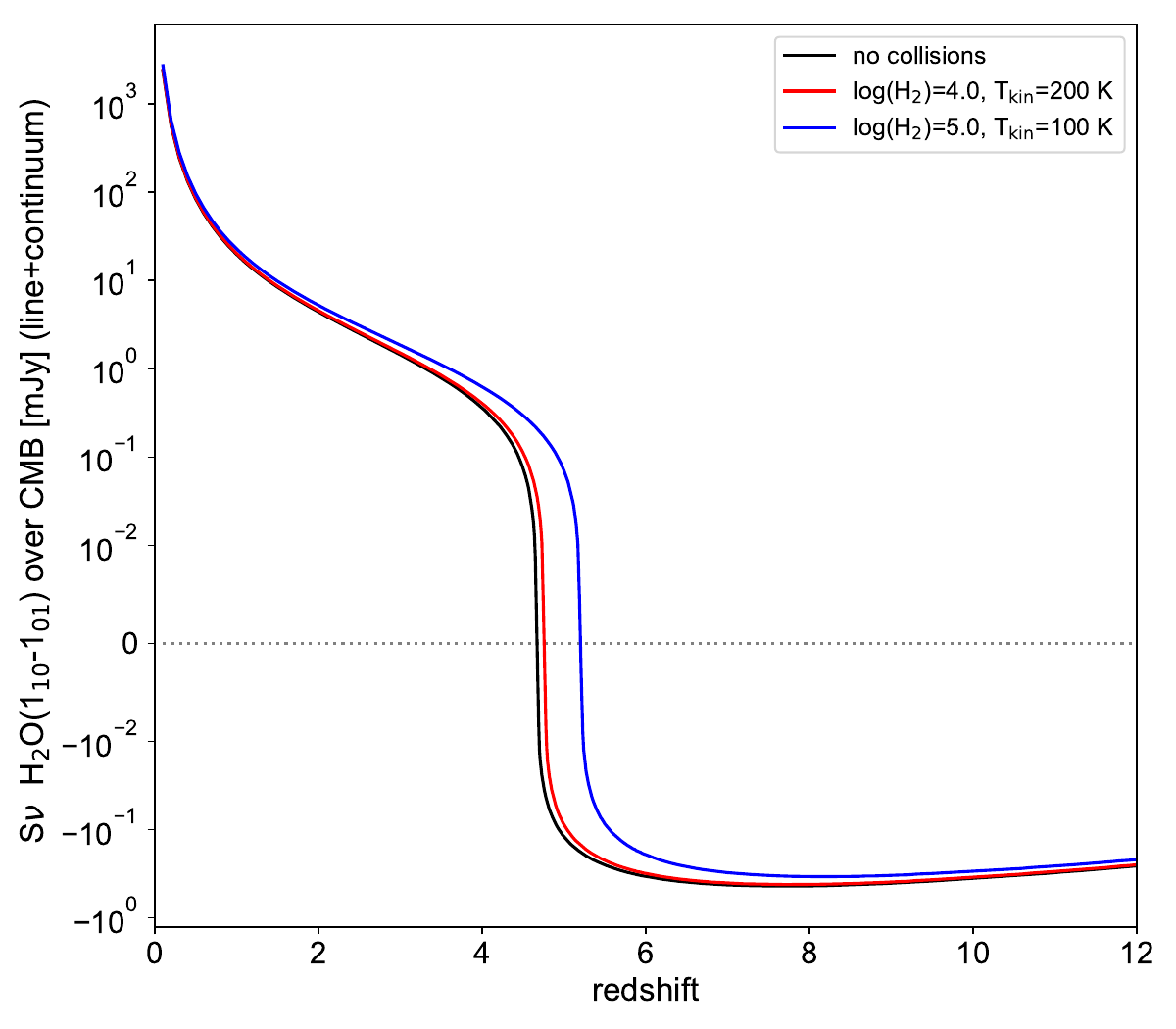}
	\caption{Left:\ Direct measurements of the CMB temperature previously existed at $z$=0--1 based on the Sunyaev-Zel'dovich effect toward galaxy clusters, and model-dependent measurements existed for quasar absorption line systems out to $z$$\sim$3. H$_2$O 1$_{10}$--1$_{01}$ absorption-based studies offer the potential to provide such measurements to much higher redshifts. The largest source of error for the measurement in HFLS3 is the knowledge of the size of the galaxy. The orange line and shaded area show the best combined fit to all current data, which is consistent with standard $\Lambda$CDM cosmology (see \cite{riechers22} and references therein). Right:\ Depending on the contribution from collisional excitation, models suggest that for starbursts like HFLS3, the CMB absorption should be detectable at $z$$>$4.5, and be almost independent of redshift out to $z$$>$12. The ngVLA will provide the sensitivity to carry out such measurements for the most evolved galaxies at $z$$>$10.}
	\label{f2}
\end{figure}  
\subsection{VLBI probes of supernovae, supernova factories and starburst galaxies}
\RaggedRight\label{ros01}
\vspace*{\baselineskip}

\noindent \textbf{Thematic Areas:} \linebreak $\checked$ Stellar Astrophysics \linebreak $\square$ Solar System, Planetary Systems and Habitability \linebreak
$\square$ Circuit of Cosmic   Matter (incl. star formation) \linebreak $\square$ The Galaxy and the Local Group \linebreak
  $\square$   Galaxies and AGN \linebreak $\square$  Cosmology, Large Scale Structure and Early Universe \linebreak
  $\checked$    Extreme conditions in the cosmos, fundamental astrophysics    \linebreak
  $\square$ Interdisciplinary research and technology \linebreak
  
\textbf{Principal Author:}

Name:	Eduardo Ros
 \linebreak						
Institution:  Max-Planck-Institut für Radioastronomie
 \linebreak
Email: ros@mpifr-bonn.mpg.de
 \linebreak
 
\textbf{Co-authors:} Miguel Á. Pérez Torres (Inst. de Astrof\'{\i}sica de Andaluc\'{\i}a), Antxon Alberdi  (Inst. de Astrof\'{\i}sica de Andaluc\'{\i}a), Jos\'e C. Guirado (Univ. de Val\`encia)
  \linebreak

The incorporation of long baselines into the Next Generation Very Large Array (ngVLA) will enhance its operational capabilities, allowing it to operate in a manner similar to Very Long Baseline Interferometry (VLBI). 
This enhancement will be characterised by significantly improved $(u,v)$-coverage, due to both the increased number of antennas and the broader frequency range. 
Known for its unique ability to study compact radio emission in exceptional detail, VLBI focuses primarily on stellar surfaces, young stars, supernovae and their aftermath, such as neutron stars and black holes with companions.

The long baseline options within the ngVLA represent a major leap forward in structural sensitivity at larger scales compared to current VLBI methods. This enhancement will allow unprecedented flux density sensitivity on compact angular scales up to a frequency of about 100\,GHz---a significantly unexplored area of astrophysics for these objects.

With its expanded capabilities, the ngVLA will make a major contribution to the study of stellar physics. In addition to the instrument's role in studying gamma-ray bursts, microquasars, pulsars and supernova remnants, as described in other sections of this document, it will be essential for studying radio supernovae and nearby starburst galaxies.

The radio emission from young supernovae is thought to result from the shock interaction between the supernova ejecta and the circumstellar medium (see e.g., \cite{2009ARA&A..47...63S}. Different physical parameters of this medium influence the timing of the radio peak of supernovae, ranging from early emission, as observed in SN 1993J in Messier 81 \cite{1993Natur.364..600S,1995Natur.373...44M}, to later emission after ten years, as expected for superluminous Type I supernovae \cite{2012Sci...337..927G,2021ApJ...912...21E,2023ApJ...954L..45M}, or even many years later, as seen in 'regular' supernova remnants.

In the absence of recent supernova explosions in the Milky Way, starburst galaxies, particularly Luminous Infrared Galaxies (LIRGs), become prime targets for supernova observation. LIRGs serve as supernova factories, providing the opportunity to observe individual supernova explosions across the electromagnetic spectrum \cite{2021A&ARv..29....2P}.

Leveraging the sensitivity of the ngVLA will significantly increase the number of core-collapse supernovae detected and observed compared to current studies. Galaxies such as IC 10, less than 1 Mpc away, NGC 253 (Sculptor) at 3.4\,Mpc, and Messier\,82 at about 4 Mpc can have their radio structures resolved. This is achieved by studying their angular expansion over long periods of time, providing insights into the morphology and deceleration parameters of the ejected material. Such observations provide valuable information about the circumstellar and interstellar environments, as well as the progenitor star. In addition, the detection of individual radio supernovae and remnants in circumstellar starbursts \cite{2009A&A...507L..17P} provides a reliable estimate of the star formation rate in nearby galaxies, assists in the identification of Type I-a radio supernovae, and offers clear insights into the nature of progenitor systems.  

Resolving young supernovae in starburst galaxies at longer distances, such as Arp\,299 at 50\,Mpc, NGC\,1615 at 68\,Mpc, Mrk\,321 at 75\,Mpc, and Arp\,220 at 83\,Mpc, will be challenging, but catching supernovae at the moment of explosion, when they peak at higher frequencies, may help.

It is noteworthy that the study of radio supernovae is closely linked to the observation of their most energetic multi-wavelength counterparts, namely gamma-ray bursts and fast radio bursts. This synergy is particularly relevant in the current era of multi-messenger, transient astronomy (see e.g., \cite{2021ARA&A..59..155M}).
\subsection{Zooming into Feedback Engines}
\RaggedRight\label{schilke01}
\vspace*{\baselineskip}

\noindent \textbf{Thematic Areas:} \linebreak $\square$ Stellar Astrophysics \linebreak $\square$ Solar System, Planetary Systems and Habitability \linebreak
$\checked$ Circuit of Cosmic   Matter (incl. star formation) \linebreak $\square$ The Galaxy and the Local Group \linebreak
  $\square$   Galaxies and AGN \linebreak $\square$  Cosmology, Large Scale Structure and Early Universe \linebreak
  $\square$    Extreme conditions in the cosmos, fundamental astrophysics    \linebreak
  $\square$ Interdisciplinary research and technology \linebreak
  
\textbf{Principal Author:}

Name:	Peter Schilke
 \linebreak						
Institution:  University of Cologne
 \linebreak
Email: schilke@ph1.uni-koeln.de
 \linebreak
 
%\textbf{Co-authors:} (names and institutions)
%  \linebreak

High-mass stars and clusters provide most of the energy input shaping clouds on scales from a few~pc to few 100~pc, and galaxies in the form of mechanical
(\eg outflows, stellar winds, expanding \HII\ regions, and eventually, on larger scales, supernova explosions) and radiative (e.g., thermal, UV, X-ray, cosmic ray) feedback. 
A complete understanding of these processes requires a detailed study of the
small-scale structure (i.e.\ a few 1000\,AU) of the different feedback engines, since this is the scale where the feedback originates and its power and properties are determined. 
The physical and chemical structure of  cluster-forming habitats is studied using submm/mm and radio observations sensitive to the dense gas structure and the presence of embedded newly formed massive stars that inject energy into their surroundings. 

In the last few years, these studies have undergone a major quantum leap, thanks to new powerful facilities like ALMA and JVLA, which provide unique higher-resolution and higher-sensitivity observations. 
This has enabled detailed studies at 1000\,AU scales for some of the most active and energetic regions in the Galaxy (\eg SgrB2, W51, NGC6334: \cite{SanchezMonge2017}; \cite{Ginsburg2017}; \cite{Sadaghiani2020}). 
The results show highly fragmented structures fed by dense filaments, with multiple massive stars forming in a reduced volume and powering energetic outflows and \HII regions. 
Using different molecules/isotopologues/excitation stages and exploiting a combination of optically thick and thin transitions, it is possible to go beyond the restriction of 2D images and perform a tomography to restore at least part of the line-of-sight structure.
% current observations have much more potential that is worth exploiting.or the emission from relatively simple molecules
% (\eg \ce{CO, CH3CN}), novel observations (including \eg highly-excited states) open new possibilities that we want to exploit.
%

Using sophisticated excitation analysis of various molecules, in particular of high and vibrationally excited states, it is possible to determine the local IR field produced by the embedded star-forming population, otherwise difficult to determine due to high opacity and extinction (\eg \cite{Veach2013, Bruderer2015}), and thus to better constrain the thermal structure of these massive and energetic regions. These lines have the potential to locate embedded sources that warm the surrounding dense gas. 
\begin{figure}[h]
\centering
\includegraphics[width=0.7\textwidth]{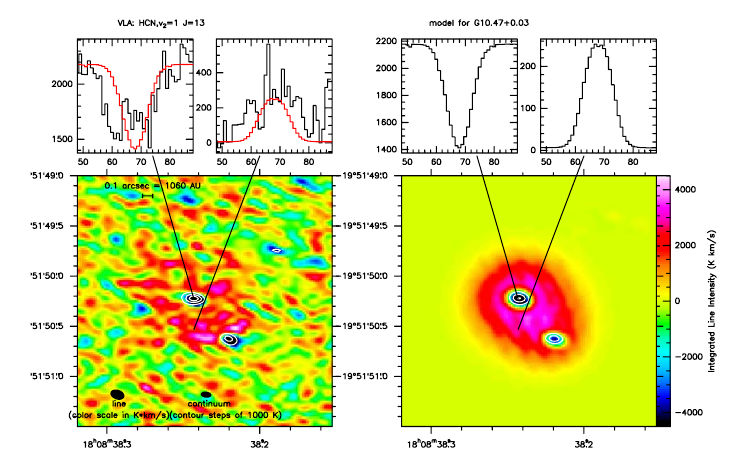}
\caption{VLA observations of $\ell$-type HCN($v_2=1,J=13$) lines toward G10.47 from \citealt{Rolffs2011}.}
\label{fig:G10.47}
\end{figure}

This is possible because the vibrationally excited states are pumped by IR radiation and thus, are sensitive to a specific wavelength corresponding to the energy of the vibrational level.
Complex organic molecules are particularly well suited for this task because they selectively trace the inner, hotter regions where the dust is hot enough to release those species. 
Also, they have, by virtue of their complexity, many rotational lines, which 
means that even in a moderately wide spectrum there are many lines that can be fitted, better constraining the solution;
and many vibrational levels spanning many excitation energies, which allows a finely grained exploration of the IR field.

While in principle this can be done with ALMA, this method suffers in very high column density regions, particularly toward disks around high-mass stars, from the dust being optically thick, even in ALMA Band 3.  The solution is to go to lower frequencies, domain of ngVLA, where \eg $\ell$-type  lines of vibrationally excited HCN or vibrational satellite lines of \ce{HC3N}, \ce{CH3OH}, \ce{C2H5CN} and many others are readily observable (see Fig.~\ref{fig:G10.47} for observations with the VLA in 2009).
\subsection{The chemical characterisation of young protostars: reveal the heritage of forming planets}
\RaggedRight\label{deSimone01}
\vspace*{\baselineskip}

\noindent \textbf{Thematic Areas:} \linebreak $\square$ Stellar Astrophysics \linebreak $\checked$ Solar System, Planetary Systems and Habitability \linebreak
$\square$ Circuit of Cosmic   Matter (incl. star formation) \linebreak $\square$ The Galaxy and the Local Group \linebreak
  $\square$   Galaxies and AGN \linebreak $\square$  Cosmology, Large Scale Structure and Early Universe \linebreak
  $\square$    Extreme conditions in the cosmos, fundamental astrophysics    \linebreak
  $\square$ Interdisciplinary research and technology \linebreak
  
\textbf{Principal Author:}

Name:	Marta De Simone
 \linebreak						
Institution:  European Southern Observatory (ESO), Garching
 \linebreak
Email: marta.desimone@eso.org
 \linebreak
 
%\textbf{Co-authors:} (names and institutions)
%  \linebreak

Chemically characterizing the early stages of planet formation is crucial to understand what chemical content future forming planets may inherit. 
Indeed, several recent evidences suggest that planetesimal formation may already begin in the disks of young protostars \citep[$\lesssim$ 0.5 Myr; e.g.,][]{Manara2018,sheehan_2018,tychoniec_2020}. 
However, these objects are deeply embedded in their natal environment, making their characterization difficult. 
The high sensitivity and angular resolution at centimeter wavelength provided by the ngVLA will be of paramount importance to retrieve the chemical history of these objects while eliminating observational biases due to optically thick dust. \\

From a chemical point of view, the protostellar stage is characterised by high density and temperature conditions. When the dust temperature reaches the water sublimation one ($\sim$100 K) the icy dust mantles evaporates releasing into the gas all the hydrogenated species formed on their surface, and triggering a new complex chemistry. Indeed, some protostars posseses a compact ($<$100 au) region rich in interstellar Complex Organic Molecules \citep[iCOMS;][] {Herbst-vanDishoeck09,Ceccarelli2023}, called hot corino \citep{ceccarelli_hot_2004}. These species are saturated C-bearing molecules with more than six atoms and including heteroatoms (such as O, N), and constitute the bricks to form more complex and prebiotic species.
However, not every protostar possesses a hot corino, and significant differences have been observed in their millimeter (mm) spectra. Indeed, apart from hot corinos (rich in iCOMs), there are sources, known as Warm Carbon Chain Chemistry (WCCC) sources, poor in iCOMs and rich in carbon chains and rings \citep{sakai_warm_2013}.
%Some sources also show a hybrid nature, containing both WCCC and hot corino signatures. 
The origin of this chemical diversity may be related to a different collapse timescale which lead to a different composition of the icy dust grain mantles, to external environmental effects, or to observational biases \citep[e.g.,][]{aikawa_chemical_2020, DeSimone2020, de_simone_tracking_2022}. 
%In addition, the dust surrounding these objects may be  optically thick enough to obscure the molecular emission biasing our comprehension \citep{de_simone_hot_2020}. On the other hand, complex carbon species (with at least seven C atoms) are challenging to observe at (sub-)mm wavelengths, but have been found to be abundant in few sources using radio observations \citep{ McGuire2020, Cernicharo2021,Bianchi2023a}.\\

\begin{figure}
\centering
\includegraphics[width=0.28\textwidth]{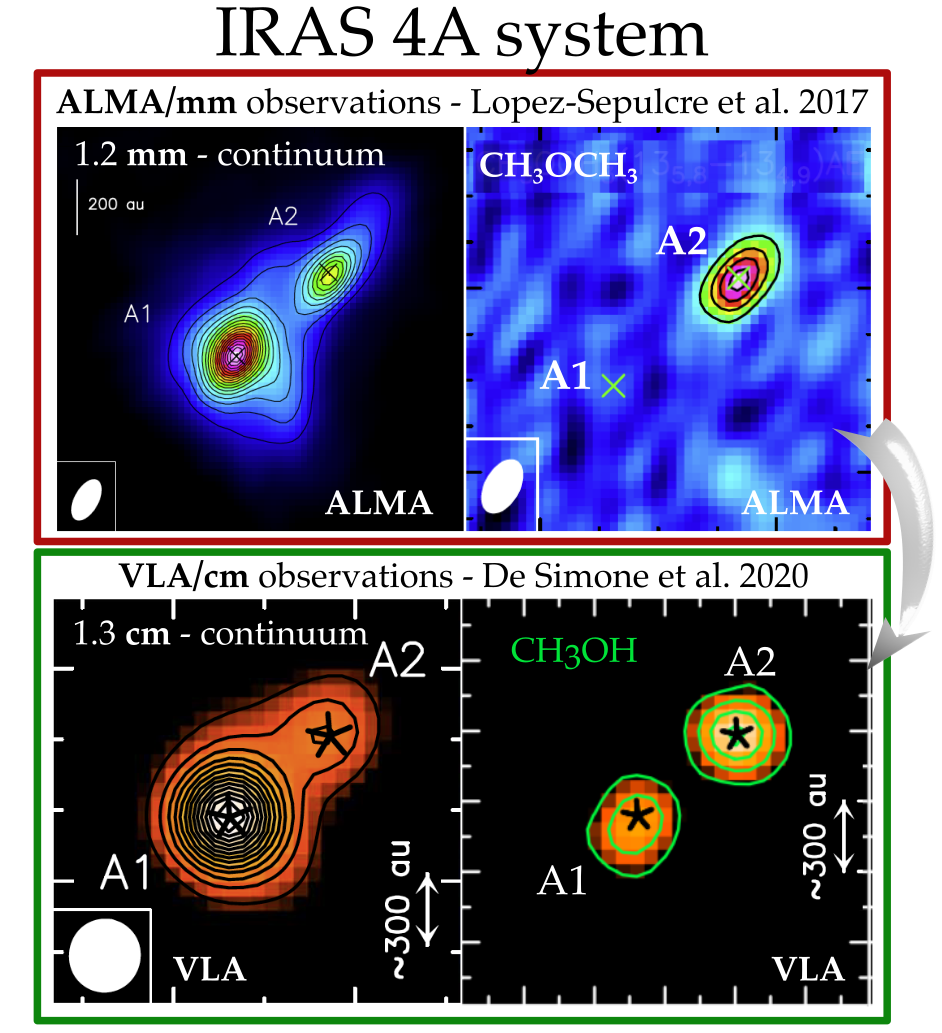}
%\hspace{1cm}
\includegraphics[width=0.30\textwidth]{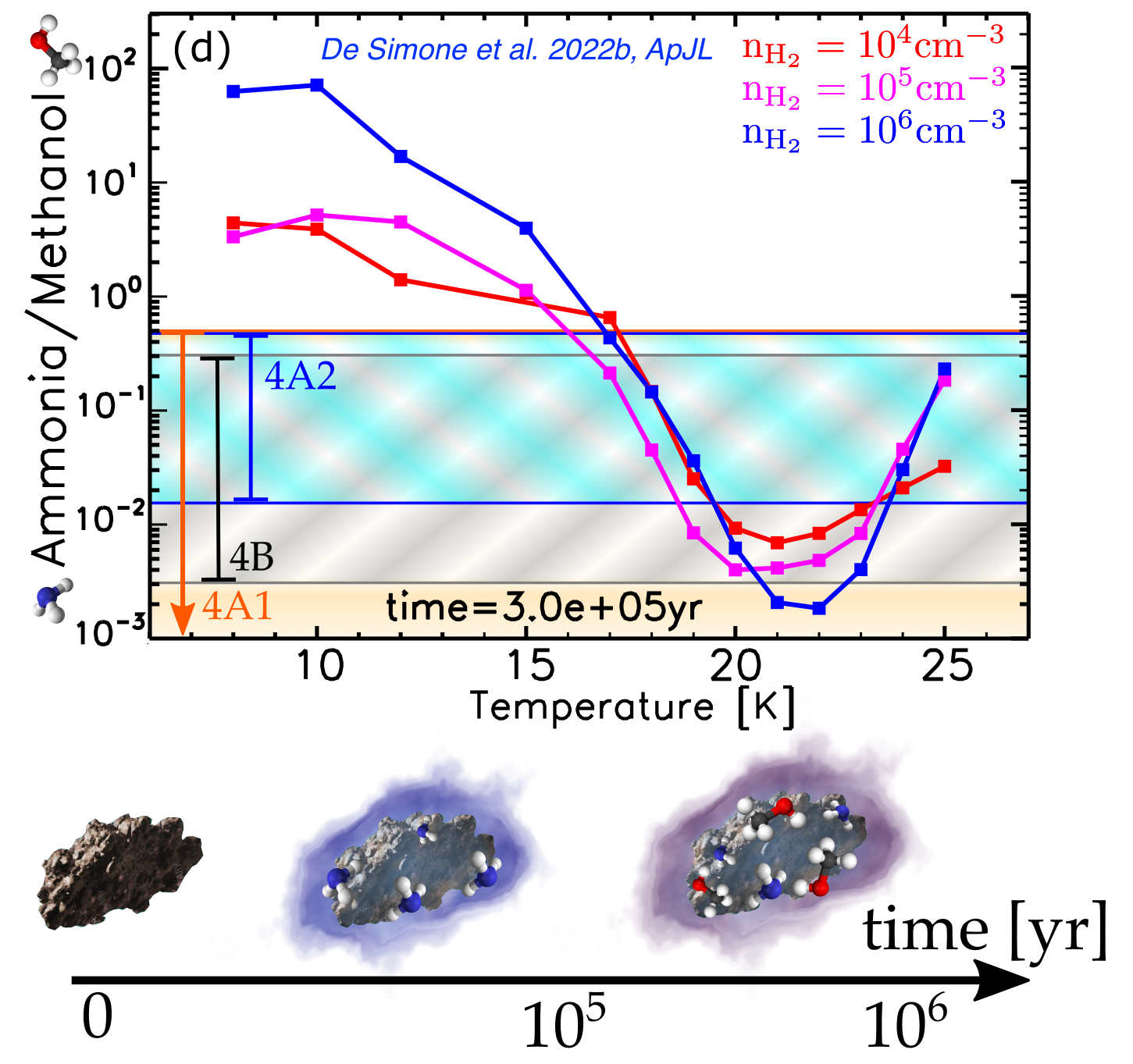}
\includegraphics[width=0.35\textwidth]{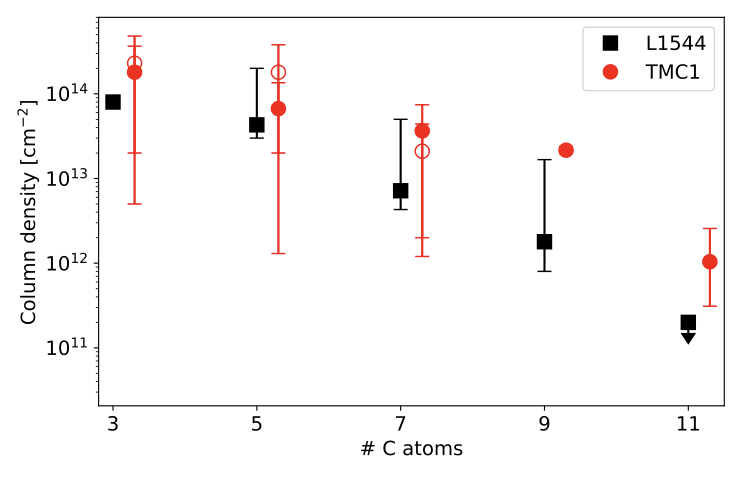}
\caption{\small
\textbf{Pivotal findings of recent VLA, GBT, and YEBES observations. }
\textit{Left:} \textbf{Obscuring dust.} IRAS 4A at mm (\textit{upper panels}) and cm (\textit{lower panels}) wavelengths. The mm dust is so thick to obscure the 4A1 hot corino that pops up in the cm \citep{DeSimone2020}.
\textit{Middle:} \textbf{Tracing the ice mantle history.} %The NH$_3$/CH$_3$OH ratio depends on the cloud physical conditions at the time of the mantle formation (gas density, dust temperature, formation timescale).
NH$_3$/CH$_3$OH ratio versus dust temperature at different prestellar core densities after 0.3 Myr. This ratio depends on the gas density, dust temperature and formation timescale. The colored bands show the observed ratio in 4A2, 4A1 and 4B.
%These can be constrained comparing the observations (colored bands) with model predictions (curves). 
Comparing observations and model prediction is possible to constrain the temperature of the cloud at the time of the ice mantles formation to be $>$17 K \citep{de_simone_tracking_2022}.
\textit{Right:} Comparison between the cyanopolyyne abundances from HC$_3$N to HC$_{11}$N
measured in TMC-1 and L1544 with YEBES and GBT \citep{cernicharo_2020, loomis_2021,cabezas_2022,Bianchi2023a}.}
\label{fig:VLA_results}
%\vspace{-0.5cm}
\end{figure}

In this respect, the pivotal findings of recent VLA, GBT, and YEBES observations are that (see Figure \ref{fig:VLA_results}):
i) dust around protostars can be optically thick enough to completely (or partially) obscure hot corino emission \citep[][]{DeSimone2020}, leading to an underestimation of the molecular abundances estimated so far;
ii) Observations of CH$_3$OH (methanol) and NH$_3$ (ammonia) in protostars can be used as critical tracers of the ice mantle composition to (indirectly) retrieve the physical condition of the cloud at the moment of their formation \citep[][]{de_simone_tracking_2022};
iii) complex carbon species (with at least seven C atoms) are challenging to observe at (sub-)mm wavelengths, but have been found to be abundant in few sources using radio observations \citep{McGuire2020, Cernicharo2021,Bianchi2023a}.
These findings opened a new avenue in the protostellar chemical characterization where the radio wavelength regime is the major actor.\\

However, we are pushing the current facilities to their limit. For example, the only iCOM detected so far at cm wavelength has been methanol (CH$_3$OH), that is actually one of the simplest. The other more complex, such as formamide (NH$_2$CHO), acetaldehyde (CH$_3$CHO), or glycolaldehyde (HCOCH$_2$OH), have lower abundances, so their emission lines lie below the detection threshold. Therefore, detecting more complex and less abundant molecules with respect to methanol and performing an unbiased spectral survey with enough spectral resolution and large bandwidth is still challenging in terms of sensitivity.
ngVLA will provide great angular resolution to explore the inner regions of protostars (including the mid-plane), and great sensitivity to detect heavier and larger molecules as their rotational spectrum is shifted toward lower frequencies. Additionally, in the radio regime, the spectrum will be less affected by line confusion being less populated by rotational transitions of lighter and smaller molecules, making easier the identification of these less abundant and more complex species. 
Moreover, the wide frequency coverage provided by ngVLA will allow to perform large unbiased spectral surveys, going beyond the optically thick dust veil. With a multi-line analysis, it will be then possible to retrieve the real chemical content of young protostellar region, and to understand what forming planets can inherit. 
Finally, a new hunt for amino acids and other pre-biotic species can begin. We will be able, at last, do a comparison with what found in comets and meteorites, the pristine objects of our Solar System, taking a significant step towards understanding the origins of Life.
\subsection{Studies of fast radio Bursts above 2\,GHz with the ngVLA}
\RaggedRight\label{spitler01}
\vspace*{\baselineskip}

\noindent \textbf{Thematic Areas:} \linebreak $\square$ Stellar Astrophysics \linebreak $\square$ Solar System, Planetary Systems and Habitability \linebreak
$\square$ Circuit of Cosmic   Matter (incl. star formation) \linebreak $\square$ The Galaxy and the Local Group \linebreak
  $\square$   Galaxies and AGN \linebreak $\square$  Cosmology, Large Scale Structure and Early Universe \linebreak
  $\checked$    Extreme conditions in the cosmos, fundamental astrophysics    \linebreak
  $\square$ Interdisciplinary research and technology \linebreak
  
\textbf{Principal Author:}

Name:	Laura G. Spitler
 \linebreak						
Institution:  Max-Planck-Institut f\"{u}r Radioastronomie
 \linebreak
Email: lspitler@mpifr-bonn.mpg.de
 \linebreak
 
\textbf{Co-authors:}
Florian Eppel (JMU Würzburg)
\linebreak

Fast radio bursts (FRBs) are radio transients emitted by so-far unidentified sources at cosmological distances \cite{cc19}. Their unknown astrophysical origin (or origins) is currently one of the biggest mysteries in astronomy, although the majority of models involve a neutron star \cite{pww+19}. Furthermore, they are also useful astronomical tools with applications for, e.g., studying the intergalactic medium (e.g. \cite{mpm+20}), measuring cosmological parameters such as the Hubble constant (e.g. \cite{jgp+22}), and testing fundamental physics (e.g. \cite{bem+16}). 
Observationally, FRBs are characterized by micro- to millisecond-timescale emission that has, so far, only been detected at radio frequencies and exhibit complex and varying time-frequency structure and polarization properties \cite{phl22}. A few percent of the known sources have been confirmed to emit multiple bursts, the so-called ``repeaters", which enable a wide-range of follow-up observations \cite{chime23rep, jsn+23}. 

Identifying an FRB's host galaxy is critically important for understanding their astrophysical origins, as well as most of the astrophysical and cosmological applications. Unambiguous localization requires a precision of $\sim$arcsecond, and therefore an interferometric radio telescope is necessary \cite{eb17}. Several blind, FRB surveys are running commensally on interferometers including the Jansky Very Large Array (VLA) \cite{lbb+18} and the Australian Square Kilometer Array Pathfinder (ASKAP) \cite{2010PASA...27..272M}. Meanwhile, new interferometers are being designed and built with the FRBs as a key science case (e.g. DSA-110\footnote{https://www.deepsynoptic.org/}, CHIME/FRB+CHORD \cite{2019clrp.2020...28V}, Square Kilometer Array \cite{mkg+15}). Note, ASKAP and DSA-110 operates solely around 1.4 GHz, while CHIME/FRB+CHORD observes at $400-800~$MHz. What is missing in most current and future observatories is the ability to discover and localize FRBs above $\sim$2~GHz. Beyond targeted follow-up of active repeaters, the nature of FRBs at higher radio frequencies is entirely unexplored. A commensal FRB search system on the ngVLA would help fill in this gap \cite{2018ASPC..517..773L}. The ngVLA will be the only high-frequency radio instrument with a large field of view and, simultaneously, the capability to localize FRBs down to $<$arcsecond precision upon detection. This will enable systematic multi-wavelength (MWL) surveys of FRBs in synergy with current and future X-ray missions, especially for the population of once-off FRBs. MWL information is key in understanding the nature of FRBs, since several FRB models predict associated high-energy emission \cite{pww+19}. Due to the small beam sizes and/or inaccurate localization capabilities of the current generation of telescopes, especially once-off events are difficult to target in today's MWL programs which focus on the few known repeaters \cite{scholz2020,piro2021}. The ngVLA will therefore play a key role in constraining different FRB models and ultimately in revealing the progenitor of FRBs.

Perhaps even more important for FRB science is the ngVLA's widefield, high resolution, high sensitivity imaging capabilities. The presence or absence of radio emission in the host galaxy coincident with the position of the FRB source is an important clue in understanding the source's local environment. Of particular interest are the so-called ``persistent radio sources" (PRS) that have been associated with two repeaters \cite{mph+17,2023arXiv230716355Z}. In the case FRB~20121102A, the diameter of the PSR is $<0.7$~pc and detected over a wide frequency range \cite{clw+17}. The nature of these PRSs is also unclear but may be powered by the interaction of the FRB source with its environment \cite{mm18}. The ngVLA's high sensitivity will allow the detection of weaker PRSs or provide deeper constraints over a wider frequency range than other observatories. 
\subsection{Study of the Dense ISM with the ngVLA}
\RaggedRight\label{toth01}
\vspace*{\baselineskip}

\noindent \textbf{Thematic Areas:} \linebreak $\square$ Stellar Astrophysics \linebreak $\square$ Solar System, Planetary Systems and Habitability \linebreak
$\checked$ Circuit of Cosmic   Matter (incl. star formation) \linebreak $\checked$ The Galaxy and the Local Group \linebreak
  $\square$   Galaxies and AGN \linebreak $\square$  Cosmology, Large Scale Structure and Early Universe \linebreak
  $\square$    Extreme conditions in the cosmos, fundamental astrophysics    \linebreak
  $\square$ Interdisciplinary research and technology \linebreak
  
\textbf{Principal Author:}

Name:	L. Viktor TÓTH
 \linebreak						
Institution:  Eötvös University Budapest, Institute of Physics and Astronomy
 \linebreak
Email: l.v.toth@astro.elte.hu
 \linebreak
 
\textbf{Co-authors:}   Mika Juvela, University of Helsinki
  \linebreak

As a recent evaluation of ISM simulations by \cite{Juvela2022} showed, ngVLA observations may uncover the density and velocity structure of ISM in early phases of massive protostar formation up to 3kpc.  

%Use the \cite command for references, e.g., \cite{Kadler2012}
%Add entries to bibliography.bib

%% MJ

In studies of star formation (SF), many questions remain open regarding the formation and filamentary structures of clouds, their fragmentation, and the mass accretion and the balance between gravity, turbulence, and magnetic fields across different scales \citep{Motte2018,Hacar2023}. One also needs a better understanding of the components of the interstellar medium (ISM) itself, the gas and dust, which are used as tracers of the SF process and which themselves undergo significant evolution along the
SF process. The understanding of the high-mass star formation (HMSF) is particularly important, since massive stars are crucial to many astrophysical processes even at galactic scales, via their ionising radiation, stellar winds, and the production of heavy elements \citep{Kennicutt2005}. High-mass stars
are formed in massive ($M>1000$\,M$_{\odot}$, $\Sigma \sim 0.02$\,g\,cm$^{-2}$) and mainly cold ($T\sim$10–40\,K) clouds \citep{Peretto2010,Tan2014,Lim2016}, which tend to reside at kiloparsec distances. Therefore, one needs sensitive radio observations of dust and of molecules on low excitation
levels. One also needs the coverage of a wide range of size scales, ending up at the scale of protostellar cores (down to disk scales, $\sim$100\,au), possible only with ground-based interferometry.

ngVLA meets many of these requirements. It covers the millimeter range that includes the low-lying transitions of many key molecules, such as ${\rm NH}_3$, ${\rm HCO^{+}}$ and ${\rm N_2H^{+}}$ as well as many deuterated molecules that are abundant in pre-stellar cores (e.g. DCO$^{+}$ and ${\rm N_2D^{+}}$). Although single-dish observations are still needed at the largest scales, the compact ngVLA configurations provide already an almost complete picture of the more compact almost complete picture of the more
compact HMSF regions (Fig.~\ref{fig:simulation}). High sensitivity is also crucial in kinematic studies, to study the mass flow from cloud scales to individual protostars, that require a high velocity resolution of $\sim$0.1\,km\,s$^{-1}$. Simulations have shown that ngVLA can provide an accurate picture of both the mass distribution and the velocity field in HMSF regions \citep{Juvela2022}. For example, for an target at 4\,kpc distance, one obtains a good coverage of the scales $\sim$0.01-0.3\,pc by using just the core antenna configuration, while the more extended configurations can reveal the structure of individual protostellar systems at the disk scales.

\begin{figure}[h]
{
\centering
\includegraphics[width=0.9\textwidth]{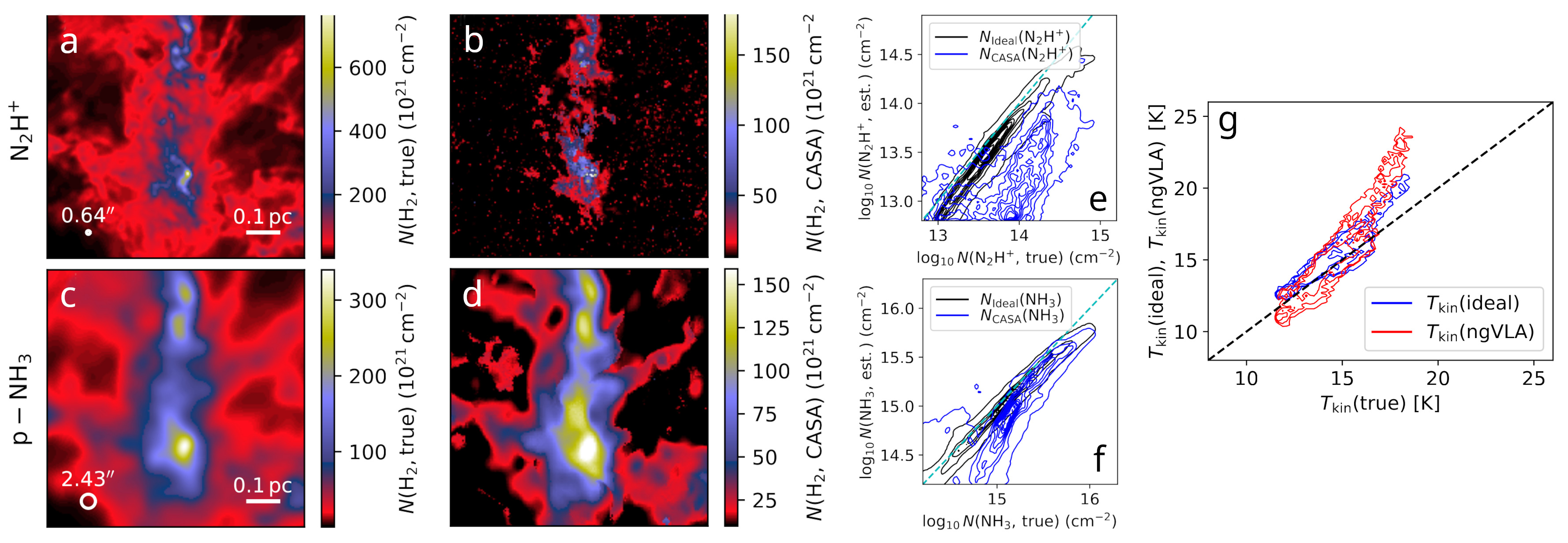}}
\caption{Synthetic ngVLA observations of a massive cloud filament, 6\,h integrations with the core configuration. Shown are line area maps for the true and ngVLA-observed
${\rm N_2H}^{+}$ and ${\rm NH}_3$ spectral lines (frames a-d), the same as correlations
shown for ideal observations (including radiative transfer effects) and synthetic ngVLA observations vs. true values (frames e-f), and the accuracy of ${\rm NH}_3$ kinetic temperature estimates (frame g; for ideal and for simulated ngVLA line observations).}
\label{fig:simulation}
\end{figure}  
\subsection{Star and Planet Formation with ngVLA}
\RaggedRight\label{ueda03}
\vspace*{\baselineskip}

\noindent \textbf{Thematic Areas:} \linebreak $\square$ Stellar Astrophysics \linebreak $\checked$ Solar System, Planetary Systems and Habitability \linebreak
$\checked$ Circuit of Cosmic   Matter (incl. star formation) \linebreak $\square$ The Galaxy and the Local Group \linebreak
  $\square$   Galaxies and AGN \linebreak $\square$  Cosmology, Large Scale Structure and Early Universe \linebreak
  $\square$    Extreme conditions in the cosmos, fundamental astrophysics    \linebreak
  $\square$ Interdisciplinary research and technology \linebreak
  
\textbf{Principal Author:}

Name: Takahiro Ueda
 \linebreak						
Institution: Max Planck Institute for Astronomy
 \linebreak
Email: ueda@mpia.de
 \linebreak
 
\textbf{Co-authors:} Sebastian Wolf, Univ. Kiel, Mario Flock, MPIA, Tilman Birnstiel, LMU M\"{u}nchen
 \linebreak

The process of star and planet formation is a complex and fascinating journey that begins with the gravitational collapse of a molecular cloud core. 
The collapse gives rise to the formation of a disk of gas and dust around the protostar, establishing the initial conditions for planet formation.
\vspace*{2mm}

% ====
\textbf{Star formation in molecular clouds}\hspace*{2mm}
The magnetic field is anticipated to play a pivotal role in the core collapse, with its morphology traceable through the linear polarization of the thermal reemission radiation from aligned dust grains. 
The ngVLA is poised to observe dust polarization, providing insights into 
the magnetic field morphology and strength
on scales of $10^2$ to $10^3$\,au in nearby star-forming regions
along with
dust grain properties such as size and aspect ratio, 
as well as the involved grain alignment efficiency and underlying mechanisms.
The young forming-disks have moved into the focus as planet formation was shown to happen early on.
The small sizes and high optical depths \citep[][]{Ohashi+23} require the capabilities of the ngVLA to reveal the early presence of structure and their origins.

% ====
\textbf{Planet formation in protoplanetary disks}\hspace*{2mm}
%Global structure of circumstellar disks:
%Physical conditions in the future habitable zone around solar-type stars
After disk formation, 
an intriguing focus of the ngVLA lies 
in observing their future habitable zone 
($\sim$ few au for solar-type stars) 
with an angular resolution beyond that achievable with ALMA, 
providing constraints on the physical conditions for terrestrial planet formation.
Its unprecedented resolution allows for probing the months-scale time variability of substructures induced by giant planets like Jupiter \citep[][]{2020A&A...642A.104K}.
In particular, the ngVLA will offer insights complementary to those obtained with optical/infared long-baseline interferometry, tracing the upper disk layers and constraining the vertical disk structure
\citep[e.g., VLTI;][]{2022A&A...658A..81H}.
Furthermore, as the potential planet-forming regions of selected protoplanetary disks were found to be optically thick even in the submillimeter regime
\citep[e.g.,][]{2008ApJ...674L.101W}, the lower optical depth in the wavelength range targeted by the ngVLA will allow for a less ambiguous data analysis.

%ALMA has revealed diverse substructures in the outer regions ($\lesssim10$ au) of disks, including gaps, rings, spirals, and asymmetries \citep{Andrews+18}.  Despite these revelations, our understanding of the future habitable zone in disks remains sparse due to the limited spatial resolution. The inner (sub-)au scale regions have been studied with VLTI using infrared wavelengths which mostly traces disk upper layers. By utilizing the (sub-)cm wavelengths, the ngVLA will observe deep interior layers, demonstrating strong synergy with VLTI (see Science case 2.32 and 2.33).

(Sub-)cm wavelengths are also critical for investigating the dust-size evolution beyond the millimeter size regime. 
For example, Rosetta/MIRO measurements of the comet 67P/Churyumov-Gerasimenko indicate an upper pebble size of 3-6\,mm \citep{buerger-glissmann2023}. 
Given the wavelength range covered by the ngVLA 
it will not only allow tracing this particle size regime,
but also exploit the strong wavelength dependence of the degree and even orientation of the linear polarization on particle size and particle internal structure
\citep{2019A&A...627L..10B}.

%Given that the observed intensity at a given wavelength is most sensitive to the thermal emission from dust whose size is comparable to the observing wavelength, observing disks at cm wavelengths becomes imperative for probing the building blocks of planets and small bodies.

Another pivotal topic involves resolving ambiguities in interpreting substructures in protoplanetary disks. 
While ALMA has shown the ubiquity of these structures, 
their origin is still under debate. Both, the sensitivity to the thermal reemission radiation and thus spatial distribution of millimeter/centimeter-sized pebbles as well as the possibility to constrain the magnetic field via the Zeeman effect will allow distinguishing between the potential roles of instabilities, magnetic fields, or embedded planets 
\citep[see science case 2.14;][]{2017A&A...607A.104B, 2016A&A...590A..17R}. 
Eventually, the long-wavelength emission of circumplanetary disks will provide constraints on the final phase of giant planet formation, i.e., the gas-accretion phase. 

%it remains unclear whether they stem from forming planets or other mechanisms, such as magnetic fields and/or disk instabilities (see Science case 2.14). The distribution of mm/cm-sized pebbles relative to substructures emerges as a key factor in resolving these ambiguities, given that different mechanisms propose distinct dust-size distributions in relation to the substructures (see Science case 2.4).

\textbf{Debris disks: Signposts of the evolution of planetary systems}\hspace*{2mm}
The ngVLA also holds substantial promise for debris disk studies, i.e., the characterization of the planetary systems through the observations of the dust released from comets and minor body collisions. 
As shown by \citep{Stuber+23}, it will provide particularly valuable constraints on the origin of the phenomenon of hot exozodiacal dust \citep{2006A&A...452..237A,2020MNRAS.499L..47K}.
Understanding this phenomenon represents an essential prerequisite for future detection and characterization of exoplanets in the habitable zone using, e.g., the Habitable Exoplanets Observatory (HabEx).
Moreover, the wavelength dependence of the spatial brightness distribution of debris disks will not only provide constraints on the location of planetesimal belts and embedded planets, traced through planet-debris disk interaction, but even on physical properties of the planetesimals themselves, e.g., their collisional strength \citep[][]{2018A&A...618A..38K}.

%particularly in observing hot exozodiacal dust \citep{Absil+13}. Despite ongoing debates about the origin of exozodiacal emission, it is expected to be linked to dust trapping near the central star and/or in situ supply from exocomets. 
%The ngVLA has the potential to detect exozodiacal emission at millimeter wavelengths, offering an opportunity to provide constraints on its origin and contributing valuable insights to this ongoing scientific discourse \citep{Stuber+23}.

\vspace*{2mm}
In summary, the ngVLA stands poised to revolutionize the study of star and planet formation.
By probing magnetic field and grain properties, the ngVLA will unravel the intricacies of gravitational collapse and disk formation. 
Its unprecedented resolution enables exploration of the future habitable zone, providing valuable insights into the conditions during the formation of terrestrial planets like our Earth. 
By resolving ambiguities in interpreting disk substructures and delving into debris disk studies, the ngVLA emerges as a powerful tool, offering new perspectives 
on ongoing, most decisive scientific debates in the field. 
\subsection{Planet Formation at the Innermost Region of Protoplanetary Disks}
\RaggedRight\label{ueda01}
\vspace*{\baselineskip}

\noindent \textbf{Thematic Areas:} \linebreak $\square$ Stellar Astrophysics \linebreak $\checked$ Solar System, Planetary Systems and Habitability \linebreak
$\square$ Circuit of Cosmic   Matter (incl. star formation) \linebreak $\square$ The Galaxy and the Local Group \linebreak
  $\square$   Galaxies and AGN \linebreak $\square$  Cosmology, Large Scale Structure and Early Universe \linebreak
  $\square$    Extreme conditions in the cosmos, fundamental astrophysics    \linebreak
  $\square$ Interdisciplinary research and technology \linebreak
  
\textbf{Principal Author:}

Name: Takahiro Ueda
 \linebreak						
Institution: Max Planck Institute for Astronomy
 \linebreak
Email: ueda@mpia.de
 \linebreak
 
\textbf{Co-authors:} Mario Flock, MPIA,
Tilman Birnstiel, LMU München, 
Sebastian Wolf, Univ. Kiel
  \linebreak

Close-in exoplanets are known to be common around Solar-type stars.
How are they formed? Is their formation process different from that of solar system terrestrial planets?
Understanding the formation of these planets is of great importance in planetary science. 
One preferential site of rocky planetesimal/planet formation is the inner edge of the so-called dead zone \citep{Kretke+09,Ueda+19,Ueda+21}.
The dead zone is the location where magneto-rotational instability (MRI \cite{BH98}) is suppressed because of poor gas ionization \citep{Gammie96}.
The inner edge of the dead zone is characterized as the radial location where disk temperature reaches $\sim1000$ K, above which thermal ionization of the gas is effective enough to activate MRI (e.g., \cite{DT15}).

It is challenging to directly observe the dead-zone inner edge of disks around few-Myr-old T-Tauri stars because it is too close to the central star.
However, disks around more luminous stars (e.g., Herbig Ae stars) or young disks with inner regions heated by disk accretion offer ideal laboratories to test the hypothesis that close-in planets form at the dead-zone inner edge.

Figure \ref{fig:1} (left) shows a simulation of dust evolution around a Herbig Ae star with a stellar luminosity of $56L_{\odot}$ \citep{Ueda+22}.
The high luminosity positions the dead-zone inner edge at $\sim1$ au in the model.
The dead-zone inner edge traps dust particles drifting from outer region because of steep drop in the gas surface density.
The radial dust accumulation results in a mid-plane dust-to-gas mass ratio reaching unity, triggering the formation of rocky planetesimals. 
Figure \ref{fig:1} (right) shows the vertical optical depth at different observing wavelengths.
The optical depth is almost unity at $\lesssim3$ mm, while it is fully optically thin at 10 mm except at the peak.

Figure \ref{fig:2} shows the synthetic images of ngVLA observations for the disk model depicted in Figure \ref{fig:1} \citep{Ueda+22}.
The images include rms noise corresponding to an integration time of 10 hours.
The expected angular resolution is $\sim0.002$ arcsec and $\sim0.006$ at Band 6 (100 GHz) and 4 (30 GHz), respectively.
A distinct ring-like structure is visible at $\sim1$ au with the provided integration time, highlighting the ngVLA's potential as a powerful tool to investigate planet formation in the innermost regions of disks.

\begin{figure*}[h]
\centering
\includegraphics[width=\textwidth]{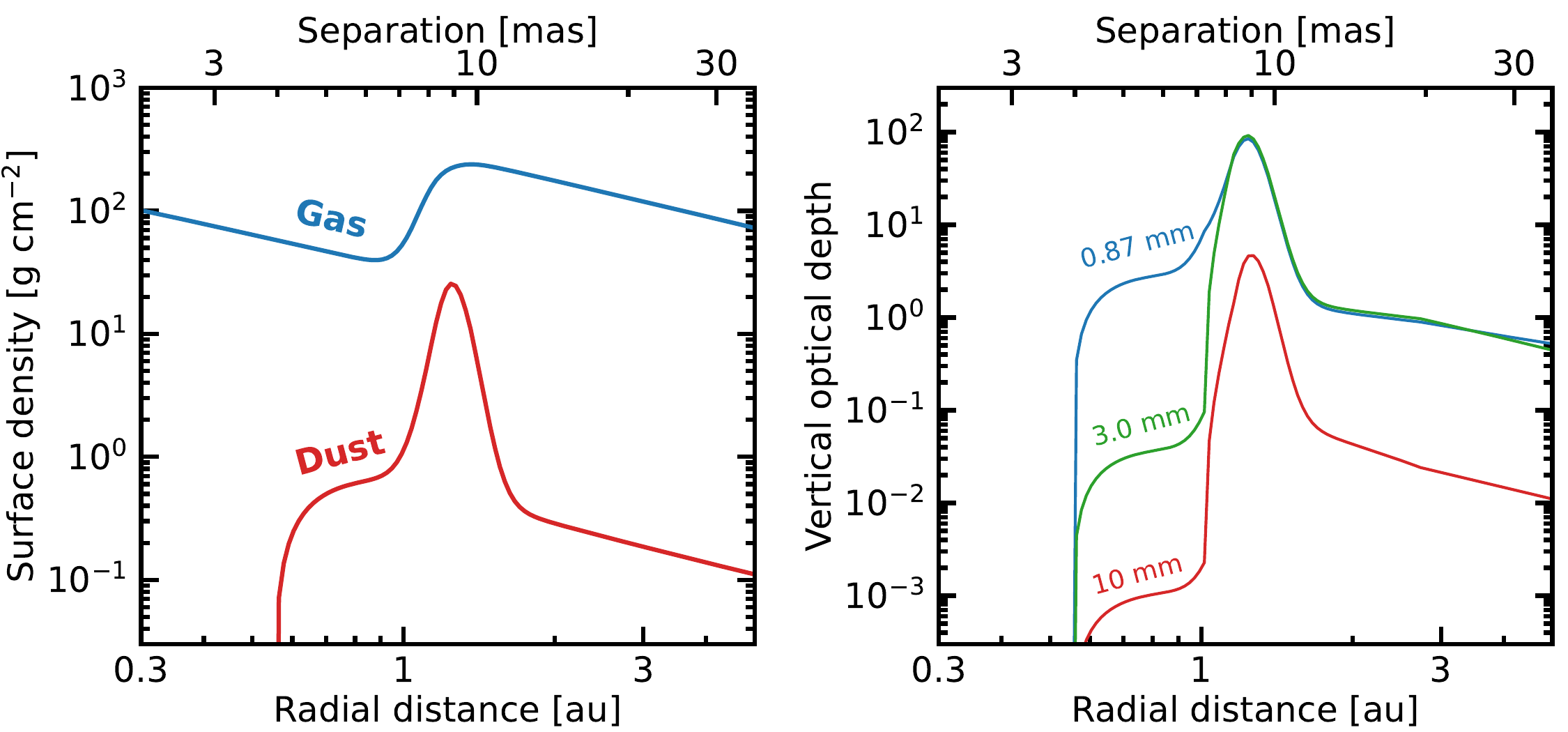}
\caption{
Simulated dust distribution (left) and corresponding optical depth (right) of the disk around a Herbig Ae star with stellar luminosity of $56L_{\odot}$. See \cite{Ueda+22} for more details.
}
\label{fig:1}
\end{figure*}

\begin{figure*}[h]
\centering
\includegraphics[width=\textwidth]{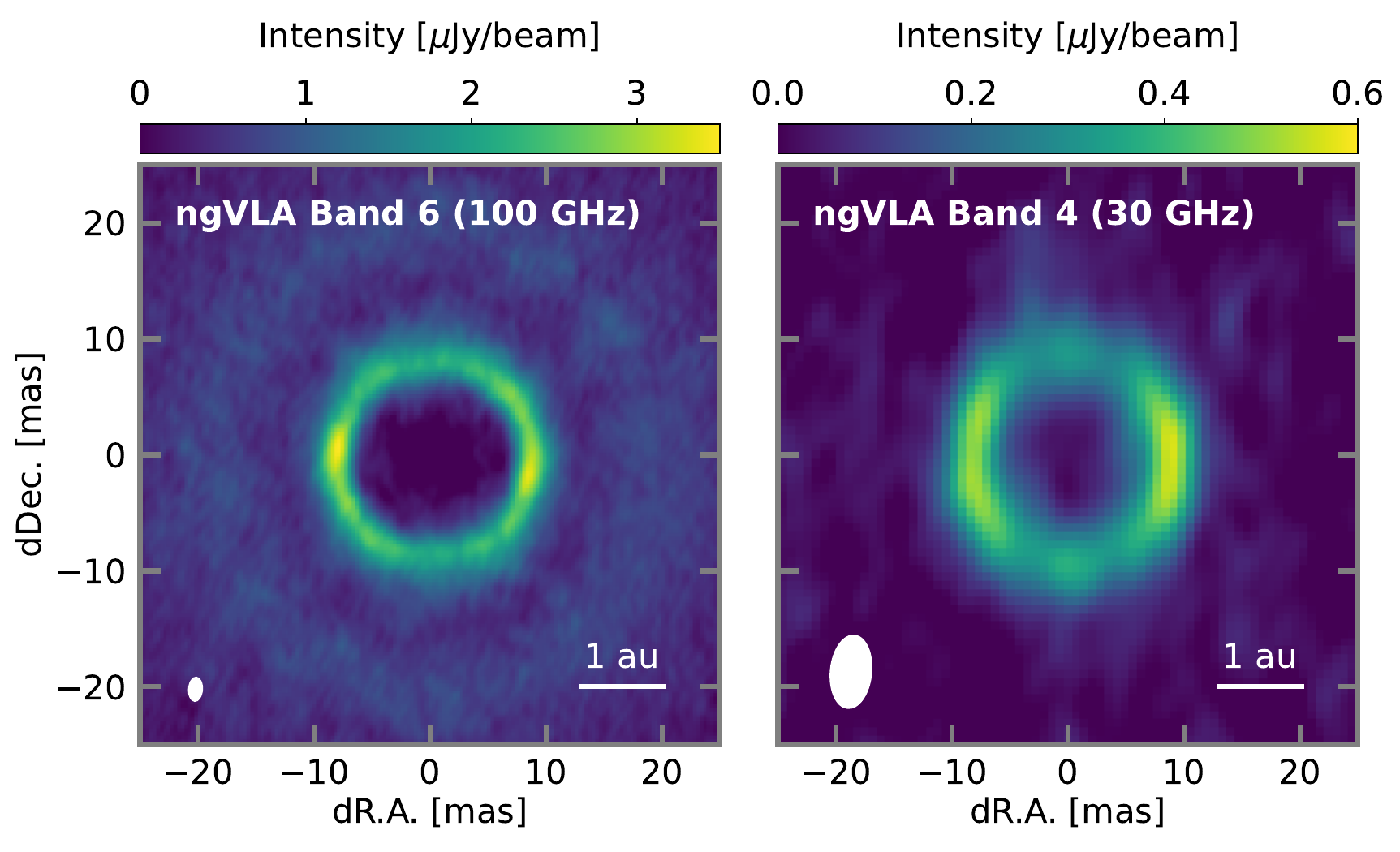}
\caption{
Synthetic images of the ngVLA observations on the disk model shown in Figure \ref{fig:1}. The on-source time is assumed to be 10 hrs.
See \cite{Ueda+22} for more details.
}
\label{fig:2}
\end{figure*}  
\subsection{From near-infrared to centimeter: comprehensive understanding of the inner planet-forming region}
\RaggedRight\label{ueda02}
\vspace*{\baselineskip}

\noindent \textbf{Thematic Areas:} \linebreak $\square$ Stellar Astrophysics \linebreak $\checked$ Solar System, Planetary Systems and Habitability \linebreak
$\square$ Circuit of Cosmic   Matter (incl. star formation) \linebreak $\square$ The Galaxy and the Local Group \linebreak
  $\square$   Galaxies and AGN \linebreak $\square$  Cosmology, Large Scale Structure and Early Universe \linebreak
  $\square$    Extreme conditions in the cosmos, fundamental astrophysics    \linebreak
  $\square$ Interdisciplinary research and technology \linebreak
  
\textbf{Principal Author:}

Name: Takahiro Ueda
 \linebreak						
Institution: Max Planck Institute for Astronomy
 \linebreak
Email: ueda@mpia.de
 \linebreak
 
\textbf{Co-authors:} Mario Flock, MPIA, Tilman Birnstiel, LMU M\"{u}nchen, Sebastian Wolf, Univ. Kiel, Karine Perraut, Univ. Grenoble Alpes
 \linebreak

The inner regions of protoplanetary disks are of great interest in astronomical research. 
The inner regions of these disks have been actively studied by the Very Large Telescope Interferometer (VLTI) operated by the European Southern Observatory.
The VLTI observes the inner disk regions with an effective angular resolution down to approximately 2 milliarcseconds at infrared wavelengths, enabling us to probe the inner edge of dust disks in typical nearby star-forming regions.

One of the pivotal findings of the VLTI observations is that the emission sizes at mid-infrared wavelengths (N band; $10.7~{\rm \mu m}$) are much larger than the expectations derived from the full disk model. 
In contrast, the observed emission size aligns with the model at the near-infrared wavelength (Figure \ref{fig:0} right) \citep{Perraut+19}. 
One plausible interpretation for this inconsistency is dust filtration: forming planets trap large dust, depleting mid-infrared emission within their orbits.
The dust filtration is anticipated to play an important role even in the formation of the Solar system, as geochemical evidence suggests distinct populations among the building blocks of Solar system asteroids, potentially spatially separated by proto-Jupiter \citep{Kruijer+20}.
However, from a theoretical perspective, the occurrence of strong dust filtration remains uncertain due to uncertainties in dust coagulation/fragmentation processes \citep{Stammler+23}.

The ngVLA holds the promise of directly imaging the inner disk structure, as depicted in Figure \ref{fig:0} showcasing synthetic images of a disk  hosting a $10M_{\oplus}$ planet at 1 au ($\sim7$ mas at 150 pc). 
The lower panels show the disk model with the turbulence strength parameter $\alpha_{\rm turb}$ of $10^{-3}$ in which no strong dust filtration occurs.
Conversely, the upper panels exhibit strong dust filtration at $\sim1$ au due to low turbulence ($\alpha_{\rm turb}=10^{-5}$).
The model with dust filtration shows a larger emission size at the N band. 
At ngVLA wavelengths, the expectation is to observe the cavity structure if dust filtration is indeed occurring. 
The ngVLA will offer crucial insights into the presence of inner planets, shedding light on the formation of both the Solar system and exoplanets.

\begin{figure*}[h]
\centering
\includegraphics[width=1\textwidth]{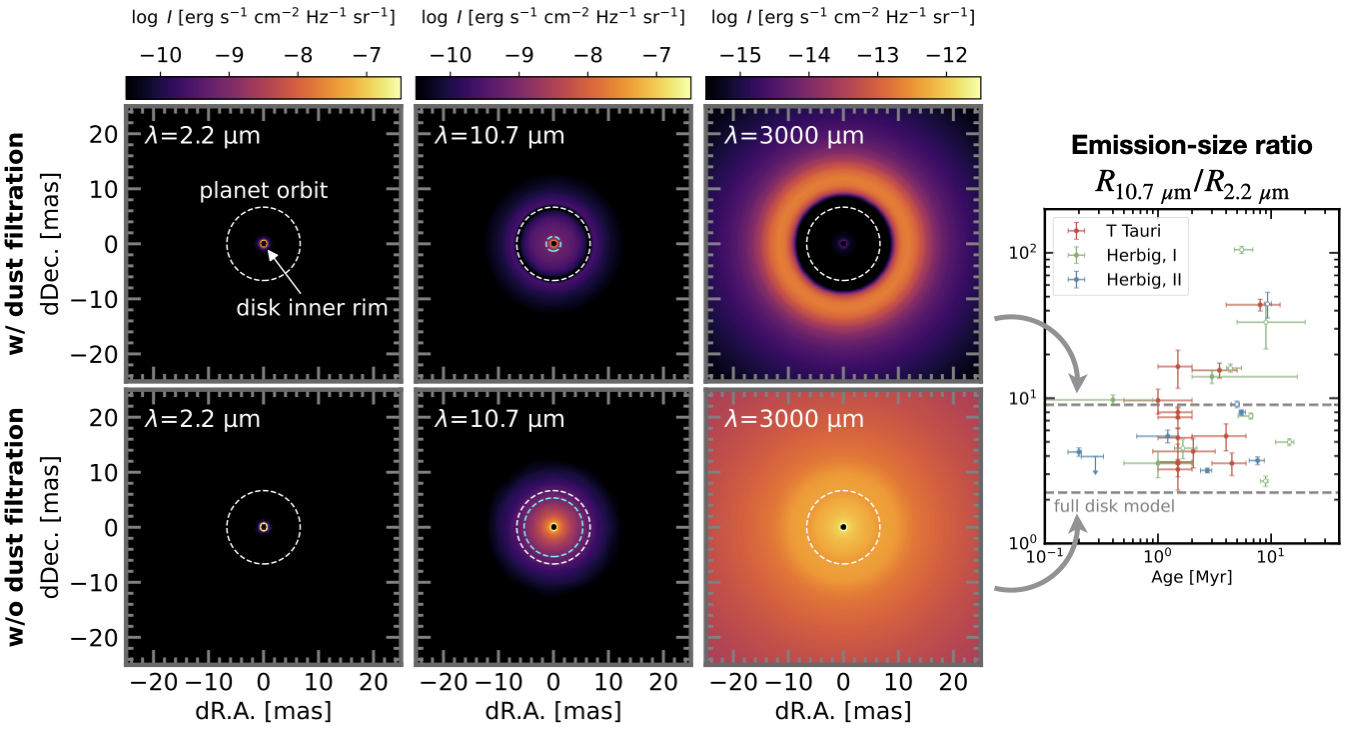}
\caption{
Synthetic images of the inner planet-forming disks with a $10M_{\oplus}$ planet at 1 au.
The upper panels show the case of $\alpha_{\rm turb}=10^{-5}$ in which strong dust filtration occurs at $\sim1$ au, while the lower panels show the case of $\alpha_{\rm turb}=10^{-3}$ where no strong dust filtration occurs.
The white dashed line denotes the planet orbit.
The cyan dashed line in the middle panels denotes the half-emission radius.
The rightmost figure shows the observed ratio between half-emission radii at K ($2.2~{\rm \mu m}$) and N ($10.7~{\rm \mu m}$ ) bands \citep{Menu+15,Varga+18,Perraut+19,Perraut+21}.
}
\label{fig:0}
\end{figure*}

\subsection{Big Data, surveys, classification}
\RaggedRight\label{vardoulaki01}
\vspace*{\baselineskip}

\noindent \textbf{Thematic Areas:} \linebreak $\square$ Stellar Astrophysics \linebreak $\square$ Solar System, Planetary Systems and Habitability \linebreak
$\square$ Circuit of Cosmic   Matter (incl. star formation) \linebreak $\square$ The Galaxy and the Local Group \linebreak
  \checked   Galaxies and AGN \linebreak $\square$  Cosmology, Large Scale Structure and Early Universe \linebreak
  $\square$    Extreme conditions in the cosmos, fundamental astrophysics    \linebreak
  $\checked$ Interdisciplinary research and technology \linebreak
  
\textbf{Principal Author:}

Name:	Eleni Vardoulaki
 \linebreak						
Institution:  Thüringer Landessternwarte Tautenburg
 \linebreak
Email: elenivard@gmail.com
 \linebreak
 
\textbf{Co-authors:} Etienne Bonnassieux, JMU W\"urzburg, Yuri Y. Kovalev, MPIfR
  \linebreak

Radio-astronomy has, in the past decades, worked towards building the infrastructural and technical capacity necessary to reliably create high-resolution (sub-arcsec), high-sensitivity (collecting areas up to a square kilometer), and high-fidelity maps of our Universe. In the Southern hemisphere, this is exemplified in the SKA; the ngVLA will provide a complementary and worthy successor in the Northern hemisphere. The ultimate aim is to generalise, over the entire extent of the sky, the work done over the past decades, from the Cambridge surveys \citep[e.g. ][]{Willott2001} which revealed the large radio sources that inhabit our universe, all the way to multi-wavelength \citep[e.g. COSMOS; ][]{Scoville2007} and large sky-surveys \citep[e.g. FIRST, VLASS, EMU, LoTSS; ][]{Becker1994,Lacy2020, Norris2011,Shimwell2019,Shimwell2022}, that peered into the deep radio universe probing the faint active and dominating star-forming galaxy populations. The ngVLA will provide an unprecedented ability to constrain the underlying physics of this separation, not on ad-hoc observational criteria but physically-significant observational tracers.

This ambition comes, of course, with a cost: namely, the size of the associated data. From antenna voltage correlation, to the computing requirements for data calibration \& reduction, to the storage of the final data-products with the appropriate quality metrics, metadata and compression algorithm, as well as the automation and distribution of as much of the subsequent analysis as possible, the result is the exponential increase of the requirements for processing capabilities (high-performance computers - HPC; federated computing) and storage (order of Peta-to-Exabytes). The above is finally in the context of a paradigm shift in data sharing and knowledge distribution, with science-ready products and code-sharing following open science policies \citep[findability, accessibility, interoperability, and reusability - FAIR; ][]{Wilkinson2016}. 
In this section, we will consider the analysis of Observatory data-products, e.g. the next-generation ngVLA surveys. By extending at least three axes of the existing parameter space (resolution, sensitivity, spectral coverage), the complexity of source-finding, association/discrimination and finally classification is also increased.  
Not only is the classification problem to be solved for a \textit{much greater} source density, the complexity of \textit{each} individual source is increased, including with the 
introduction of new source categories as sensitivity thresholds are reached (e.g. star-formation emission becoming detectable).

To rise to this challenge, three interrelated approaches have proven fruitful to identify radio sources and their host galaxies, and classify them based on their properties: a) statistical methods of population study \cite[e.g.][]{Sabater2019}, which require a pre-existing, reliable multi-wavelength source association catalogue - itself a monumental task; b) machine learning (ML) \cite[e.g. ][]{Aniyan2017, lukic2019convosource, Alegre2022}, vital to rise to the challenge of next-generation (SKA, ngVLA) catalogues with their millions of complex, resolved, multi-category radio sources; and c) citizen science, critical to compile robust training samples, thus enabling discoveries. The future of radio-astronomy lies in the joint deployment of all three approaches.

\subsubsection{Population Statistics:}

With high-quality, multi-wavelength data of large populations of astronomical sources, it is possible to robustly identify meaningful subsets within these populations. Better spectral coverage makes it more likely to discern underlying physical mechanisms \citep{Dickinson2003,Scoville2007,Driver2009,Driver2011,Koekemoer2011,Grogin2011,Smolcic2017b,Blandford2019,Mingo2019, Vardoulaki2019,Vardoulaki2021,Hardcastle2020}. This approach has proved fruitful in panchromatic fields like COSMOS \citep[e.g. ][]{Smolcic2017a,Smolcic2017b,Vardoulaki2019,Vardoulaki2021}, to address key aspects of galaxy evolution, probing back to the epoch of reionisation. These fields' relatively small sky area ($\sim$2 deg$^{2}$) allows for methods of visual inspection to create robust, enhanced catalogues (including science-ready products (e.g. position of radio source and host, fluxes, sizes, host properties, environmental parameters). Reaching deeper sensitivity \citep[2$\mu$Jy/beam, MIGHTEE; ][]{Heywood2022} reveals a key issue in radio source identification and classification: the wealth of unrelated, faint and point-like radio sources around  and within larger, diffuse radio sources impose the need for robust spectro-photometric coverage and efficient source deblending \citep[e.g. super-deblend][]{Jin2018}. Additionally, as the resolution and sensitivity of our surveys increase, new, complex emission becomes detected, while existing point sources might resolve into complex structures themselves. The performance of current-generation automatic algorithms for source identification \citep[e.g. BLOBCAT, PyBDSF; ][]{mohan2015} is thereby mitigated, hence the necessity for machine-learning approaches.
Formation of deep parsec-scale-selected complete samples of AGN \citep[e.g.,][and references therein]{Kovalev2007,Poplov2021} will allow to address pivotal questions related to highly energetic phenomena in AGN cores including jet formation and collimation, energy extraction, particle acceleration and neutrino production, and also aid with AGN/SFG disentangling. 
\textit{Current, now-conventional methods remain vital in this data-intensive radio era. Even approaches as basic as visual inspection are still needed to identify, deblend, assemble and classify more complex radio structures, but quickly show their limits as sample sizes increase. For surveys of the order of hundreds of thousands to million of radio sources \citep[e.g. EMU, LoTSS;][]{Norris2011,Norris2021,Shimwell2019,Shimwell2022,Mingo2019,Hardcastle2023}, the robust training of reliable ML/AI algorithms is crucial.}

\textbf{Astrometry and geodesy:}
A very important aspect of ngVLA long baseline surveys and AGN studies is that they will enable precise astrometry and geodesy measurements, linking radio astronomy and the study of astrophysical objects to interdisciplinary science by providing detailed characterisation of the Earth (e.g. shape, rotation, gravity, tectonic plate motion). VLBI-compact AGN form the basis of the most accurate reference frames in radio \citep{2020A&A...644A.159C} and optical \citep{2023A&A...674A...1G} bands and are used for precise geodesy measurements \citep[e.g.,][]{2009JGeod..83..859P}.
Deep surveys of ngVLA including its long baselines will allow us to significantly extend the sample of compact extragalactic radio sources forming the frame. Detailed imaging information will be used to compensate for effects of AGN structure at parsec-scales which currently provide one of the limitations of the VLBI reference frame accuracy
\citep[e.g.,][]{Xu2022}.

\subsubsection{Machine learning:} Automated algorithmic tools have already been developed for radio source identification and classification \citep[e.g. PyBDSF, Selavy, Aegean, Ceasar, Profound; ][]{mohan2015,Whiting2012,Hancock2012,Riggi2016,Riggi2019,Robotham2018,Hale2019}, but they all face limitations when dealing with complex radio structures, as recently revealed by comparative studies for multi-source finders \citep[Hydra; ][]{Boyce2023a,Boyce2023b}.
Algorithms featuring two-channel identification and classification using radio and optical/infrared observations of the field are promising: they do not require prior training, instead relying on dimensionality reduction and sorting methods. These techniques of unsupervised cataloguing, known as self-organising maps \citep[e.g. PINK; ][]{Polsterer2015, Galvin2020} can efficiently match radio sources to host galaxies, and are very useful for discovering complex radio structures. 
Advances in ML/AI include supervised deep learning convolutional neural networks \citep[CNN, e.g. CLARAN; ][]{Wu2019} which have been used extensively to automatically identify and classify large samples of radio sources. These methods are, however, strongly reliant on robust training samples. 
Although supervised CNN have been used in the field since 2017, they do come with their own limitations and costs, such as re-training and adjusting for different telescopes. 
Self-supervised learning \citep[e.g.][]{Slijepcevic2023} or eXplainable Artificial Intelligence (XAI) techniques \citep[e.g. LIME; ][]{Tang2023}, are on the rise because they aim to address the issues faced by CNNs, i.e. lacking a proper analysis for overfitting \citep{Becker1994}. Given that radio-astronomy applications lie in the weakly supervised regime ($N_{\rm unlabelled} >> N_{\rm labelled}$), learning using the significant quantity of unlabelled data could solve some of the shortcomings in current models - e.g. using physical properties, using both labelled and unlabelled data, and pre-training without labels \citep{Slijepcevic2023}. Additionally, XAI techniques \citep[e.g. LIME][]{Tang2023} can now ensure that the chosen algorithm is optimal for the task at hand.

\textit{In summary, ML/AI methods are a promising tool to handle large samples of radio sources dealing with millions of galaxies. They remain computationally expensive, and future technical work should also focus on optimising their CPU efficiency.
}

\subsubsection{Citizen science:} Complementing our advanced algorithms, visual inspection has proven very valuable to find unexpected and complex sources in large surveys, and valuable in deblending nearby and complex sources: after all, we wish not only to associate and classify, but also handle unexpected discoveries. 

Projects such as the radio galaxy zoo (RGZ) have been very successful in training and engaging citizen scientists (zooters), who classified over 100,000 sources, resulting in 14 publications within 10 years. Their successors for the EMU (Tang \& Vardoulaki in prep.) and LoTSS \citep{Hardcastle2023} will use modern approaches \citep{Bowles2022,Bowles2023,Segal2023} to build the workspaces, aiming to provide an enjoyable experience to the zooters, with the potential for VR developments allowing grater accessibility of ngVLA images \citep{Deg2023}. 
The results form the basis of training samples for millions of radio sources, in addition to science-ready products for research. Such projects have a tremendous educational value, directly involving students in schools and universities via dedicated programs.

\subsubsection{Summary:} To play the classification game one needs a winning hand: a multi-wavelength and multi-frequency approach, based on ML techniques calibrated through citizen science engagement.  
\subsection{Constraining the nature of dark matter with strong gravitational lensing}
\RaggedRight\label{vegetti01}
\vspace*{\baselineskip}

\noindent \textbf{Thematic Areas:} \linebreak $\square$ Stellar Astrophysics \linebreak $\square$ Solar System, Planetary Systems and Habitability \linebreak
$\square$ Circuit of Cosmic   Matter (incl. star formation) \linebreak $\square$ The Galaxy and the Local Group \linebreak
$\square$   Galaxies and AGN \linebreak $\checked$  Cosmology, Large Scale Structure and Early Universe \linebreak
$\checked$    Extreme conditions in the cosmos, fundamental astrophysics    \linebreak
$\square$ Interdisciplinary research and technology \linebreak
  
\textbf{Principal Author:}

Name: Simona Vegetti
 \linebreak						
Institution: Max-Planck-Institut f\"ur Astrophysik, Garching
 \linebreak
Email: svegetti@mpa-garching.mpg.de
 \linebreak
 
\textbf{Co-authors:} Devon M. Powell, MPA
  \linebreak

The nature of dark matter is one of the most fundamental puzzles of contemporary physics, with far-reaching implications, from particle physics to cosmology and astrophysics in general. As of now, we do not know what dark matter is. However, we do know that it is not one of the particles in the Standard Model of particle physics, and several extensions have been proposed \citep{Bertone2018}. 

Different dark matter models make different predictions on the distribution of dark matter on sub-galactic scales. For example, the cold dark matter (CDM) model predicts that the Universe should be clumpy down to sub-solar-mass scales and that a relatively large population of low-mass haloes should exist \citep[e.g.][]{Springel2008}. On the other hand, warm dark matter (WDM) models predict less low-mass structure, and in particular, the presence of a minimum halo mass below which the dark matter distribution should be smooth \citep[e.g.][]{Lovell2020a}. In fuzzy dark matter (FDM, i.e. a form of dark matter made of ultra-light bosonic particles) the dark matter distribution within galaxies is characterised instead by a stochastic and oscillating granular structure \citep[e.g.][]{Hui2021}. Different dark matter models also make different predictions on the structure of low-mass haloes, which are on average less concentrated in WDM and FDM than in CDM. Self-interacting dark matter (SIDM) predicts the existence of very-dense core collapsed objects. Measuring the dark matter distribution on sub-galactic scales is therefore one of the cleanest and most direct way to distinguish between dark matter models. As most of these structures (either in the form of low-mass haloes or FDM granules) are expected to be completely dark, they are best detected and characterised via their gravitational signature on the images of strongly lensed galaxies and radio jets \citep[see][for a review]{Vegetti2023}. 

The sensitivity of observations of strong lens systems containing resolved sources, to the dark matter distribution within the lens and along the line of sight, strongly depends on the angular resolution of the data. The ngVLA in its long baseline configuration ($>8000$ km), will allow us to probe the dark matter distribution on sub-kpc scales and detect dark matter haloes with masses as low as $10^6 M_\odot$, where different dark matter models make significantly different predictions on the number density of these objects. In Fig. \ref{fig:Vegetti-fig2} we show mock ngVLA observations of the gravitational lens system MG0751+2716 for different array configurations. The long baseline configuration provides an angular resolution of a few milli-arcseconds, as required to constrain the dark matter distribution on sub-kpc scales. In Fig. \ref{fig:Vegetti-fig1} we show the subhalo mass function and the expected mass sensitivity of the long baseline configuration of the ngVLA to the presence of these objects. It can be seen that these observations are expected to provide an increase in sensitivity, compared to other existing facilities, by several orders of magnitude.  

\begin{figure}[h!] 
\begin{center}
\includegraphics[width=1.0\textwidth]{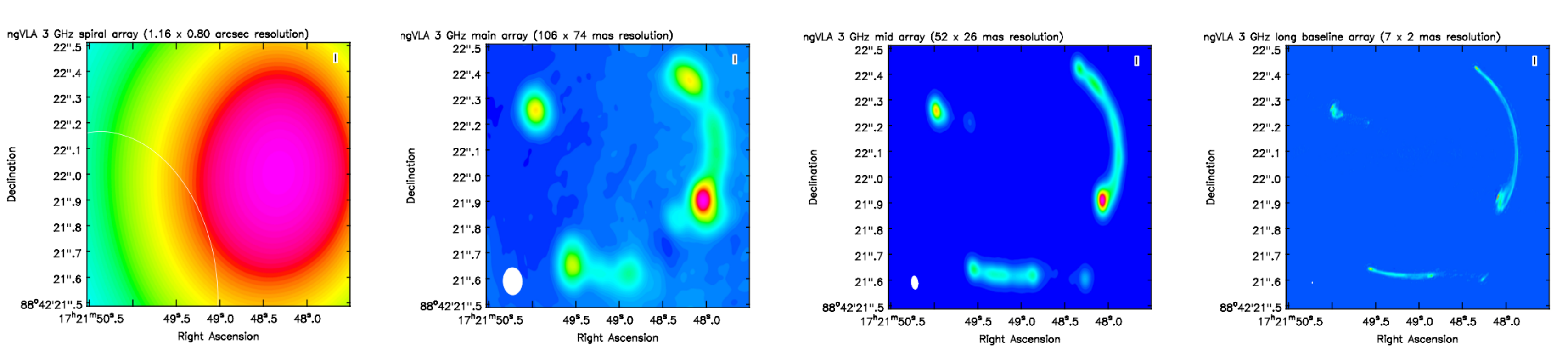} 
\end{center}
\caption{Mock ngVLA observations of the gravitational lens system MG0751+2716 for different array configurations. Credits: images created by A. Lategan (U. Pretoria) and J.~P. McKean (U. Pretoria, RUG)}  
\label{fig:Vegetti-fig2} 
\end{figure}

\begin{figure}[h!] 
\begin{center}
\includegraphics[width=0.8\textwidth]{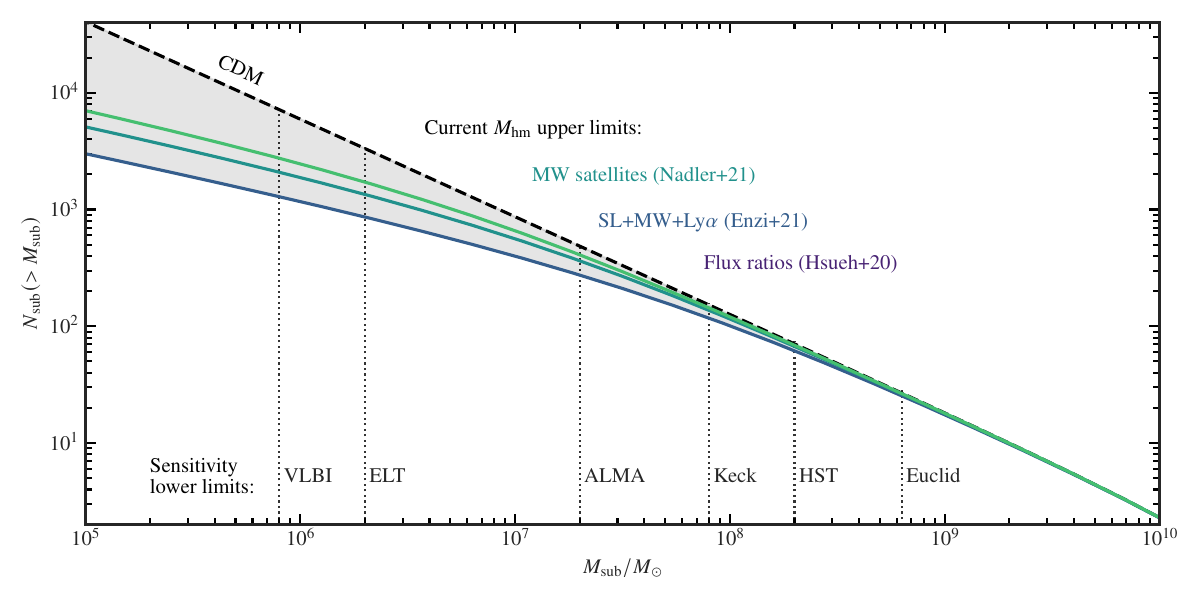} 
\end{center}
\caption{Subhalo mass function predicted from CDM (dashed black line) and still viable WDM models (blue and green solid lines). Vertical lines represents the mass sensitivity from different observing facilities. Very Long Baseline Interferometric observations with the ngVLA are expected to be sensitive to the presence of low-mass haloes with masses as low as $10^6 M_\odot$. Credits: image created by C.~M O'Riordan (MPA).}  
\label{fig:Vegetti-fig1} 
\end{figure}  
\subsection{Circumstellar Nebulae and Mass Loss}
\RaggedRight\label{weis01}
\vspace*{\baselineskip}

\noindent \textbf{Thematic Areas:} \linebreak 
$\checked$ Stellar Astrophysics \linebreak 
$\square$ Solar System, Planetary Systems and Habitability \linebreak
$\checked$ Circuit of Cosmic   Matter (incl. star formation) \linebreak 
$\checked$ The Galaxy and the Local Group \linebreak 
$\square$ Galaxies and AGN \linebreak 
$\square$  Cosmology, Large Scale Structure and Early Universe \linebreak
$\square$    Extreme conditions in the cosmos, fundamental astrophysics  \linebreak
  $\square$ Interdisciplinary research and technology \linebreak
  
\textbf{Principal Author:}

Name:	
Kerstin Weis \linebreak						
Institution:  
Ruhr University Bochum, Faculty of Physics and Astronomy, Astronomical Institute (AIRUB)
\linebreak
Email: kweis@astro.rub.de
 \linebreak
 
\textbf{Co-authors:}
Dominik J. Bomans (AIRUB)
\linebreak

%Insert your white paper text here (max of 1.5 pages including figures and references).
Stellar mass loss and radiation pressure affect the interstellar medium around a massive star from the very beginning of its live to the last phase before going supernovae. This is true for single, double and multiple stars, even if their 
detailed evolution runs somewhat differently.  First, the massive star produces a bubble in the interstellar medium with its strong wind during the main sequence, an 'interstellar bubble' \citep{Weaver1977}.  These are difficult to observe, since most OB stars form in groups or clusters, so that the  interstellar medium becomes rapidly very turbulent due to the energy input of many stars.  In the later stellar evolutionary phases, the mass loss rate and wind velocities change back and forth leading to compression of material from earlier wind phases.  Containing mainly older stellar material, these bubles are circumstellar nebulae.  Luminous Blue Variable (LBV) stars and Wolf-Rayet stars are notorious for their circumstellar bubbles \citep{Weis2020}. LBVs can also show eruptive mass loss (giant eruptions), here outer layers of the stars are ejected and forms a circumstellar nebula \citep{Weis2020}. In all cases these small, circumstellar nebulae trace at least part of the mass loss history of the star (size, mass, expansion velocity, morphology, and chemistry), but also can provide information on the properties of the interstellar medium the star resides in.  
If the star is moving with supersonic speed (as run-away star) even more information 
can be traces due the formation of a distinctive bow-shock shaped nebula \citep{Baalmann2021}.

Due to the small physical sizes (only a few parsecs) of these nebula, high spatial resolution is essential and detailed studies of the nebular are mosty limited to the Milky Way and the Magellanic Clouds.  Here the ngVLA will provide new windows for the study of the nebulae (and therefore stellar evolution) with its high spatial resolution, high sensitivity and the access to high frequencies.  With radio continuum observations the thermal emission of the ionized gas of the nebular will be accessible uneffected by dust absorption, gas kinematics of the nebulae can be studies via radio recombination lines, and via molecular lines.  The current mass loss rate can be directly measured by the free-free emission of the stellar wind \citep{Umana2015}.  Additionally, in the case of run-away stars the possibility of particle acceleration in the bow shocks can be studied by searching the non-thermal emission \citep{vandenEijnden2022} and by tracing the magnetic field with polarisation.
 
A preview (despite of lower resolution and sensitivity is the MeerKAT galactic plane survey \citep{Goedhart2024}, as shown in  Figure \ref{fig_AG_Car}.

\begin{figure}[h]
    \centering
	\includegraphics[angle=0.0,width=0.9\textwidth]{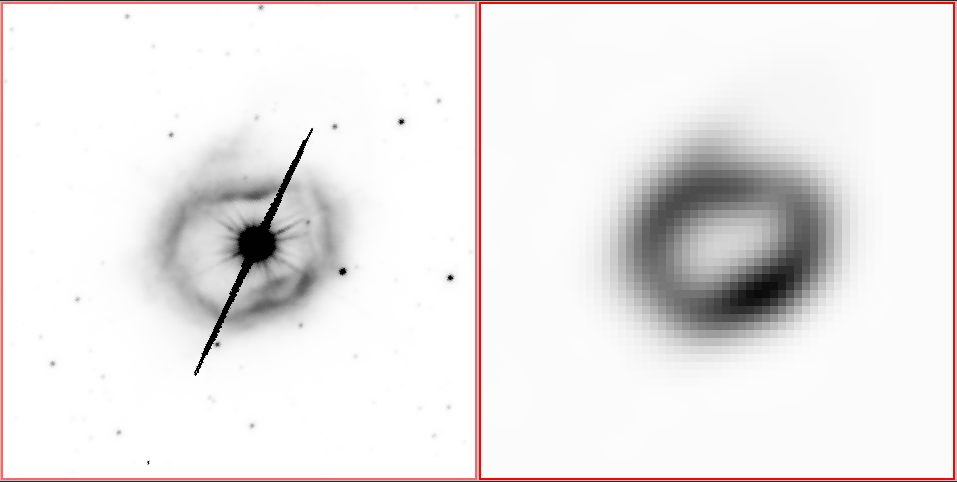}
	\caption{Ground based H$\alpha$ image of the LBV nebulae of AG Carinae (left) extracted from the IPHAS+ survey \citep{Drew2014}, and 1.28 GHz radio continuum image (right) from the MeerKAT galacitic plane survey.  The spatial resolution is about 1 arcsecond for the optical and 8 arcseconds for the radio image.}	\label{fig_AG_Car}
\end{figure}

{\bf Acknowledgements}
\linebreak
KW and DJB acknowledge funding from the German Science Foundation DFG, via the Collaborative Research Center SFB1491 ‘Cosmic Interacting Matters – From Source to Signal’.
\subsection{Sub-Structures in Blazar Jets and Synergetic TeV Observations}
\RaggedRight\label{wendel01}
\vspace*{\baselineskip}

\noindent \textbf{Thematic Areas:} \linebreak $\square$ Stellar Astrophysics \linebreak $\square$ Solar System, Planetary Systems and Habitability \linebreak
$\square$ Circuit of Cosmic   Matter (incl. star formation) \linebreak $\square$ The Galaxy and the Local Group \linebreak
  $\checked$   Galaxies and AGN \linebreak $\square$  Cosmology, Large Scale Structure and Early Universe \linebreak
  $\checked$    Extreme conditions in the cosmos, fundamental astrophysics    \linebreak
  $\square$ Interdisciplinary research and technology \linebreak
  
\textbf{Principal Author:}

Name:	Christoph Wendel
 \linebreak						
Institution:  Julius-Maximilians-Universität Würzburg
 \linebreak
Email: christoph.wendel@uni-wuerzburg.de
 \linebreak
 
\textbf{Co-authors:} Karl Mannheim (JMU Würzburg), Matthias Kadler (JMU Würzburg)
 \linebreak

Jets in active galactic nuclei (AGN) are known to accelerate particles to very high energies as demonstrated by their bright $\gamma$-ray emission up to TeV energies. The mechanism of particle acceleration remains disputed but it is likely to be related to dynamical processes inside an inhomogeneous jet medium. Indeed, stratified or multi-layered jets are readily motivated by theory \cite{1989MNRAS.237..411S} such as the spine-sheath model \cite{Ghisellini2005}, i.e. a fast-moving, highly relativistic inner spine that might be surrounded by a mildly relativistic outer sheath.\\

When jets propagate and expand, the energy stored in the magnetic field carried along the jet can be converted via relativistic magnetic reconnection into kinetic energy of the particles \cite{2013MNRAS.431..355G,2016MNRAS.462.3325P}. Due to tearing instabilities, the reconnection current sheets fragment into plasmoids which confine heated, highly magnetised and compressed jet medium. The ejected plasmoids radiate mainly in the X-ray and TeV regime and are embedded in the mini-jet current sheets along the jet. The superposition of the beamed emission from the small plasmoids is attributed to the background of a flaring period and is thought to be lasting for several hours to days. However, plasmoids can merge and grow to monster plasmoids, whose luminosity is comparable to the entire envelope luminosity with a variability timescale of several minutes. This mechanism provides an explanation for the observed ultra-fast, strong flares on top of broader, weaker envelope emission as shown in the left-hand side panel of Fig. \ref{Figure3C279LightcurveJet} \cite{2020NatCo..11.4176S,2023MNRAS.521L..53A}.\\

Relativistic magnetic reconnection can convert a magnetically dominated jet into a kinetically dominated one, meanwhile changing the velocity and collimation profile of a jet, which is observable with radio interferometry. In such a transition region,  mini-jets with streets of multiple plasmoids should be detectable as separate components due to their high brightness temperatures. However, current very-long-baseline-interferometry arrays lack either the angular resolution or the dynamic range (cf. right-hand side panel of Fig. \ref{Figure3C279LightcurveJet}). The ngVLA will provide superior information in this respect to study such filamentary mini-jet sub-structures at the energetic transition region of AGN jets.\\

In rare cases, it is already today possible that high-quality spectra at high or very high energies can put constraints on the origin of the $\gamma$ rays in blazar jets, e.g. by  searching for spectral cut-offs due to absorption or for narrow spectral bumps or dips due to pair cascades \cite{2021ApJ...917...32W}. Major advances in terms of TeV $\gamma$-ray sensitivity and energy coverage will come from the imminent Cherenkov Telescope Array \citep[CTA; ][]{CTAScienceCase,diPierro2019}. Future coordinated ngVLA and CTA observations can thus reveal a causal connection of ultra-fast TeV variability and mini-jet activity in the transition region and locate the TeV $\gamma$-ray emission region of blazar jets as well as inform about the dominant radiation and particle acceleration processes.\\

\begin{figure}[h]
%\vspace{-10pt}
\centering
\includegraphics[width=\textwidth]{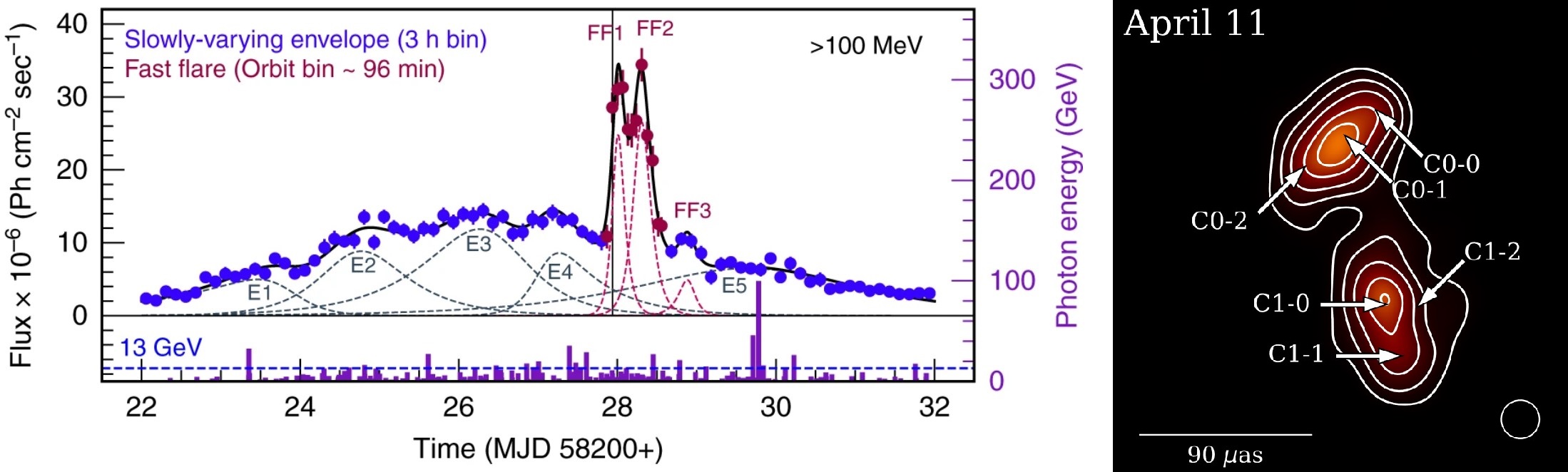}
\centering
 \vspace{-1.5\baselineskip}
\caption{Left: $\gamma$-ray lightcurve of a flaring episode of 3C\,279 detected by the \textit{Fermi Large Area Telescope} in April 2018 \cite{2020NatCo..11.4176S}. A strong minute-scale flare arises out of the envelope emission. Right: Event-Horizon-Telescope image of 3C\,279  with compact sub-components marked \cite{2020A&A...640A..69K}. At comparable effective angular resolution (lower frequency but higher signal-to-noise ratio), the ngVLA will be able to measure also the low-intensity regions (including their polarisation) between the components.}
\label{Figure3C279LightcurveJet}
%\vspace{-10pt}
\end{figure}

\subsection{Antenna positions with a sub-mm repeatability}
\RaggedRight\label{xu01}
\vspace*{\baselineskip}

\noindent \textbf{Thematic Areas:} \linebreak $\square$ Stellar Astrophysics \linebreak $\square$ Solar System, Planetary Systems and Habitability \linebreak
$\square$ Circuit of Cosmic   Matter (incl. star formation) \linebreak $\square$ The Galaxy and the Local Group \linebreak
  \checked   Galaxies and AGN \linebreak $\square$  Cosmology, Large Scale Structure and Early Universe \linebreak
  $\square$    Extreme conditions in the cosmos, fundamental astrophysics    \linebreak
  $\checked$ Interdisciplinary research and technology \linebreak
  
\textbf{Principal Author:}

Name: Ming Hui Xu
 \linebreak						
Institution:  German Research Centre for Geosciences GFZ
 \linebreak
Email: minghui.xu@gfz-potsdam.de
 \linebreak
 
\textbf{Co-authors:} Simone Bernhart, Reichert GmbH \& MPIfR, Bonn, Hayo Hase, BKG Wettzell, U. Hugentobler (TUM)
      \linebreak

Observations from ngVLA will be able to provide geodetic accuracy at the sub-mm level so that the antennas are connected to each other with an accuracy better than 1 mm. For the densely distributed antennas in the ngVLA network, two major advantages can be used to achieve this high accuracy: 1) phase delays can be routinely obtained and used for geodetic purposes, and 2) systematic errors due to troposphere and ionosphere are largely reduced. The results based on this idea are shown in Fig. \ref{fig:wettzell} for the 120-meter-long baseline between \texttt{WETTZ13N} and \texttt{WETTZELL}, demonstrating the repeatability at the sub-mm level \cite{2023JGRB..12825198X}. The ngVLA network can be entirely connected by using only short baselines, and a dedicated solution of these short-baseline observations will allow us to determine their positions with the sub-mm repeatability.

Because the effects related to source structure are one of the dominating systematic errors in geodetic VLBI observations \cite{2021JGeod..95...51X}, another major topic for geodesy is to study, monitor and model these effects to have really point-like anchors in the sky, by using both the high-quality image products and the astrophysical knowledge of modeling the structure of AGNs from ngVLA.

The invariant points of the ngVLA antennas and the axis offsets (between azimuth and elevation axes) will need to be studied and calibrated based on geodesy. More opportunities in developing geodesy will be available if there will be joint observations of ngVLA with existing global geodetic networks to contribute to the global terrestrial reference frame (TRF). One could select a sub-network of ngVLA to regularly participate in geodetic observations, and they could be core stations in the future TRF.

Geodetic positions of the antennas in ngVLA will provide an essential support for its high-accuracy astrometric/astronomical applications where positions and motions of the celestial sources are of concern.

\begin{figure}
    \centering
    \includegraphics[width=0.455\textwidth]{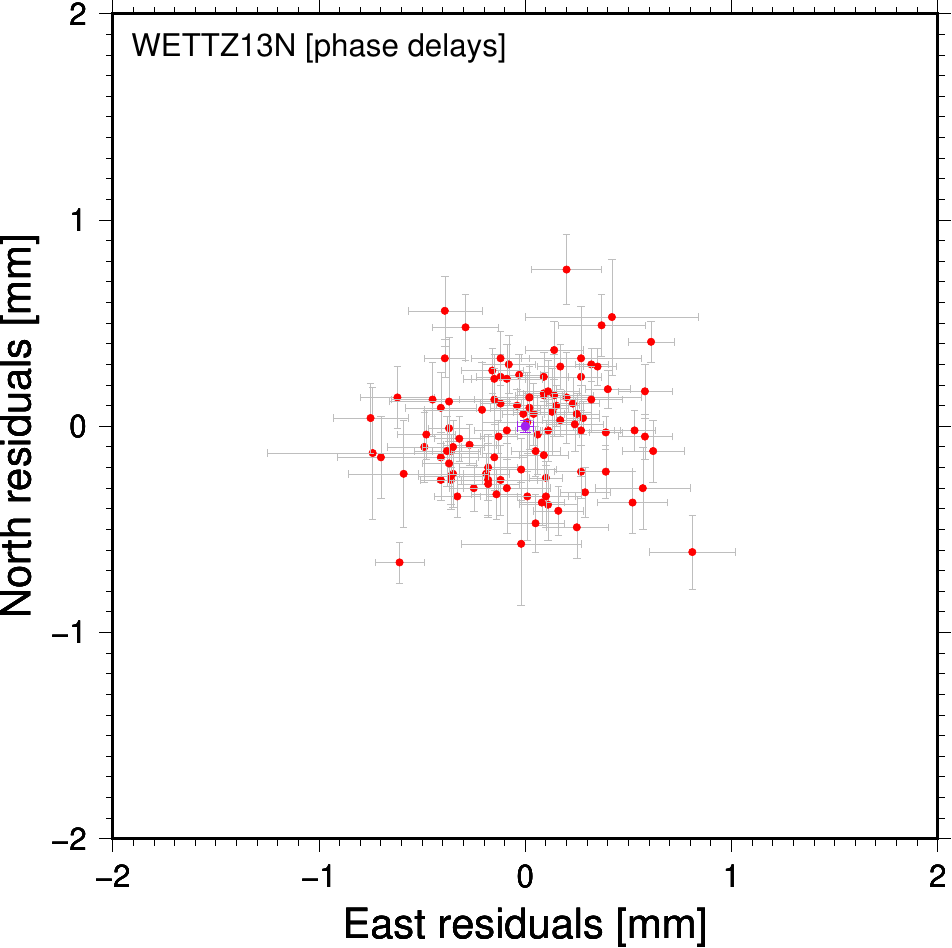}
    \includegraphics[width=0.53\textwidth]{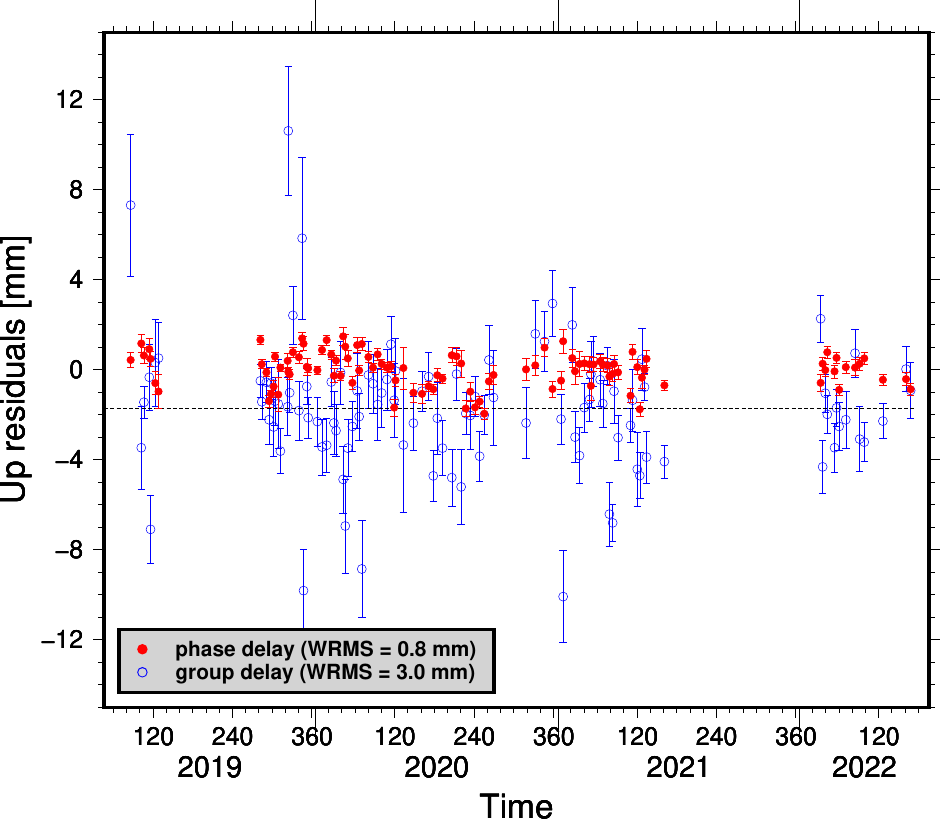}
    \caption{Relative position residuals of antenna \texttt{WETTZ13N} with respect to \texttt{WETTZELL} from three-year VLBI observations. Left: North coordinates v.s. East coordinates. Right: Up coordinates from phase delays (red) and group delays (blue).}
    \label{fig:wettzell}
\end{figure}

\subsection{High-precision $\lambda$-astrometry at mm-wavelengths with the ngVLA}
\RaggedRight\label{zhao01}
\vspace*{\baselineskip}

\noindent \textbf{Thematic Areas:} \linebreak $\square$ Stellar Astrophysics \linebreak $\square$ Solar System, Planetary Systems and Habitability \linebreak
$\square$ Circuit of Cosmic   Matter (incl. star formation) \linebreak $\square$ The Galaxy and the Local Group \linebreak
  $\checked$   Galaxies and AGN \linebreak $\square$  Cosmology, Large Scale Structure and Early Universe \linebreak
  $\square$    Extreme conditions in the cosmos, fundamental astrophysics    \linebreak
  $\square$ Interdisciplinary research and technology \linebreak
  
\textbf{Principal Author:}

Name: Guang-Yao Zhao
 \linebreak						
Institution: Max-Planck-Institut f\"ur Radioastronomie, Bonn
 \linebreak
Email: gyzhao@mpifr-bonn.mpg.de
 \linebreak
 
\textbf{Co-authors:}  Andrei Lobanov, MPIfR
 \linebreak

 Very long baseline interferometry (VLBI) at millimetre wavelengths is more challenging than their centimetre counterparts, largely due to the limitations on the coherence time imposed by the turbulent troposphere~\citep[e.g., see review by][]{dodson2017}.
while the ngVLA long baseline array (LBA) will greatly improve the sensitivity of VLBI observations at frequencies up to 116 GHz, achieving astrometry using the conventional phase-referencing method would still be unreachable at the highest frequencies, because this would require fast switching between sources within a cycle shorter than the coherence time. 
To mitigate such an issue, a dedicated method, so-called source-frequency phase referencing (SFPR), has been established~\citep{rd2011}. 
A typical SFPR calibration includes a frequency-domain phase solution transfer (FPT), i.e., using a centimetre observation to calibrate a close-in-time millimetre observation~\citep{middelberg2005}. 
Recent successful observations have demonstrated that FPT and SFPR are ideal for extending the coherence time and achieving relative astrometry at millimetre wavelengths~\citep[e.g.,][]{rioja2015, RD2020, yoon2018, zhao2019, jiang2018}. 

SFPR requires observations at two or more frequencies to be carried out (quasi-) simultaneously. This can be achieved by implementing the "quasi-optical" multi-band receivers~\citep[e.g. the Korean VLBI Network][]{rioja2015,zhao2019}, fast frequency switching \citep[e.g., with the VLBA][]{rd2011, jiang2018}, or via the paired antenna method (subarraying)~\citep[e.g.,][]{asaki1998, jung2015}. 
The technical design of the ngVLA should be feasible to perform fast frequency switching or subarraying, and thus enable SFPR observations. Together with the high angular resolution and high sensitivity of the ngVLA, this will enable astrometric measurements with an unprecedented level of precision~\citep[e.g.,][]{dodson2017}.
These measurements will have important impacts on a number of scientific areas including AGN and jet physics, stellar astrophysics, transient phenomenon, etc~\citep[see e.g.,][]{2023arXiv230604516D} 

Below we list a few scientific applications in studying AGNs and relativistic jets:
\begin{compactitem} 
\item {\bf Linking black holes to their relativistic jets:} Thanks to the high dynamic range of the ngVLA images, it is possible to detect the black hole shadows and photo rings in nearby galaxies at the higher frequencies of the ngVLA~\citep[][]{2005astro.ph..3225L, lobanov2024}. The emissions from the relativistic jets are usually more prominent at lower frequencies and show steep spectra. Astrometry across frequency bands (i.e., $\lambda$-astrometry) using SFPR will enable robust alignments of the images of the black hole shadow and the relativistic jet. Such alignments would be critical to discern between different theoretical models for jet launching in the vicinity of the supermassive black holes (SMBH), whether the outflows are anchored from the magnetosphere of the black hole~\citep{Bla77} or from the accretion disk~\citep{Bla82}.

\item {\bf Studying the magnetic field structure in the ultracompact AGN jets:} The VLBI core of an AGN jet often corresponds to the $\tau$=1 surface at a given frequency. Due to opacity effects, the location of the core changes with frequency~\citep[e.g.,][]{kovalev2008}. Measuring this effect, the so-called core shift could provide information on the physical conditions (e.g., magnetic field strength, pressure) in the ultracompact jet~\citep{Lobanov1998}. Besides, robust alignment of images at different frequencies is required to perform Faraday Rotation studies, which is key to probing the magnetic field geometry in the jet~\citep[e.g.,][]{dodson2017}.

\item {\bf Detection of close binary SMBH systems: } 
 Binary SMBHs are natural products of hierarchical galaxy formation.  Close binaries with less than a parsec separation would be strong nano-Hz gravitation wave emitters~\citep[e.g.,][]{dorazio2023}. Yet such systems are difficult to identify, especially when only one SMBH in the system is radio loud. In this case, the orbital motion of the binary system is still expected to leave signatures in the astrometric measurements~\citep[][]{dodson2017,Jiang2023}. The high-accuracy astrometric results of the ngVLA at high frequencies could allow the identification of binary SMBH candidates that will be potential targets of the next-generation gravitation wave detectors. Detailed follow-up studies of binary SMBHs will be helpful in understanding how galaxies evolve~\citep[e.g.,][]{dodson2017,2023arXiv230604516D}.
\end{compactitem}

\section{Bibliography}

\bibliographystyle{mkbib}
\bibliography{Bibliography/foreword,Bibliography/agarwal01,Bibliography/baczko01,Bibliography/beuther01,Bibliography/bigiel01,Bibliography/birnstiel01,Bibliography/boccardi01,Bibliography/bomans01,Bibliography/boogaard01,Bibliography/boogaard02,Bibliography/braun01,Bibliography/britzen01,Bibliography/brueggen01,Bibliography/brunthaler01,Bibliography/caselli01,Bibliography/elsaesser01,Bibliography/vonFellenberg01,Bibliography/flock01,Bibliography/fromm01,Bibliography/fromm02,Bibliography/fuhrmann01,Bibliography/hartogh01,Bibliography/hoeft01,Bibliography/kadler01,Bibliography/kadler02,Bibliography/keenan01,Bibliography/kovalev01,Bibliography/kreckel01,Bibliography/kovalev02,Bibliography/livingston01,Bibliography/lobanov01,Bibliography/lobanov02,Bibliography/mueller_mus01,Bibliography/riechers01,Bibliography/riechers02,Bibliography/ros01,Bibliography/schilke01,Bibliography/deSimone01,Bibliography/spitler01,Bibliography/toth01,Bibliography/ueda01,Bibliography/ueda02,Bibliography/ueda03,Bibliography/vardoulaki01,Bibliography/vegetti01,Bibliography/weis01,Bibliography/wendel01,Bibliography/xu01,Bibliography/zhao01}

\end{document}